%% file: main.tex
\let\mypdfximage\pdfximage
\def\pdfximage{\immediate\mypdfximage}

\documentclass[letterpaper, 11pt]{article}

\usepackage{amsmath}
\usepackage{amssymb}
\usepackage{amsthm}
\usepackage[font=small]{caption} \usepackage{epsfig}
\usepackage{geometry}
\usepackage{graphicx}
\usepackage[colorlinks=true, citecolor=blue, filecolor=black, linkcolor=red, urlcolor=blue]{hyperref}
\usepackage{amsfonts}
\usepackage{epstopdf}
\usepackage{color}
\usepackage[textsize=tiny]{todonotes}

\usepackage{soul} \setstcolor{red}
\setulcolor{blue}

\usepackage{lipsum}
\usepackage{algorithm}
\usepackage{algpseudocode}
\algrenewcommand\algorithmicrequire{\textbf{Input:}}
\algrenewcommand\algorithmicensure{\textbf{Output:}}

\ifpdf
  \DeclareGraphicsExtensions{.eps,.pdf,.png,.jpg}
\else
  \DeclareGraphicsExtensions{.eps}
\fi

\usepackage[textsize=tiny]{todonotes}
\usepackage{subcaption}
\usepackage[cmyk]{xcolor}

\usepackage[capitalise,nameinlink,noabbrev]{cleveref} 
\input{macros}

\title{Mean--Variance Risk-Aware Bayesian Optimal Experimental Design for Nonlinear Models}

\author{
    Wanggang Shen\footnote{\href{mailto:wgshen@umich.edu}{wgshen@umich.edu}, University of Michigan, Ann Arbor, MI 48109, USA.},
    Xun Huan\footnote{\href{mailto:xhuan@umich.edu}{xhuan@umich.edu}, University of Michigan, Ann Arbor, MI 48109, USA. \href{https://uq.engin.umich.edu}{https://uq.engin.umich.edu}}}

\date{}

\begin{document}

\maketitle

\begin{abstract}
   We propose a variance-penalized formulation of Bayesian optimal experimental design for nonlinear models that augments the classical expected utility criterion with a penalty on utility variability, yielding a mean--variance objective that promotes robust experimental performance. To evaluate this objective, we develop Monte Carlo estimators for the expected utility, its second moment, and the resulting utility variance using prior sampling, thereby avoiding explicit posterior sampling. We then derive leading-order bias and variance expressions using conditional delta-method arguments. The objective is optimized using Bayesian optimization with common random samples to reduce noise. Numerical examples, including a linear-Gaussian benchmark, a nonlinear test problem, and contaminant source inversion in diffusion fields, demonstrate that the proposed approach identifies designs with substantially reduced variability while maintaining competitive expected utility.
\end{abstract}

\textit{Keywords:}
robust experimental design, information gain, utility variance, Monte Carlo methods

\vspace{1em}
\textit{MSC codes:}
62K05, 62F15, 65C05, 90C15, 62P30

\input{sections/1_introduction}
\input{sections/2_formulation}
\input{sections/3_methodology}
\input{sections/4_results}
\input{sections/5_conclusions}

\section*{Acknowledgments}
This research is based upon work supported in part by the U.S. Department of Energy, Office of Science, Office of Advanced Scientific Computing Research, under Award Number DE-SC0021398. 
This work relates to Department of Navy award N000142512411 issued by the
Office of Naval Research.
This research was supported in part through computational resources and services provided by Advanced Research Computing at the University of Michigan, Ann Arbor.

The authors acknowledge the use of ChatGPT (OpenAI) for assistance in editorial improvements of existing author-generated content. All technical content, theoretical contributions, experimental results, and scientific insights remain entirely the work of the human authors. 

\input{sections/6_appendix}

\bibliographystyle{abbrv}
\bibliography{references}

\end{document}

%% file: macros.tex
\definecolor{darkred}{rgb}{.7,0,0}

\definecolor{darkgreen}{rgb}{.15,.55,0}

\definecolor{darkblue}{rgb}{0,0,0.7}

\newcommand{\design}{\xi}
\newcommand{\pdf}{p}

\newcommand{\param}{\theta}
\newcommand{\Param}{\Theta}

\newcommand{\Var}{\mathbb{V}\text{ar}}

\newcommand{\tr}{\text{tr}}
\newcommand{\argmax}{\operatornamewithlimits{argmax}}

\newcommand{\RR}{\mathbb{R}}

\crefformat{equation}{#2(#1)#3}
\Crefformat{equation}{#2(#1)#3}

\newtheorem{remark}{Remark}
\crefname{remark}{Remark}{Remarks}
\Crefname{remark}{Remark}{Remarks}

%% file: sections/1_introduction.tex
\section{Introduction}
\label{sec:roed_intro}

Experiments play a central role in scientific discovery. However, conducting experiments and acquiring data can be costly and time-consuming, and not all experiments provide the same amount of information. Optimal experimental design (OED) (see, e.g.,~\cite{Huan2024} for a recent review) provides a systematic framework for selecting experiments that maximize their value with respect to a specified criterion. In Bayesian OED~\cite{Chaloner1995,Ryan2015,Alexanderian2021,Strutz2023,Rainforth2024}, this objective typically follows a decision-theoretic formulation and is expressed as the \emph{expected utility} of an experiment, often chosen to be the expected information gain (EIG)~\cite{Lindley1956} obtained from the resulting observations.

Maximizing expected utility has become the dominant paradigm in Bayesian OED. The expectation is taken with respect to the distribution of unknown parameters and future observations, reflecting the fact that experimental outcomes are inherently random and cannot be known \textit{a priori}. While this formulation yields designs that are optimal on average, it does not account for the variability of the utility across different possible experimental outcomes. A design with high expected utility may therefore still carry a substantial risk of producing a very low utility realization once the experiment is conducted.

\begin{figure}[htbp]
  \centering
  \includegraphics[width=0.65\linewidth]{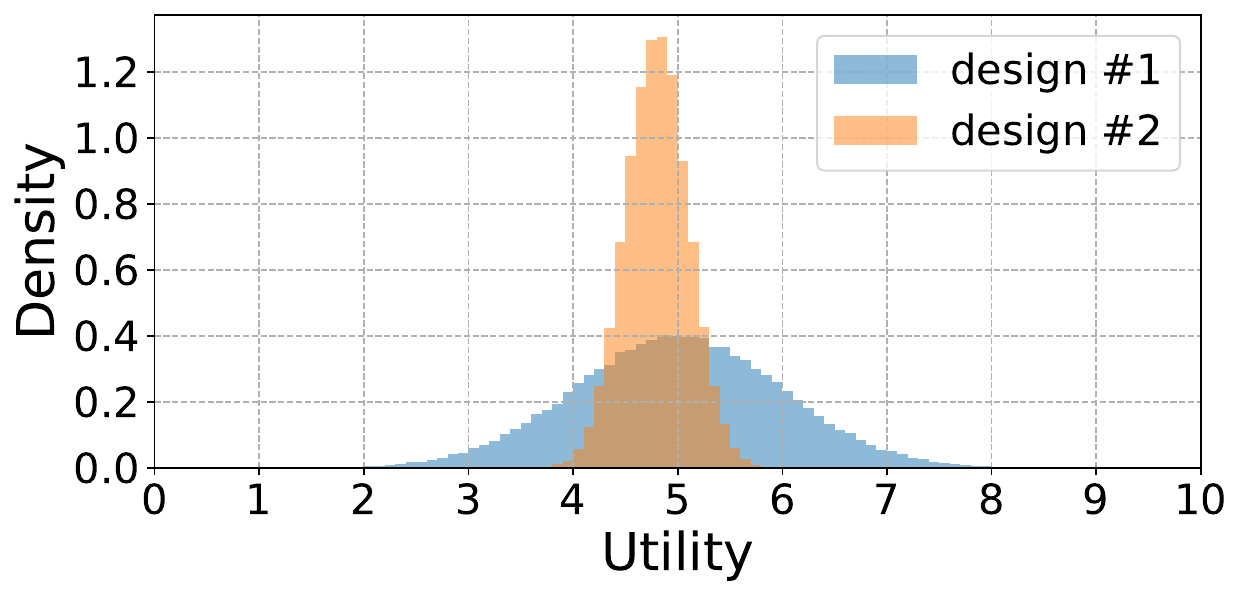}
  \caption{Utility distributions at two example designs.}
  \label{fig:roed_illustrative}
\end{figure}

\Cref{fig:roed_illustrative} illustrates this issue via two example designs. Design~1 has a slightly higher expected utility than design~2, making it preferable under the conventional expected-utility criterion. However, design~1 exhibits much larger variance in utility. Consequently, there is a non-negligible probability that the realized utility may be significantly lower (or higher) than that of design~2. In contrast, design~2 produces more stable outcomes with substantially lower variability. In practice, particularly when experiments are expensive, such stability may be desirable even at the cost of a modest reduction in average utility.

A large body of work in Bayesian OED has focused on improving the estimation of the EIG for complex nonlinear models. A commonly used approach is the nested Monte Carlo (NMC) estimator~\cite{Ryan2003}, with further improvements through techniques such as importance sampling~\cite{Beck2018,Feng2019,Englezou2022}, surrogate models~\cite{Huan2013,Huan2014,Duong2023}, variational bounds~\cite{Foster2019,Foster2020,Kleinegesse2020}, and Gaussian approximations of the posterior~\cite{Long2013,Overstall2017}. These developments have significantly improved the computational tractability of Bayesian OED.

Despite these advances, most existing work continues to focus exclusively on maximizing expected utility and does not explicitly address the variability of experimental utility. In practice, the randomness of experimental observations induces variability in the realized information gain, which can lead to highly uncertain outcomes even when designs are optimal in expectation.

These observations motivate a \emph{risk-aware} formulation of OED that accounts not only for the expected utility but also for the dispersion of its possible realizations. A wide variety of risk measures have been proposed for quantifying undesirable outcomes; see \cite{Royset2025} for a recent review. Examples include mean–variance criteria~\cite{Markowitz1952}, deviation-based measures, quantiles and superquantiles (value-at-risk and conditional value-at-risk (CVaR)), worst-case risk, and entropic risk; see \cite{Rockafellar2013} and \cite[Chapter~6]{Shapiro2021}. 

Risk-aware formulations have received relatively limited attention in experimental design. 
Early work such as~\cite{Valenzuela2015} considers coherent risk measures applied to scalar functions of the Fisher information matrix under prescribed parameter uncertainty, without posterior updating. 
Similarly,~\cite{Kusumo2022} evaluates both the expectation and CVaR of the Fisher information matrix induced by a prior distribution, leading to a bi-objective formulation that captures the trade-off between average performance and risk; however, it remains within a pre-posterior framework.
A related direction is explored in~\cite{Kouri2022}, which introduces an $R$-optimality criterion to control tail risk in predictive variance across the input domain. This approach is rooted in classical experimental design and does not involve general Bayesian utility functions or posterior-based criteria.

More recent efforts have begun to incorporate risk measures within a Bayesian OED framework. 
In particular, a series of presentations and reports~\cite{White2021,White2022,Jakeman2023,White2023} introduce risk measures such as average value-at-risk (AVaR) into Bayesian OED, primarily in a goal-oriented setting where uncertainty in quantities of interest is targeted. These approaches replace expectation-based criteria with risk measures derived from posterior or predictive distributions, enabling control over tail behavior. 
However, the development of general computational methods for estimating such objectives remains limited.

Overall, while recent work has begun to incorporate risk measures into Bayesian experimental design, a general framework for risk-aware Bayesian OED with outcome-dependent utilities---and the associated efficient estimators---remains largely unexplored.

This paper proposes\footnote{Code available at: \url{https://github.com/wgshen/rOED}.} a mean--variance Bayesian OED criterion that incorporates both the expectation and variance of the experimental utility in nonlinear model settings. The formulation allows explicit control of the trade-off between expected performance and variability of experimental outcomes. The main contributions are summarized as follows:
\begin{itemize}
\item A mean--variance Bayesian OED criterion that augments the conventional expected-utility objective with a penalty on the variance of utility.
\item A Monte Carlo estimator for evaluating the variance-penalized objective together with an analysis of its convergence properties.
\item Numerical demonstrations on a linear-Gaussian benchmark, a nonlinear synthetic example, and a contaminant source inversion problem with and without building obstacles.
\end{itemize}

To efficiently identify optimal designs under the proposed criterion, we employ Bayesian optimization as a global optimization strategy. Common random samples are used to induce correlation across design evaluations, thereby smoothing the objective landscape and improving optimization efficiency.

The remainder of this paper is organized as follows. \Cref{sec:roed_formulation} presents the formulation of the mean--variance design criterion. \Cref{sec:roed_method} introduces the proposed Monte Carlo estimator and analyzes its convergence behavior. \Cref{sec:roed_results} presents numerical experiments, including a linear-Gaussian benchmark, a nonlinear synthetic example, and a contaminant source inversion problem with and without building obstacles. Finally, \cref{sec:roed_conclusions} summarizes the paper and discusses future research directions.

%% file: sections/2_formulation.tex
\section{Problem formulation}
\label{sec:roed_formulation}

\subsection{Background of expected utility-based OED}
\label{sec:roed_expected_utility}

We begin by reviewing the notation and formulation of Bayesian OED. Let $\design \in \RR^{d}$ denote the controllable design variables, $\Param \in \RR^{p}$ the unknown model parameters, and $Y \in \RR^{n}$ the experimental observations. Their corresponding realizations are denoted by $\param$ and $y$.

When an experiment is conducted under design $\design$ and observation $y$ is obtained, uncertainty in the parameters is updated according to Bayes' rule,
\begin{align}
\pdf(\param | y,\design)
=
\frac{\pdf(y|\param,\design)\,\pdf(\param)}
{\pdf(y|\design)},
\label{e:Bayes}
\end{align}
where $\pdf(\param)$ denotes the (design-independent) prior, $\pdf(y|\param,\design)$ the likelihood, and $\pdf(y|\design)$ the marginal likelihood. Throughout, we work with probability density functions for continuous variables and assume they exist; analogous formulations for discrete variables follow similarly.

To quantify the value of an experiment, we introduce a utility function $u(\design,y,\param)$, which represents the utility associated with observing $y$ under design $\design$ when the true parameter is $\param$. Since $Y$ is unknown prior to conducting the experiment and $\Param$ is always uncertain, the utility itself is a random variable.

To compare different designs, this random utility must be summarized into a scalar objective. The standard approach adopts a risk-neutral, decision-theoretic formulation and considers the expected utility,
\begin{align}
U(\design)
=
\mathbb{E}_{Y,\Param|\design}
\left[
u(\design,Y,\Param)
\right].
\label{e:EU_general}
\end{align}
An optimal design is then obtained by solving
\begin{align}
\design^\ast
\in
\argmax_{\design \in \mathcal{D}}
U(\design).
\label{e:EU_opt}
\end{align}

In many applications, the goal is to reduce uncertainty in the parameters, and a common choice of utility is the parameter information gain. Following \cite{Lindley1956}, this is defined as the Kullback--Leibler (KL) divergence from the prior to the posterior,
\begin{align}
u_{\mathrm{KL}}(\design,y)
=
\mathbb{E}_{\Param|y,\design}
\left[
\log
\frac{\pdf(\Param|y,\design)}{\pdf(\Param)}
\right].
\label{e:uKL}
\end{align}
This utility depends only on the observation $y$ and quantifies the information gained about $\Param$ after the experiment.
With this choice, the expected utility reduces to the EIG,
\begin{align}
U_{\mathrm{KL}}(\design)
=
\mathbb{E}_{\Param,Y|\design} \left[
\mathbb{E}_{\Param|Y,\design} \left[\log
\frac{\pdf(\Param|Y,\design)}{\pdf(\Param)}\right] \right]
=
\mathbb{E}_{\Param,Y|\design} \left[
\log
\frac{\pdf(\Param|Y,\design)}{\pdf(\Param)}\right], 
\label{e:EU_uKL}
\end{align}
which is also equivalent to the mutual information between $\Param$ and $Y$ conditioned on $\design$.

\subsection{Utility variance}

The expected-utility criterion considers only the average value of the utility across possible experimental outcomes. However, the realized utility $u(\design,Y,\Param)$ is a random variable prior to conducting the experiment, since both the observation $Y$ and the parameter $\Param$ are uncertain. Consequently, two designs with similar expected utility may exhibit substantially different variability in their realized utilities.

To characterize this variability, we define the variance of the utility with respect to the joint distribution of $(\Param,Y)$ given $\design$,
\begin{align}
V(\design)
=
\Var_{\Param,Y|\design}\big[u(\design,Y,\Param)\big]
&=
\mathbb{E}_{\Param,Y|\design}
\left[
\big(u(\design,Y,\Param)-U(\design)\big)^2
\right] \\
&=
\mathbb{E}_{\Param,Y|\design}\big[u(\design,Y,\Param)^2\big]
-
U(\design)^2.
\label{e:u_variance_general}
\end{align}
This definition is fully general and applies to utilities that depend on both $y$ and $\param$.

When $u_{\mathrm{KL}}$ in \cref{e:uKL} is used, the utility depends only on the observation $y$. In this case, the variability is entirely induced by the predictive distribution of $Y$, and the variance reduces to
\begin{align}
V(\design)
=
\Var_{Y|\design}\big[u_{\mathrm{KL}}(\design,Y)\big].
\label{e:u_variance_uKL}
\end{align}

\subsection{Mean--variance design criterion}

To account for both the expected performance and the variability of experimental outcomes, we introduce a variance-penalized design objective
\begin{align}
J_{\lambda}(\design)
=
U(\design)
-
\lambda V(\design),
\label{e:meanvar}
\end{align}
where $\lambda \in \mathbb{R}$ is a tuning parameter that controls the trade-off between expected utility and its variability.
For $\lambda > 0$, the objective penalizes large utility variance and therefore favors designs with more stable outcomes. Larger values of $\lambda$ correspond to stronger risk aversion. When $\lambda = 0$, the formulation reduces to the classical expected-utility criterion. Negative values of $\lambda$ correspond to risk-seeking behavior.

While coherent risk measures such as CVaR are often used in risk-aware formulations, the mean--variance criterion adopted here is not coherent in general. 
Instead, it is chosen for its analytical simplicity and computational tractability, which enable efficient estimation within the nested structure of Bayesian OED. 
Extending the proposed framework to coherent risk measures is a natural direction for future work.

The mean--variance Bayesian OED problem therefore becomes
\begin{align}
\design^\ast_{\lambda}
\in
\argmax_{\design \in \mathcal{D}}
J_{\lambda}(\design).
\label{e:meanvar_opt}
\end{align}

%% file: sections/3_methodology.tex
\section{Numerical method}
\label{sec:roed_method}

We first develop MC estimators for the expected utility, its second moment, and the resulting mean--variance objective, then discuss practical strategies for reducing computational cost and optimizing the resulting noisy objective.

\subsection{Monte Carlo estimation of the objective}
\label{sec:roed_estimator}

The quantities $U(\design)$, $V(\design)$, and $J_{\lambda}(\design)$ generally do not admit closed-form expressions for nonlinear models. We therefore approximate them using Monte Carlo (MC) methods in the information-gain setting introduced in \cref{e:uKL} and \cref{e:EU_uKL}. 
In the remainder of this section, we focus on the information-gain utility $u_{\mathrm{KL}}$.

\subsubsection{Expected utility}

With $u_{\mathrm{KL}}$, the expected utility (i.e., EIG) in \cref{e:EU_uKL} can be further written as
\begin{align}
U_{\mathrm{KL}}(\design)
&=
\mathbb{E}_{\Param,Y|\design}
\left[
\log
\frac{\pdf(\Param|Y,\design)}{\pdf(\Param)}
\right] = \mathbb{E}_{\Param,Y|\design}
\left[
\log
\frac{\pdf(Y|\Param,\design)}{\pdf(Y|\design)}
\right] \\
&=
\iint
\pdf(\param)\pdf(y|\param,\design)
\log
\left[
\frac{\pdf(y|\param,\design)}{\pdf(y|\design)}
\right]
\,\mathrm{d}y\,\mathrm{d}\param,
\label{e:EIG_NMC}
\end{align}
which is amenable to sampling from the joint distribution by first drawing $\param \sim \pdf(\param)$ and then $y \sim \pdf(y|\param,\design)$.

A NMC estimator~\cite{Ryan2003} is then given by
\begin{align}
\widehat{U}(\design)
=
\frac{1}{N}
\sum_{i=1}^N
\left[
\log \pdf(y^{(i)}|\param^{(i)},\design)
-
\log \widehat{\pdf}(y^{(i)}|\design)
\right],
\label{e:Uhat}
\end{align}
where $\param^{(i)}\sim \pdf(\param)$ and $y^{(i)} \sim \pdf(y|\param^{(i)},\design)$ are independent samples, and
\begin{align}
\widehat{\pdf}(y^{(i)}|\design)
=
\frac{1}{M_1}
\sum_{j=1}^{M_1}
\pdf(y^{(i)}|\param^{(i,j)},\design)
\label{e:marginal_likelihood_est}
\end{align}
with ${\param}^{(i,j)} \sim \pdf(\param)$ is the MC estimator for the marginal likelihood $\pdf(y|\design)=\int \pdf(y|\param,\design) \pdf(\param)\,\textrm{d}\param$.

The estimator $\widehat{U}(\design)$ is biased because the logarithm is applied to an MC approximation of $\pdf(y|\design)$, but it becomes asymptotically unbiased as $M_1 \to \infty$. Its bias scales, to leading order, as
\begin{align}
\mathbb{E}\!\left[\widehat{U}(\design)-U_{\mathrm{KL}}(\design)\right]
=
\frac{E_1(\design)}{M_1} + o(M_1^{-1}),
\label{e:Uhat_bias}
\end{align}
and its variance scales as
\begin{align}
\Var\!\left[\widehat{U}(\design)\right]
=
\frac{A_1(\design)}{N}
+
\frac{B_1(\design)}{NM_1}+
o(N^{-1})+o((NM_1)^{-1}),
\label{e:Uhat_var}
\end{align}
where $A_1(\design)$, $B_1(\design)$, and $E_1(\design)$ are problem-dependent constants \cite{Ryan2003,Beck2018,Rainforth2018}, and the $o(\cdot)$ notation denotes higher-order remainder terms as $N, M_1 \to \infty$.

\subsubsection{Utility second moment}

To estimate the utility variance, we first consider the second moment
\begin{align}
M_2(\design)
&=
\mathbb{E}_{Y|\design}\!\left[u_{\mathrm{KL}}(\design,Y)^2\right] \\
&=
\int
\pdf(y|\design)
\left[
\int
\pdf(\param|y,\design)
\log\frac{\pdf(\param|y,\design)}{\pdf(\param)}
\,\mathrm{d}\param
\right]^2
\mathrm{d}y.
\end{align}
Since $u_{\mathrm{KL}}$ depends only on $Y$, the expectation is taken with respect to $\pdf(y|\design)$.
Applying Bayes' rule, this can be rewritten as
\begin{align}
M_2(\design)
=
\int
\pdf(y|\design)
\left[
\int
\pdf(\param|y,\design)
\log\frac{\pdf(y|\param,\design)}{\pdf(y|\design)}
\,\mathrm{d}\param
\right]^2
\mathrm{d}y.
\label{e:M2_rewrite_1}
\end{align}

Expanding the square yields three terms, which correspond to contributions involving only the marginal likelihood, cross terms, and likelihood-weighted expectations, respectively,
\begin{align}
M_2(\design)=M_{2,a}(\design)+M_{2,b}(\design)+M_{2,c}(\design),
\label{e:M2_split}
\end{align}
where
\begin{align}
M_{2,a}(\design)
&=
\int
\pdf(y|\design)\big[\log \pdf(y|\design)\big]^2
\,\mathrm{d}y,
\label{e:M2a}
\\
M_{2,b}(\design)
&=
-2
\iint
\pdf(y|\design)\pdf(\param|y,\design)
\log\pdf(y|\param,\design)\log \pdf(y|\design)
\,\mathrm{d}\param\,\mathrm{d}y,
\label{e:M2b}
\\
M_{2,c}(\design)
&=
\int
\pdf(y|\design)
\left[
\int
\pdf(\param|y,\design)\log\pdf(y|\param,\design)
\,\mathrm{d}\param
\right]^2
\mathrm{d}y.
\label{e:M2c}
\end{align}

For $M_{2,a}(\design)$, we use
\begin{align}
\widehat{M}_{2,a}(\design)
=
\frac{1}{N}
\sum_{i=1}^N
\left[
\log \widehat{\pdf}(y^{(i)}|\design)
\right]^2,
\label{e:M2a_hat}
\end{align}
where $y^{(i)} \sim \pdf(y|\design)$ are generated by first sampling $\param^{(i)} \sim \pdf(\param)$ and then $y^{(i)} \sim \pdf(y|\param^{(i)},\design)$, and $\widehat{\pdf}(y^{(i)}|\design)$ is given by \cref{e:marginal_likelihood_est}.

For $M_{2,b}(\design)$, the estimator is
\begin{align}
\widehat{M}_{2,b}(\design)
=
-\frac{2}{N}
\sum_{i=1}^N
\left[
\log \pdf(y^{(i)}|\param^{(i)},\design)
\;
\log \widehat{\pdf}(y^{(i)}|\design)
\right].
\label{e:M2b_hat}
\end{align}

For $M_{2,c}(\design)$, we rewrite the inner expectation as
\begin{align}
\int
\pdf(\param|y,\design)\log\pdf(y|\param,\design)\,\mathrm{d}\param
=
\frac{
\int
\pdf(\param)\pdf(y|\param,\design)\log\pdf(y|\param,\design)\,\mathrm{d}\param
}{
\pdf(y|\design)
},
\label{e:M2c_rewrite}
\end{align}
which avoids posterior sampling. An MC estimator is then
\begin{align}
\widehat{M}_{2,c}(\design)
=
\frac{1}{N}
\sum_{i=1}^N
\left[
\frac{
\frac{1}{M_2}\sum_{k=1}^{M_2}
\pdf(y^{(i)}|\param^{(i,k)},\design)\log\pdf(y^{(i)}|\param^{(i,k)},\design)
}{
\widehat{\pdf}(y^{(i)}|\design)
}
\right]^2,
\label{e:M2c_hat}
\end{align}
where $\param^{(i,k)} \sim \pdf(\param)$ are independent prior samples.

Combining the three components yields the estimator
\begin{align}
\widehat{M}_2(\design)
=
\widehat{M}_{2,a}(\design)
+
\widehat{M}_{2,b}(\design)
+
\widehat{M}_{2,c}(\design).
\label{e:M2_hat}
\end{align}

The bias and variance of $\widehat{M}_2(\design)$ depend on the outer sample size $N$ and the inner sample sizes $M_1$ and $M_2$. Detailed derivations are provided in \ref{app:esti_util_mu2_a} to~\ref{app:esti_util_mu2_c}. Retaining the leading-order terms,
\begin{align}
\mathbb{E}\!\left[\widehat{M}_2(\design)-M_2(\design)\right]
&=
\frac{E_2(\design)}{M_1}
+
\frac{F_2(\design)}{M_2}
+
o(M_1^{-1})+o(M_2^{-1}),
\\
\Var\!\left[\widehat{M}_2(\design)\right]
&=
\frac{A_2(\design)}{N}
+
\frac{B_2(\design)}{NM_1}
+
\frac{C_2(\design)}{NM_2}
+
o(N^{-1})+o((NM_1)^{-1})+o((NM_2)^{-1}),
\end{align}
where $A_2(\design)$, $B_2(\design)$, $C_2(\design)$, $E_2(\design)$, and $F_2(\design)$ are problem-dependent constants. 

\subsubsection{Utility variance}

The utility variance can be expressed as
\begin{align}
V(\design)=M_2(\design)-U(\design)^2.
\label{e:V_from_M2}
\end{align}
Accordingly, we estimate it using
\begin{align}
\widehat{V}(\design)
=
\widehat{M}_2(\design)-\widehat{U}(\design)^2.
\label{e:Vhat}
\end{align}

The term $\widehat{U}(\design)^2$ is itself a biased estimator of $U(\design)^2$. From \ref{app:esti_util_squared}, its leading-order bias and variance are
\begin{align}
\mathbb{E}\!\left[\widehat{U}(\design)^2-U(\design)^2\right]
&=
\frac{D_3(\design)}{N}
+
\frac{E_3(\design)}{M_1}
+
o(N^{-1})+o(M_1^{-1}),
\\
\Var\!\left[\widehat{U}(\design)^2\right]
&=
\frac{A_3(\design)}{N}
+
\frac{B_3(\design)}{NM_1}
+
o(N^{-1})+o((NM_1)^{-1}),
\end{align}
for problem-dependent constants.

Combining these results with those of $\widehat{M}_2(\design)$, the estimator $\widehat{V}(\design)$ has leading-order bias
\begin{align}
\mathbb{E}\!\left[\widehat{V}(\design)-V(\design)\right]
&=
\frac{D_4(\design)}{N}
+
\frac{E_4(\design)}{M_1}
+
\frac{F_4(\design)}{M_2}
+
o(N^{-1})+o(M_1^{-1})+o(M_2^{-1}),
\label{e:Vhat_bias}
\end{align}
and variance
\begin{align}
\Var\!\left[\widehat{V}(\design)\right]
&=
\frac{A_4(\design)}{N}
+
\frac{B_4(\design)}{NM_1}
+
\frac{C_4(\design)}{NM_2}
+
o(N^{-1})+o((NM_1)^{-1})+o((NM_2)^{-1}),
\label{e:Vhat_var}
\end{align}
where the constants collect contributions from both $\widehat{M}_2(\design)$ and $\widehat{U}(\design)^2$.

\subsubsection{Mean--variance objective}

Finally, the mean--variance objective
\begin{align*}
J_{\lambda}(\design)=U(\design)-\lambda V(\design)
\end{align*}
is estimated by
\begin{align}
\widehat{J}_{\lambda}(\design)
=
\widehat{U}(\design)-\lambda \widehat{V}(\design).
\label{e:Jhat}
\end{align}
Equivalently,
\begin{align}
\widehat{J}_{\lambda}(\design)
=
\widehat{U}(\design)
-
\lambda\widehat{M}_2(\design)
+
\lambda\widehat{U}(\design)^2,
\label{e:Jhat_expanded}
\end{align}
which is convenient in implementation since all quantities are already available.

Using the leading-order expansions of $\widehat{U}(\design)$ and $\widehat{V}(\design)$, the estimator $\widehat{J}_{\lambda}(\design)$ satisfies
\begin{align}
\mathbb{E}\!\left[\widehat{J}_{\lambda}(\design)-J_{\lambda}(\design)\right]
&=
- \frac{\lambda D_4(\design)}{N}
+
\frac{E_1(\design)-\lambda E_4(\design)}{M_1}
-
\frac{\lambda F_4(\design)}{M_2}
\label{e:Jhat_bias}\\
&\hspace{1.5em}+
o(N^{-1})+o(M_1^{-1})+o(M_2^{-1}),\nonumber
\\
\Var\!\left[\widehat{J}_{\lambda}(\design)\right]
&=
\frac{A_1(\design)+\lambda^2 A_4(\design)}{N}
+
\frac{B_1(\design)+\lambda^2 B_4(\design)}{NM_1}
+
\frac{\lambda^2 C_4(\design)}{NM_2}
\label{e:Jhat_var}\\
&\hspace{1.5em}+
o(N^{-1})+o((NM_1)^{-1})+o((NM_2)^{-1}),\nonumber
\end{align}
where all constants are problem-dependent.

\subsubsection{Sample reuse across nested loops}

The overall mean--variance MC estimator is more expensive than the standard NMC estimator for expected utility alone, since it requires repeated evaluation of the marginal likelihood estimator and the second-moment terms. 
Naively, the computational cost is dominated by inner-loop likelihood evaluations and scales as $\mathcal{O}(NM_1 + NM_2)$.

To reduce this cost, we follow the practical sample-reuse strategy of \cite{Huan2013}, in which the same prior samples are reused across the outer and inner loops. In particular, we set $N=M_1=M_2$ and reuse the outer samples in the inner estimators, i.e.,
\[
\param^{(i,j)} = \param^{(j)},
\qquad
\param^{(i,k)} = \param^{(k)}.
\]
This allows all quantities in $\widehat{U}(\design)$, $\widehat{M}_2(\design)$, and hence $\widehat{J}_{\lambda}(\design)$ to be constructed from a common set of samples.

It is important to distinguish between likelihood evaluations and forward model evaluations. As written, the estimators still require $\mathcal{O}(N^2)$ likelihood evaluations when $N=M_1=M_2$, since each outer observation $y^{(i)}$ must be paired with many parameter samples in the inner sums. However, when each likelihood evaluation is built from a forward model solve followed by a relatively cheap noise model evaluation, for example with an additive Gaussian noise data model:
\begin{align}
Y = G(\Param, \design) + \mathcal{E}
\end{align}
where $\mathcal{E}\sim \mathcal{N}(0,\Gamma)$ and hence $\pdf(y|\param,\design)=\mathcal{N}(y; G(\param,\design),\Gamma)$,
sample reuse reduces the number of \emph{forward model evaluations} $G$ to $\mathcal{O}(N)$ for each design $\design$, because the same model outputs can be reused across all inner-loop calculations.

While sample reuse introduces additional bias, this effect is typically small in practice, and the computational savings can be substantial. Under this strategy, both the bias and the variance of $\widehat{J}_{\lambda}(\design)$ remain of order $\mathcal{O}(1/N)$ when $N=M_1=M_2$, although the associated constants are larger than those of the standard NMC estimator for expected utility alone because the mean--variance objective requires estimation of higher-order moments.

\subsection{Optimization}

With an MC estimator of the objective $J_{\lambda}(\design)$ in hand, the Bayesian OED problem in \cref{e:meanvar_opt} can be solved using standard optimization methods. The key challenge is that the objective is both expensive to evaluate and corrupted by MC noise.

In principle, any optimization algorithm can be applied. When gradient information is available, either analytically or via stochastic gradient estimators, one may employ gradient-based methods such as stochastic gradient ascent or quasi-Newton schemes (e.g., L-BFGS-B)~\cite{Huan2014}. In the absence of reliable gradients, derivative-free methods such as simultaneous perturbation stochastic approximation (SPSA) or Nelder--Mead~\cite{Huan2013} can be used, although their performance may degrade in the presence of noise or in higher-dimensional settings.

In this work, we adopt Bayesian optimization (BO)~\cite{Mokus1975,Jones1998, Shahriari2016,Frazier2018,Wang2023} as a practical and robust choice. BO is a derivative-free global optimization framework that is well suited to expensive and noisy objective functions. It constructs a surrogate model (typically a Gaussian process) of $J_{\lambda}(\design)$ and selects new evaluation points via an acquisition function that balances exploration and exploitation.

It is important to distinguish the uncertainty modeled by BO from the uncertainty in the experimental utility. The surrogate model in BO represents uncertainty in the objective due to limited evaluations and can be reduced by further optimization iterations, whereas the variability of the utility arises from the stochastic nature of experimental outcomes and is intrinsic to the design problem.

We emphasize that BO is not specific to OED, but rather serves as a general-purpose optimizer for the variance-penalized objective. Other optimization strategies could be used in its place. In our implementation, we employ an existing BO library~\cite{Nogueira2014} with minor modifications to accommodate problem-specific constraints.

A summary of the overall BO-based procedure is provided in \cref{alg:meanvar_boed}.

\subsubsection{Common random sampling}
\label{sec:roed_common_random_samples}

Although BO is robust to noisy objective evaluations, its convergence may still be slowed by high MC noise in estimating the objective. To mitigate this without increasing the sample size, we employ the technique of \emph{common random sampling} (CRS) as investigated in~\cite{Huan2014}, in which the same realizations of $\Param$ and observational noise are reused across different designs $\design$.

This induces correlation in the MC estimation error across designs, effectively reducing relative noise and producing a smoother objective landscape that is easier to optimize. It can be interpreted as a variance-reduction technique that improves the signal-to-noise ratio of pairwise comparisons between designs. Importantly, this technique does not reduce the computational cost of objective evaluation, but improves optimization efficiency by stabilizing comparisons between candidate designs.

CRS is distinct from the sample-reuse strategy described earlier, which reuses samples within a single objective evaluation to reduce computational cost. In contrast, CRS reuses samples \emph{across} different design evaluations. In practice, this can be implemented by fixing the random seed when generating MC samples for each evaluation of the objective.

While this approach may introduce additional variability in the estimated optimal design under finite sample sizes, this effect diminishes as the number of samples increases. In practice, the benefit of a smoother objective function typically outweighs this drawback. We demonstrate this effect in the numerical results.

\begin{algorithm}[htbp]
\caption{Mean--variance Bayesian OED via MC estimation and Bayesian optimization}
\label{alg:meanvar_boed}
\begin{algorithmic}[1]
\Require variance-penalty parameter $\lambda$, initial design set $\{\design^{(m)}\}_{m=1}^{n_0}$, sample sizes $N,M_1,M_2$, optimization budget $T$
\Ensure approximate optimizer $\design^\ast_\lambda$

\For{$m=1,\dots,n_0$}
    \State Evaluate $\widehat{J}_{\lambda}(\design^{(m)})$ using the MC estimator in \cref{e:Jhat_expanded}
\EndFor

\For{$t=1,\dots,T$}
    \State Fit or update a BO surrogate model for $J_\lambda(\design)$ using all previously evaluated pairs $\{(\design^{(m)},\widehat{J}_{\lambda}(\design^{(m)}))\}$
    \State Construct an acquisition function $a(\design)$ from the surrogate model
    \State Select a next design
    \[
    \design^{(n_0+t)}
    \in
    \argmax_{\design\in\mathcal{D}}
    a(\design)
    \]
    \State Generate outer samples $\param^{(i)}\sim\pdf(\param)$ and $y^{(i)}\sim\pdf(y|\param^{(i)},\design^{(n_0+t)})$, for $i=1,\dots,N$
    \State Generate inner prior samples for estimating $\widehat{\pdf}(y^{(i)}|\design^{(n_0+t)})$ and $\widehat{M}_2(\design^{(n_0+t)})$
    \State Compute $\widehat{U}(\design^{(n_0+t)})$ using \cref{e:Uhat}
    \State Compute $\widehat{M}_2(\design^{(n_0+t)})$ using \cref{e:M2_hat}
    \State Compute
    \[
    \widehat{J}_{\lambda}(\design^{(n_0+t)})
    =
    \widehat{U}(\design^{(n_0+t)})
    -
    \lambda \widehat{M}_2(\design^{(n_0+t)})
    +
    \lambda \widehat{U}(\design^{(n_0+t)})^2
    \]
\EndFor

\State \Return the best evaluated design
\[
\design^\ast_\lambda
\in
\argmax_{\design^{(m)}} \widehat{J}_{\lambda}(\design^{(m)})
\]
\end{algorithmic}
\end{algorithm}

%% file: sections/4_results.tex
\section{Numerical results}
\label{sec:roed_results}

We present three examples to demonstrate the performance and benefits of the proposed mean--variance Bayesian OED formulation. 

The first example is a linear-Gaussian problem (\cref{sec:roed_ex_linGau}) with a closed-form solution, which serves as a validation benchmark. In this case, we compare the proposed MC estimators against analytical results to assess their accuracy and convergence.
The second example is a synthetic nonlinear problem (\cref{sec:roed_ex_nonlinear}), where we illustrate the effect of incorporating variability into the design criterion and highlight the differences between mean-optimal and risk-aware designs.
Finally, we consider variants of a contaminant source inversion problem (\cref{sec:roed_ex_source_wo_building,sec:roed_ex_source_w_building}) to demonstrate the applicability of the proposed approach in a more realistic, multi-dimensional setting.

\subsection{Case 1: Linear-Gaussian benchmark}
\label{sec:roed_ex_linGau}

Consider a linear observation model with scalar parameter $\Param \in \mathbb{R}$ and scalar design variable $\design \in [0,3]$:
\begin{align}
    Y 
    = G(\Param,\design) + \mathcal{E} 
    = \Param \design + \mathcal{E},
\end{align}
where $\mathcal{E} \sim \mathcal{N}(0,1^2)$. The prior is $\Param \sim \mathcal{N}(0,3^2)$. 
Under this linear-Gaussian setting, the posterior remains Gaussian and all relevant quantities can be computed in closed form, making this example a convenient benchmark for validating the proposed estimators.

The mean--variance objective $\widehat{J}_\lambda$ consists of the expected utility estimator $\widehat{U}$ and the utility variance estimator $\widehat{V}$. Since the behavior of $\widehat{U}$ has been studied extensively in~\cite{Huan2013}, we focus here on the estimation of $\widehat{V}$.

We first examine the convergence of $\widehat{V}(\design)$ at a fixed design. We select $\design=3$, which corresponds to the largest utility variance and therefore represents the most challenging setting for estimation. In all experiments, we use sample reuse with $N=M_1=M_2$.

\begin{figure}[htbp]
  \centering
  \subfloat[Estimate versus exact]{\label{fig:roed_1DLinGau_convg_utilvar}\includegraphics[width=0.32\linewidth]{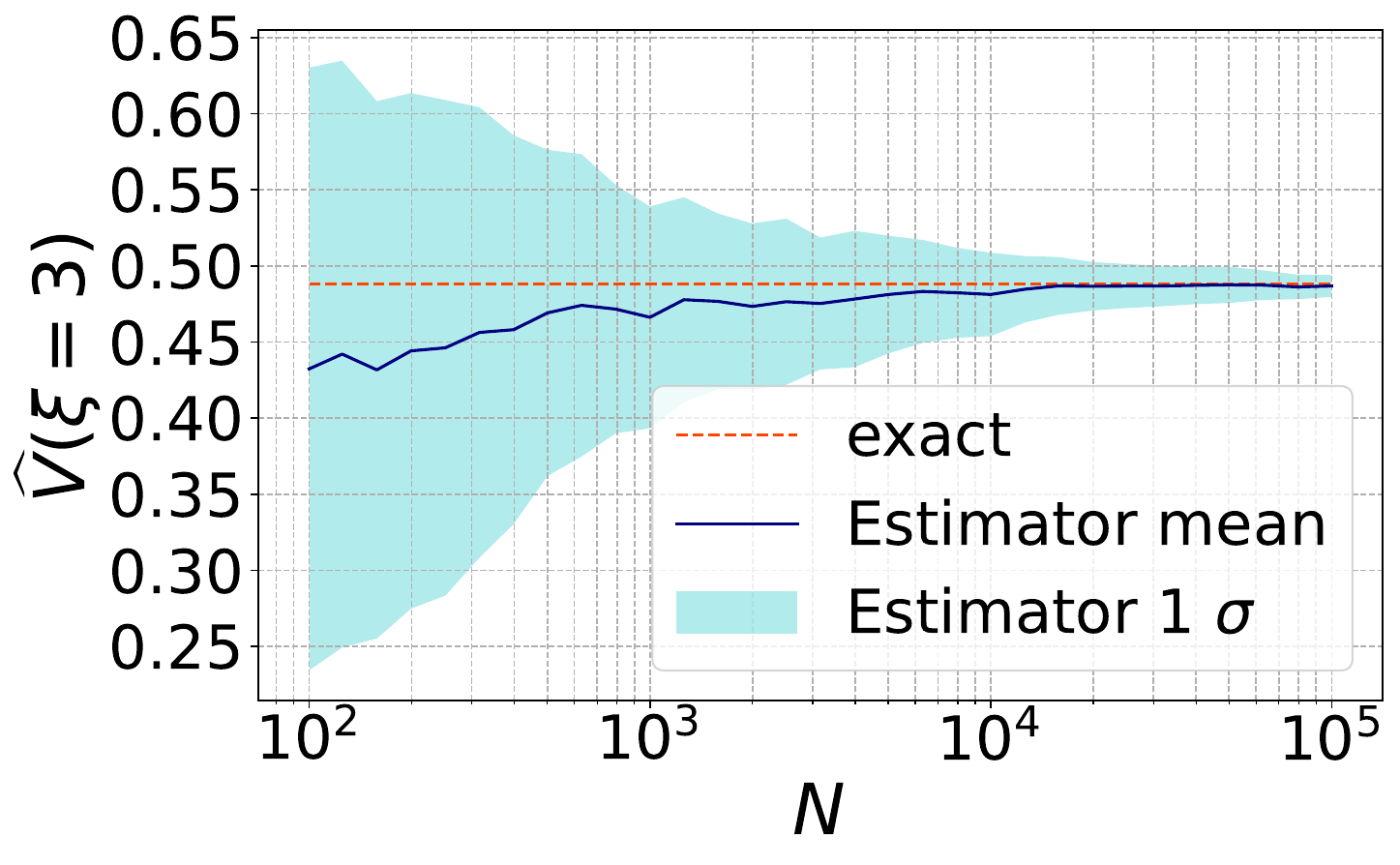}}
  \hspace{0.5em}
  \subfloat[Estimator absolute bias]{\label{fig:roed_1DLinGau_convg_bias}\includegraphics[width=0.32\linewidth]{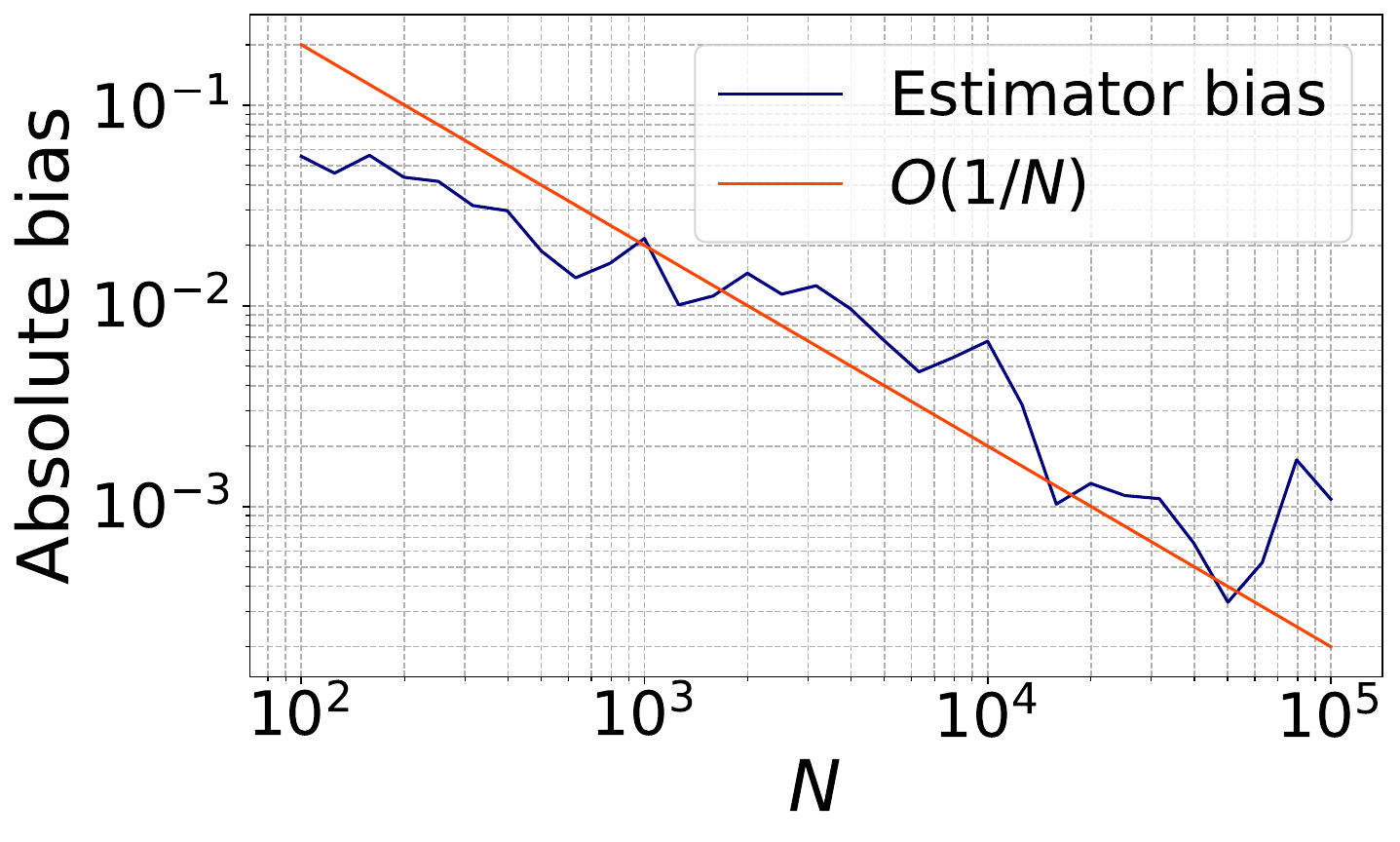}}
  \hspace{0.5em}
  \subfloat[Estimator variance]{\label{fig:roed_1DLinGau_convg_variance}\includegraphics[width=0.32\linewidth]{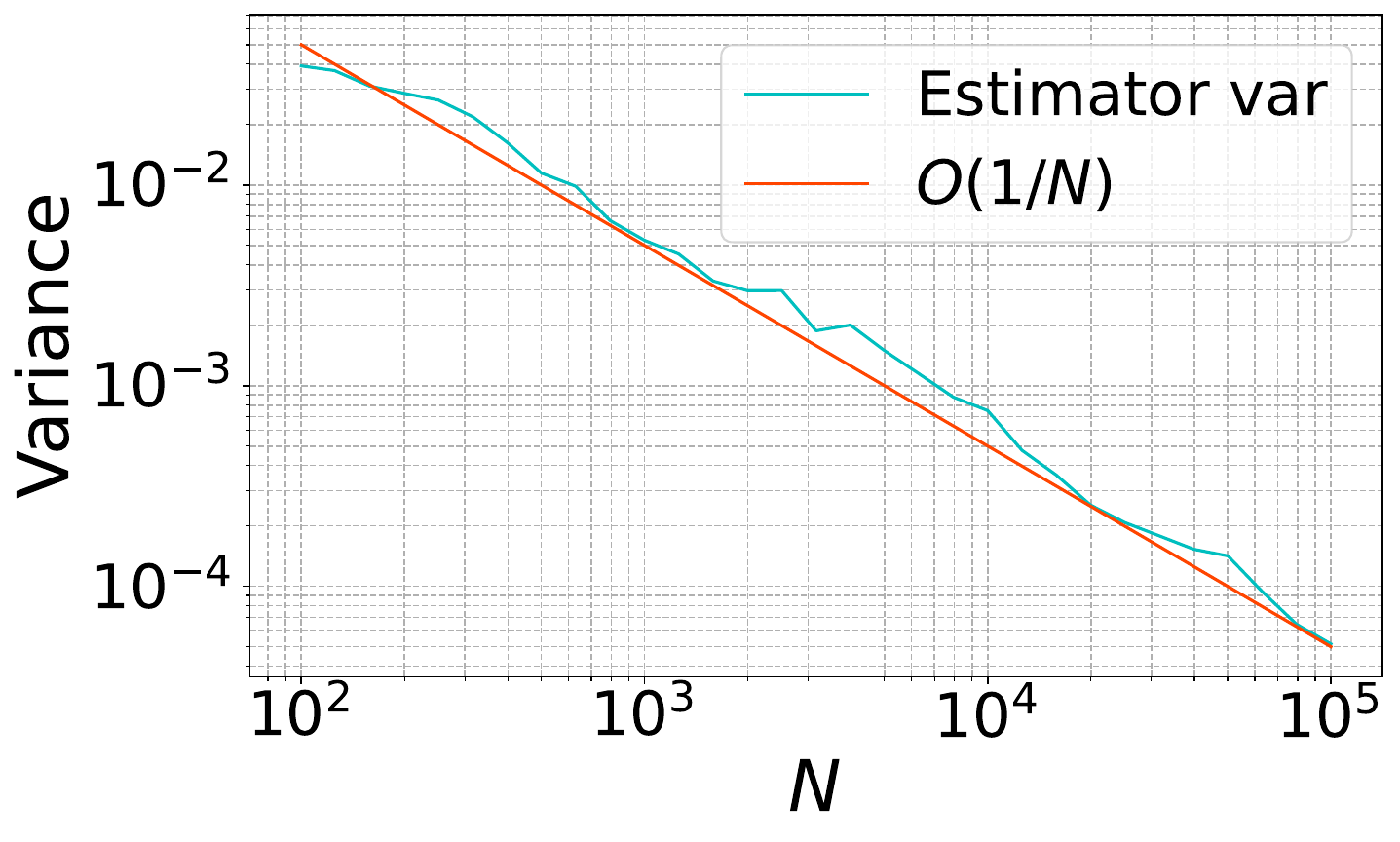}}
  \caption{Case 1: Convergence of the MC estimator for the utility variance $\widehat{V}(\design=3)$.}
  \label{fig:roed_1DLinGau_convg}
\end{figure}

\Cref{fig:roed_1DLinGau_convg} shows the convergence behavior of the estimator as $N$ increases. The estimated variance approaches the exact solution (\cref{fig:roed_1DLinGau_convg_utilvar}), while the empirical bias and variance decay approximately at rate $\mathcal{O}(1/N)$ (\cref{fig:roed_1DLinGau_convg_bias,fig:roed_1DLinGau_convg_variance}), consistent with the analysis in \cref{sec:roed_method}. The bias exhibits more fluctuation than the variance, which is attributed to higher-order terms introduced by sample reuse, whereas the variance is dominated by the leading $\mathcal{O}(1/N)$ term.

\begin{figure}[htbp]
    \centering
    \begin{subfigure}{0.32\textwidth}
        \centering
        \includegraphics[width=1.0\linewidth]{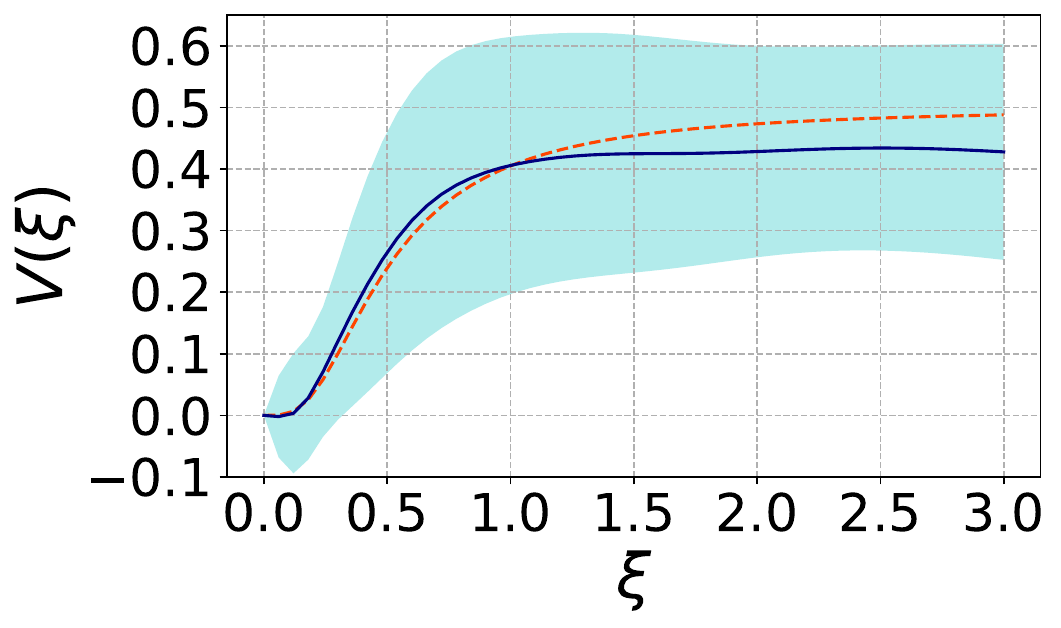}
        \\
        \includegraphics[width=1.0\linewidth]{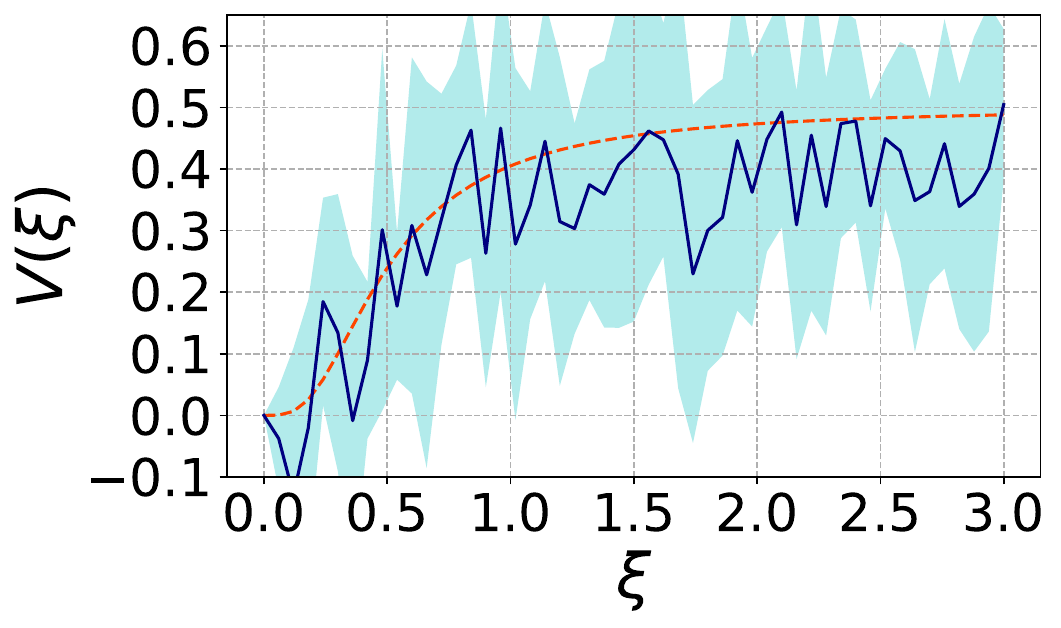}
        \caption{$N=100$}
    \end{subfigure}
    \begin{subfigure}{0.32\textwidth}
        \centering
        \includegraphics[width=1.0\linewidth]{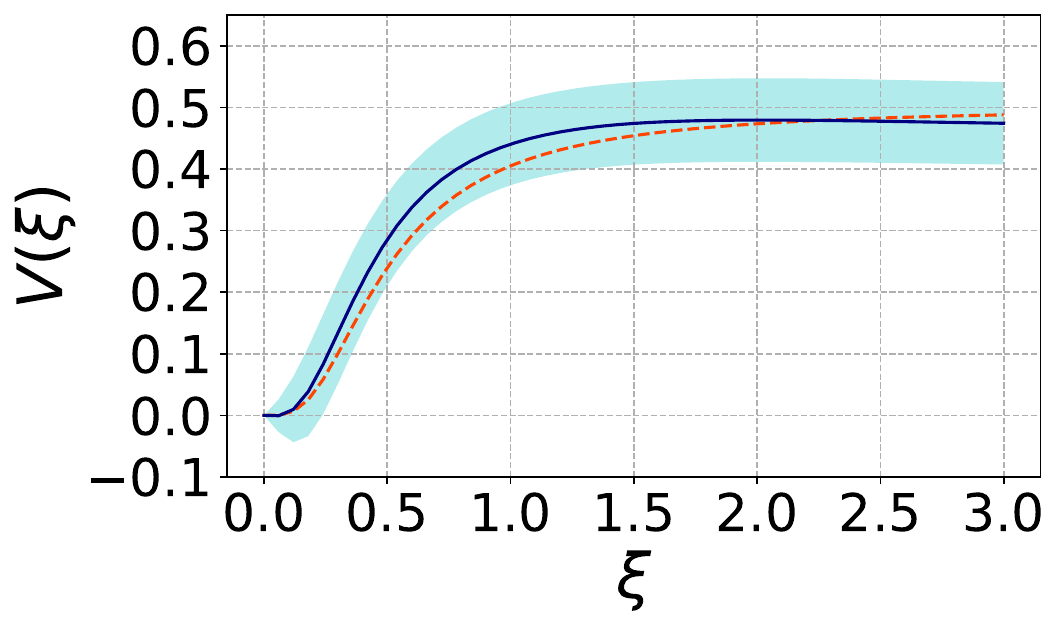}
        \\
        \includegraphics[width=1.0\linewidth]{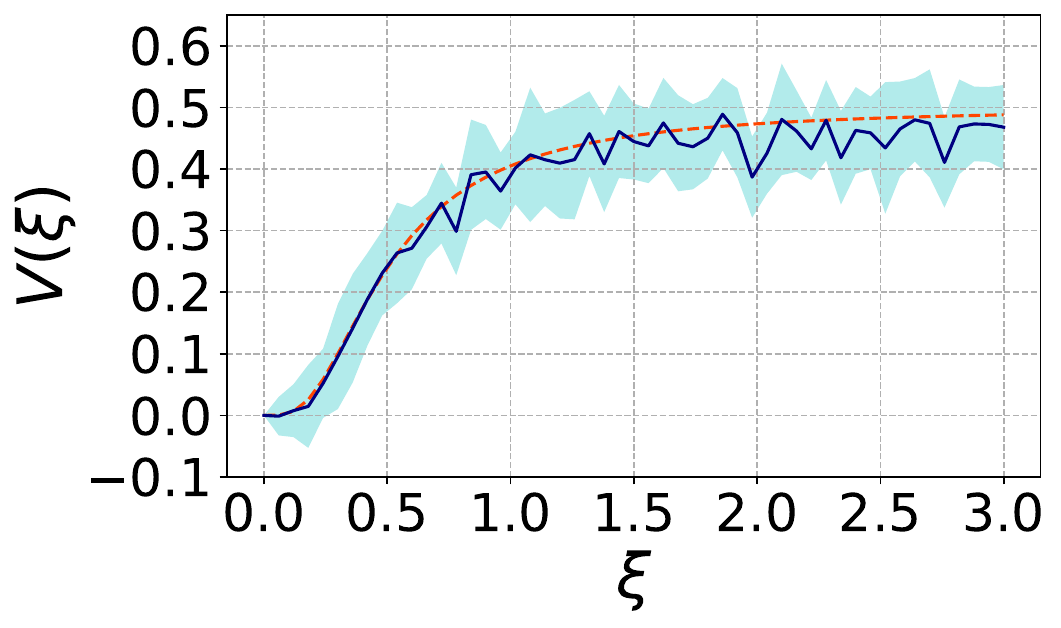}
        \caption{$N=1000$}
    \end{subfigure}
    \begin{subfigure}{0.32\textwidth}
        \centering
        \includegraphics[width=1.0\linewidth]{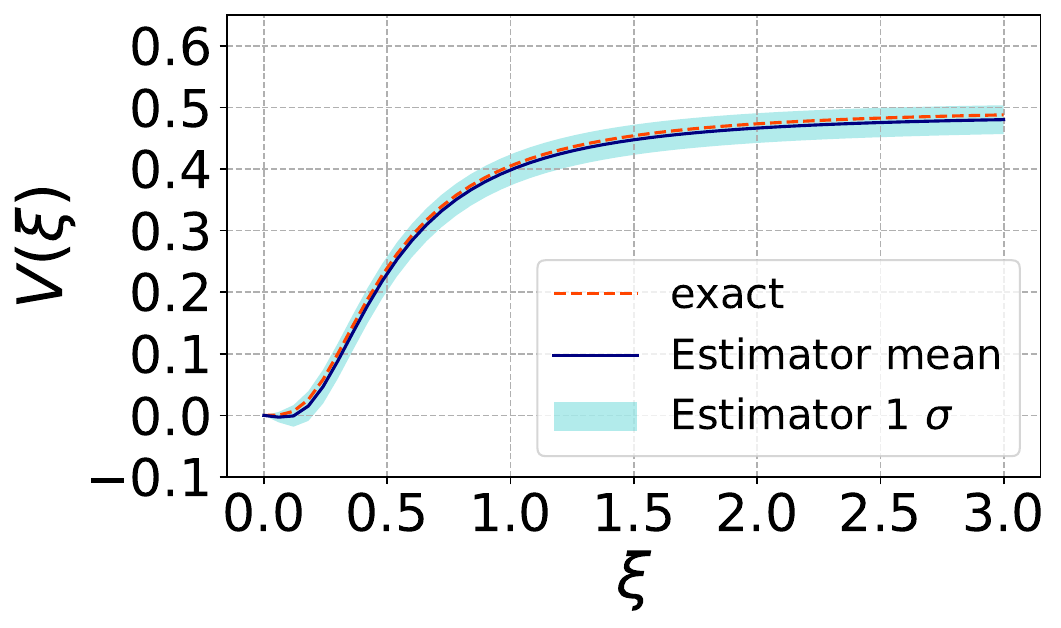}
        \\
        \includegraphics[width=1.0\linewidth]{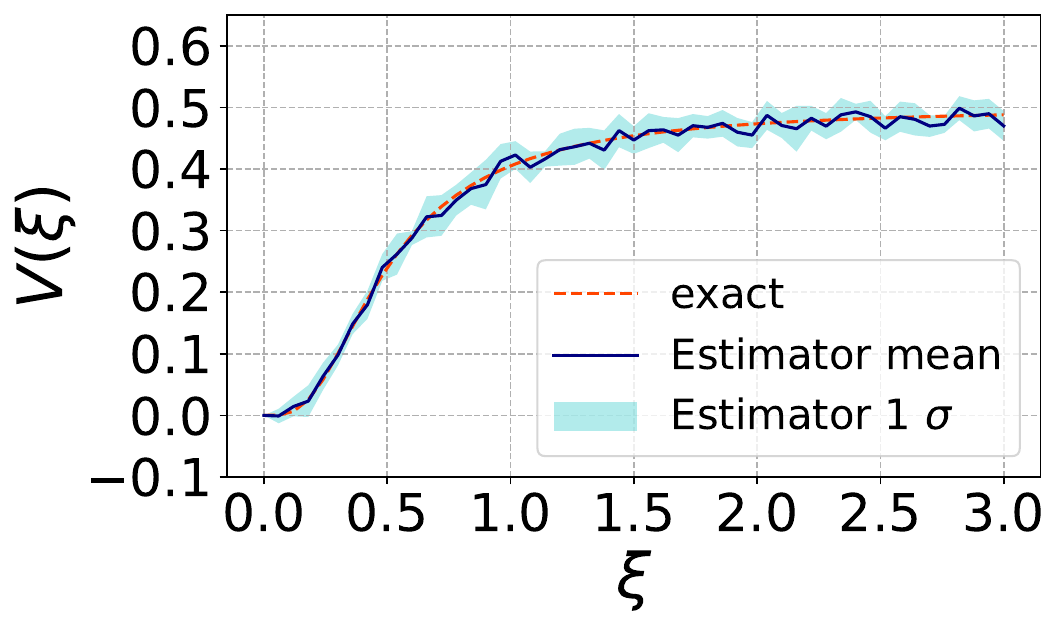}
        \caption{$N=10000$}
    \end{subfigure}
    \caption{Case 1: Estimated utility variance with (top row) and without (bottom row) CRS.}
    \label{fig:roed_1DLinGau_CRS}
\end{figure}

We next examine the effect of CRS introduced in \cref{sec:roed_common_random_samples}. \Cref{fig:roed_1DLinGau_CRS} compare the estimated utility variance $\widehat{V}$ with (top row) and without (bottom row) CRS. Each curve is averaged over 10 independent runs to estimate the mean and variability of the estimator.
Using CRS produces a significantly smoother function, as expected, due to the induced correlation in MC noise across designs. Although CRS introduces a small shift in the estimated value under finite samples, this effect diminishes as $N$ increases and is outweighed by the improved smoothness. In the following experiments, we therefore adopt CRS throughout.

\begin{figure}[htbp]
  \centering
  \begin{subfigure}{0.31\linewidth}
    \centering
    \includegraphics[width=1.0\linewidth]{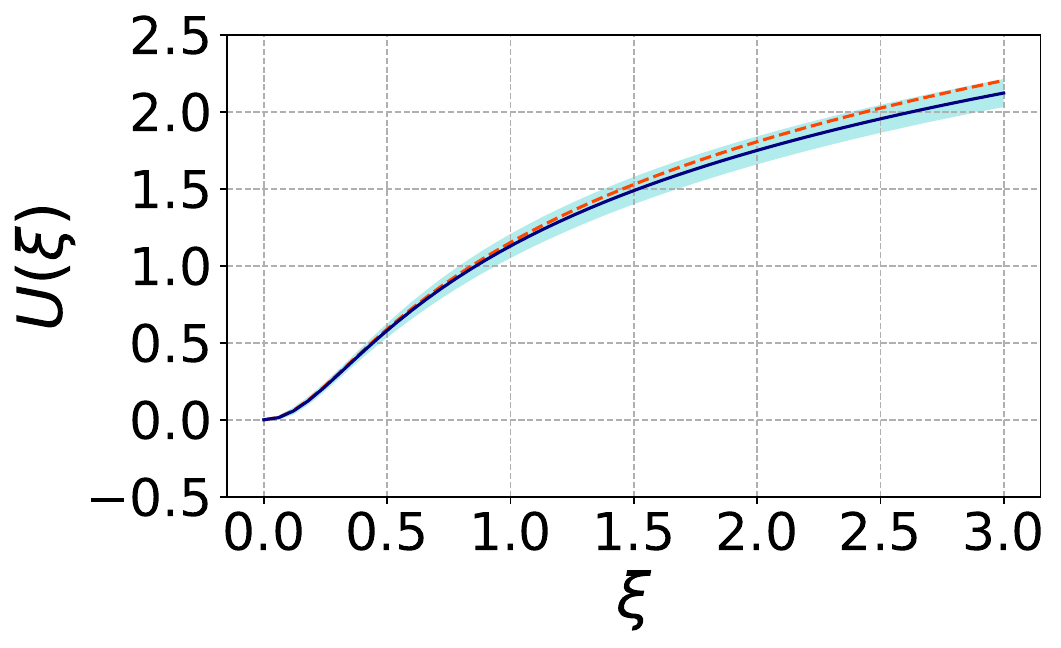}
    \caption{$N=100$}
    \label{fig:roed_1DLinGau_util_100}
  \end{subfigure}
  \hspace{0.5em}
  \begin{subfigure}{0.31\linewidth}
    \centering
    \includegraphics[width=1.0\linewidth]{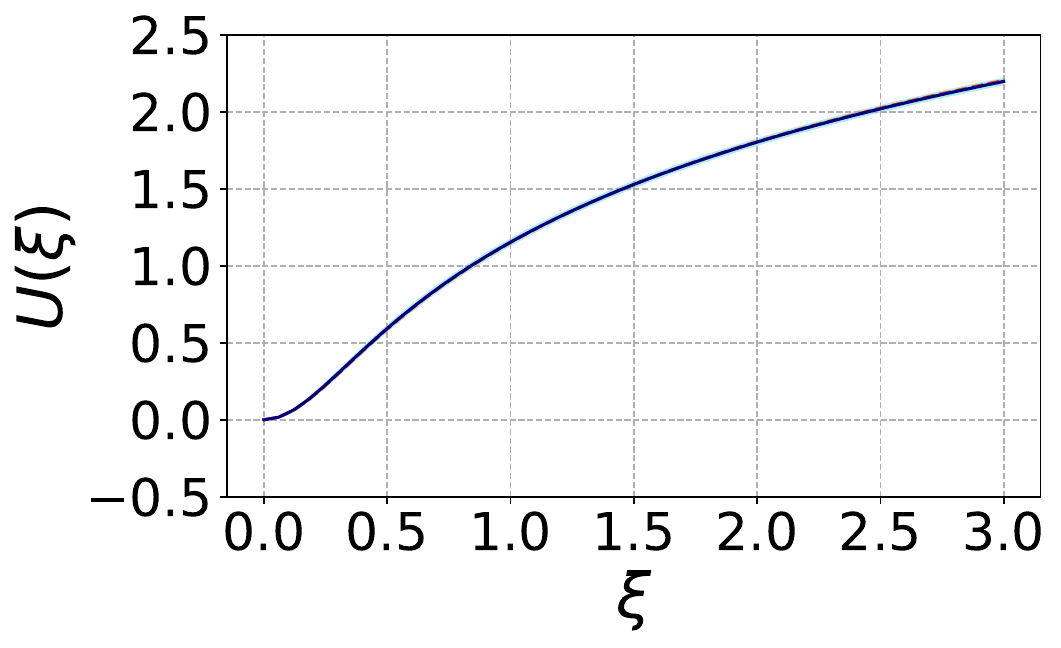}
    \caption{$N=1000$}
    \label{fig:roed_1DLinGau_util_1000}
  \end{subfigure}
  \hspace{0.5em}
  \begin{subfigure}{0.31\linewidth}
    \centering
    \includegraphics[width=1.0\linewidth]{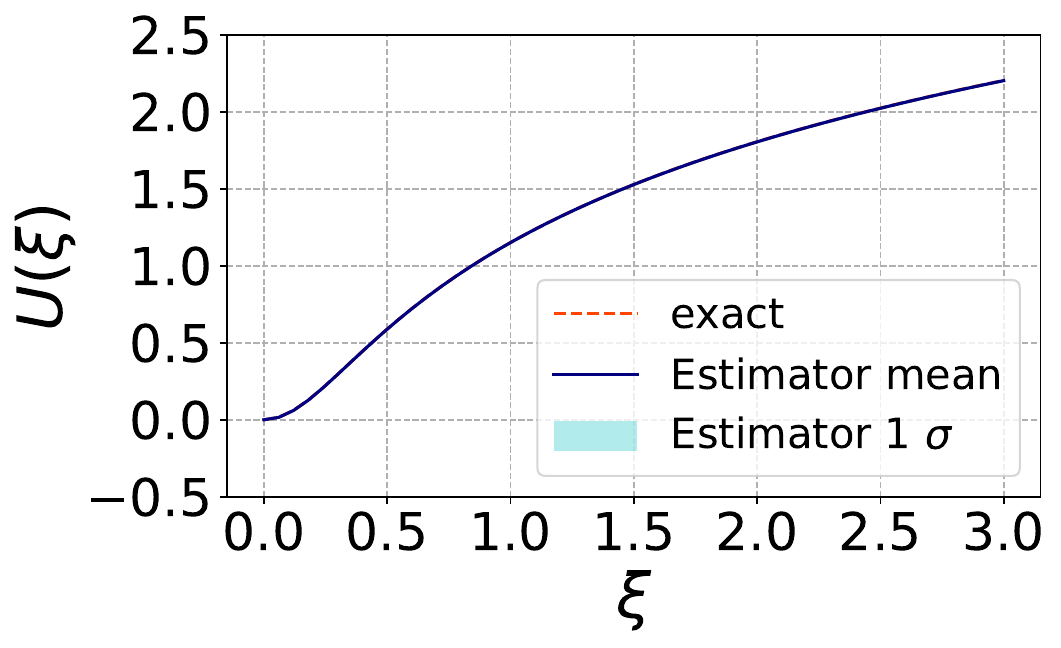}
    \caption{$N=10000$}
    \label{fig:roed_1DLinGau_util_10000}
  \end{subfigure}
  \caption{Case 1: Estimated and exact expected utility.}
  \label{fig:roed_1DLinGau_util}
\end{figure}

Finally, \cref{fig:roed_1DLinGau_util} compares the estimated expected utility with the exact solution with CRS. As $N$ increases, the estimator converges rapidly, achieving high accuracy with relatively small sample sizes (e.g., $N=1000$).

Taking \cref{fig:roed_1DLinGau_util,fig:roed_1DLinGau_CRS} together, the expected utility increases monotonically with $\design$, while the utility variance saturates beyond approximately $\design=2$. As a result, larger designs yield higher expected utility without a significant increase in variability. Consequently, for moderate values of $\lambda$, the optimal design under the mean--variance criterion coincides with the classical expected-utility optimum at the upper bound of the design domain.

\subsection{Case 2: Nonlinear test problem}
\label{sec:roed_ex_nonlinear}

We next consider a synthetic nonlinear model adapted from \cite{Huan2013}:
\begin{align}
    Y
    =
    G(\Param,\design)+\mathcal{E}
    =
    \Param^3 \design^2 + \Param \exp(-1.3|0.2-\design|) + \mathcal{E},
    \label{eq:roed_nonlin_model}
\end{align}
where $\Param \sim \mathcal{U}[0,1]$ is the scalar unknown parameter and $\mathcal{E} \sim \mathcal{N}(0,10^{-4}I)$ is additive Gaussian noise. We consider both one-dimensional (1D) and two-dimensional (2D) design spaces, namely $\design \in [0,1]$ and $\design \in [0,1]^2$. The 2D design corresponds to performing the experiment twice, with each component of $\design$ specifying the design variable for one experiment.

\begin{figure}[htbp]
  \centering
  \begin{subfigure}[t]{0.4\linewidth}
    \centering
    \includegraphics[width=\linewidth]{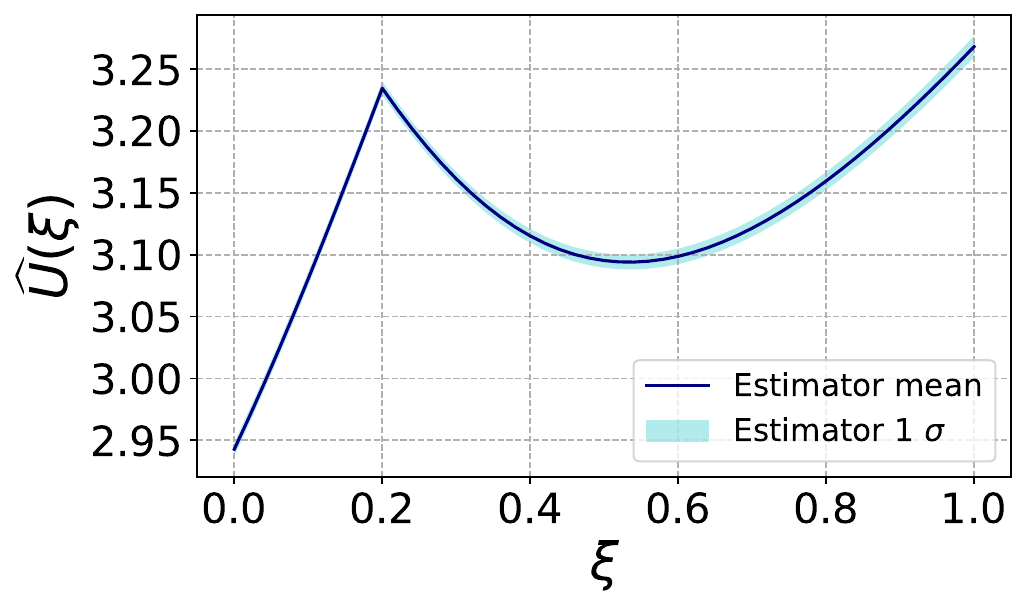}
    \caption{$\widehat{U}(\design)$}
    \label{fig:roed_1Dnonlin_util}
  \end{subfigure}
  \hspace{1em}
  \begin{subfigure}[t]{0.4\linewidth}
    \centering
    \includegraphics[width=\linewidth]{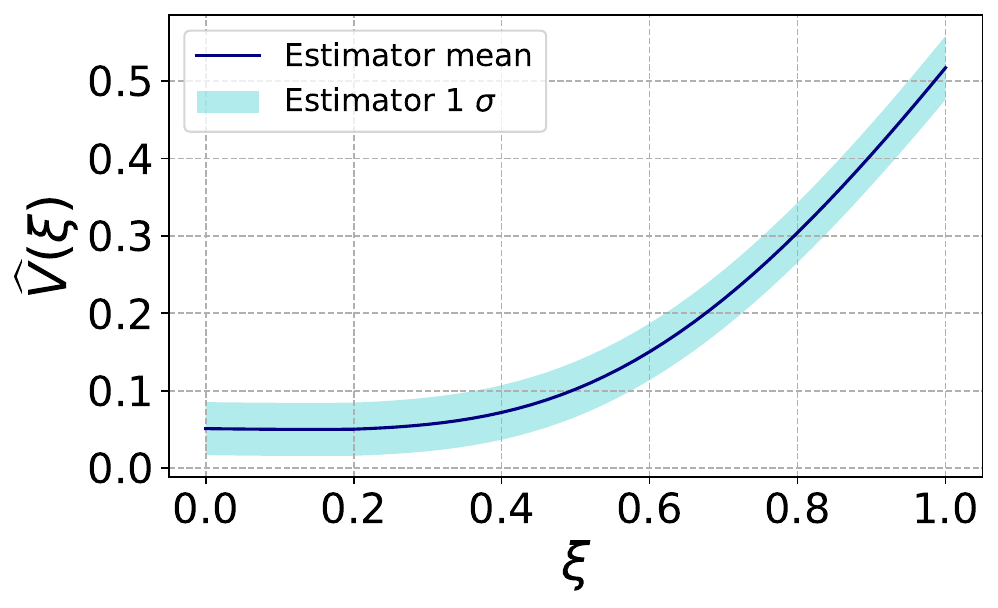}
    \caption{$\widehat{V}(\design)$}
    \label{fig:roed_1Dnonlin_utilvar_10000}
  \end{subfigure}
  \caption{Case 2: Estimated expected utility and utility variance.}
  \label{fig:roed_1Dnonlin_estimate}
\end{figure}

We begin with the 1D case. \Cref{fig:roed_1Dnonlin_estimate} shows the estimated expected utility and utility variance using $N=10000$ samples, averaged over 10 independent runs. As in the linear-Gaussian benchmark, the variance estimator is noisier than the expected-utility estimator at the same sample size, but $N=10000$ is sufficient to resolve the structure of the objective reliably.

\begin{figure}[htbp]
  \centering
  \begin{subfigure}[t]{0.4\linewidth}
    \centering
    \includegraphics[width=\linewidth]{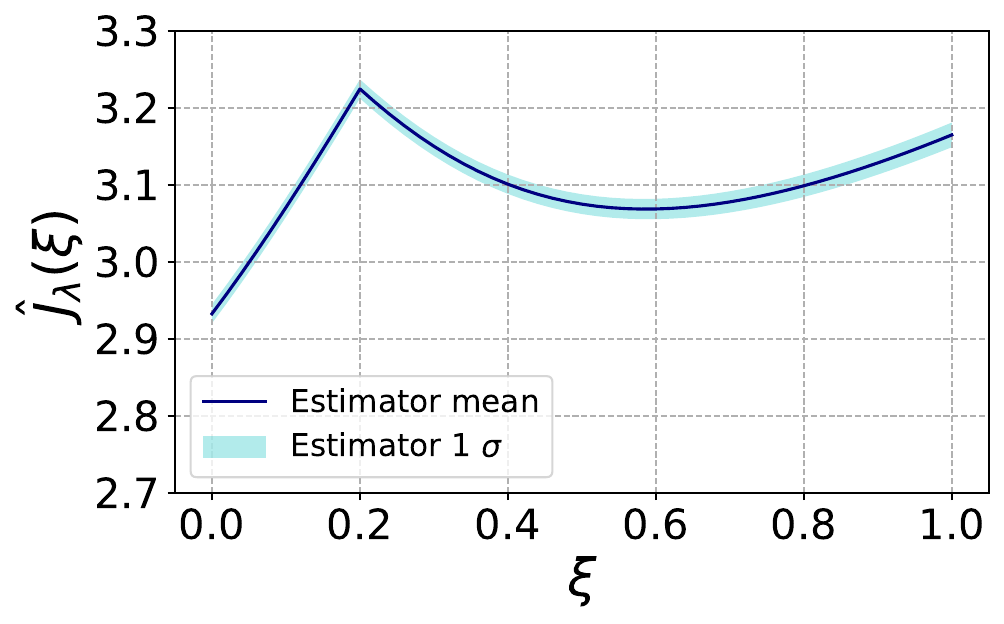}
    \caption{$\lambda=0.2$}
    \label{fig:roed_1Dnonlin_uvp_lambda_02}
  \end{subfigure}
  \hspace{1em}
  \begin{subfigure}[t]{0.4\linewidth}
    \centering
    \includegraphics[width=\linewidth]{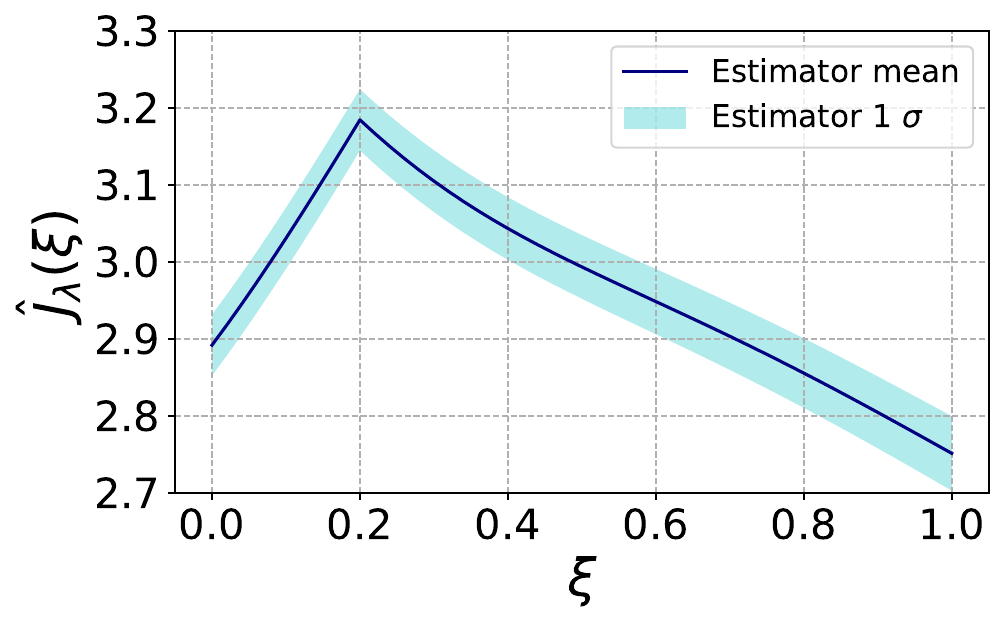}
    \caption{$\lambda=1$}
    \label{fig:roed_1Dnonlin_uvp_lambda_1}
  \end{subfigure}
  \caption{Case 2: Estimated mean--variance objective $\widehat{J}_\lambda(\design)$.}
  \label{fig:roed_1Dnonlin_uvp}
\end{figure}

A key feature of this example is that the design maximizing expected utility is not the same as the design preferred by the mean--variance criterion. Based on the expected utility alone, $\design=1$ is slightly better than $\design=0.2$, and would therefore be selected by standard Bayesian OED. However, the utility variance at $\design=1$ is much larger than at $\design=0.2$, indicating that $\design=1$ is substantially riskier. This difference is reflected in \cref{fig:roed_1Dnonlin_uvp}: for $\lambda=0.2$, the mean--variance objective already favors $\design=0.2$ over $\design=1$, and for $\lambda=1$ the preference becomes much stronger, with $\design=1$ becoming the worst choice in the design domain.

\begin{figure}[htbp]
  \centering
  \begin{subfigure}[t]{0.4\linewidth}
    \centering
    \includegraphics[width=\linewidth]{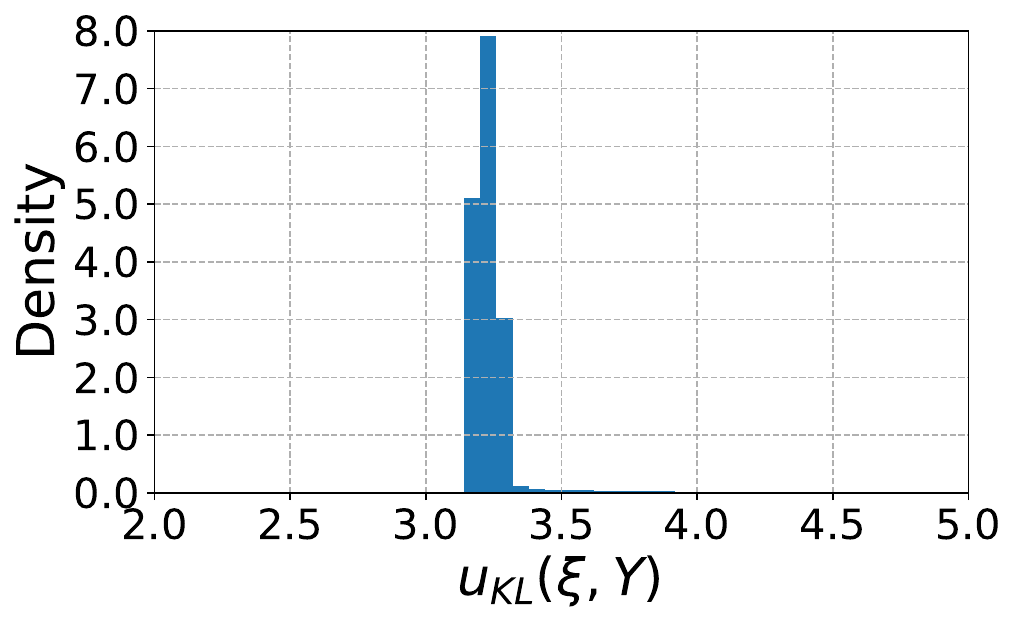}
    \caption{$\design=0.2$}
    \label{fig:roed_1Dnonlin_hist_d_02}
  \end{subfigure}
  \hspace{1em}
  \begin{subfigure}[t]{0.4\linewidth}
    \centering
    \includegraphics[width=\linewidth]{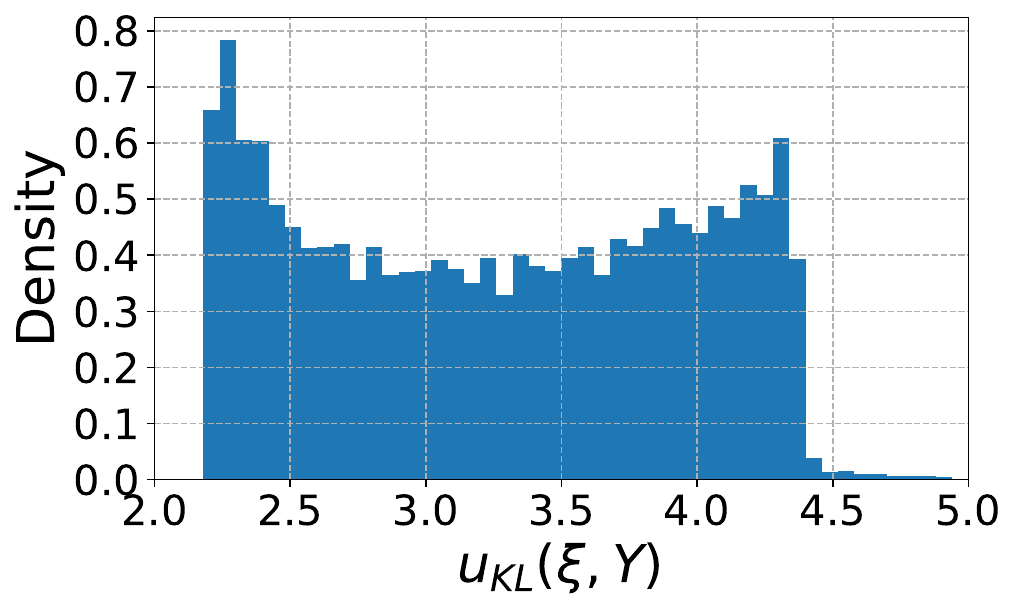}
    \caption{$\design=1$}
    \label{fig:roed_1Dnonlin_hist_d_1}
  \end{subfigure}
  \caption{Case 2: Histogram distributions of utility $u_{\mathrm{KL}}(\design,Y)$.}
  \label{fig:roed_1Dnonlin_hist}
\end{figure}

To better understand this behavior, \cref{fig:roed_1Dnonlin_hist} shows the histogram distributions of $u_{\mathrm{KL}}(\design,Y)$ at $\design=0.2$ and $\design=1$, where the KL-divergence utility is computed by grid discretization over the parameter space. The utility at $\design=0.2$ is tightly concentrated, while the utility at $\design=1$ is much more broadly spread. Thus, although $\design=1$ can yield larger information gain on average, it also carries much greater risk of poor outcomes.

\begin{figure}[htbp]
  \centering
  \begin{subfigure}[t]{0.4\linewidth}
    \centering
    \includegraphics[width=\linewidth]{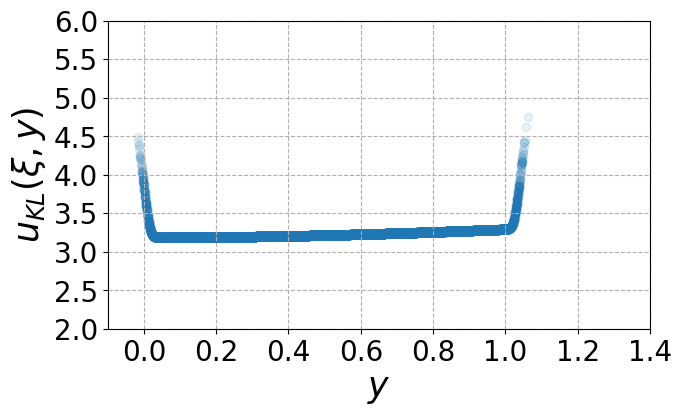}
    \caption{$\design=0.2$}
    \label{fig:roed_1Dnonlin_UvsY_d_02}
  \end{subfigure}
  \hspace{1em}
  \begin{subfigure}[t]{0.4\linewidth}
    \centering
    \includegraphics[width=\linewidth]{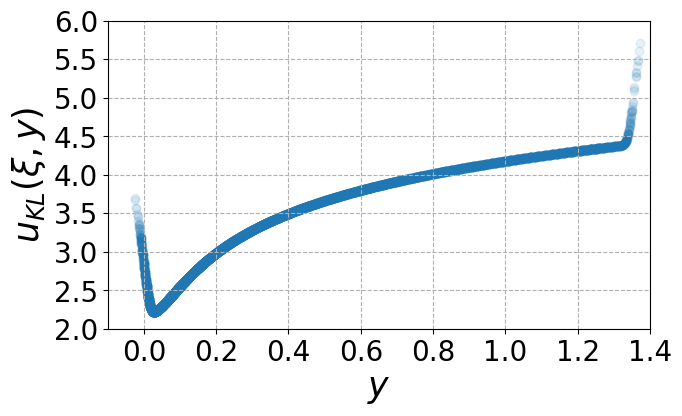}
    \caption{$\design=1$}
    \label{fig:roed_1Dnonlin_UvsY_d_1}
  \end{subfigure}
  \caption{Case 2: Scatter plots of $u_{\mathrm{KL}}(\design,y)$ versus $y$.}
  \label{fig:roed_1Dnonlin_UvsY}
\end{figure}

\begin{figure}[htbp]
  \centering
  \begin{subfigure}[t]{0.4\linewidth}
    \centering
    \includegraphics[width=\linewidth]{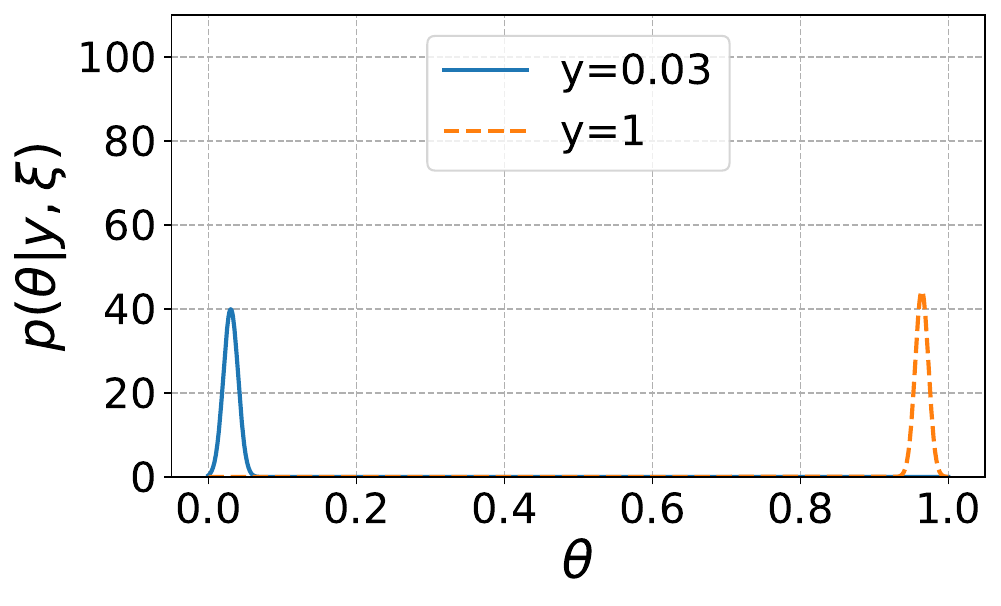}
    \caption{$\design=0.2$}
    \label{fig:roed_1Dnonlin_post_d_02}
  \end{subfigure}
  \hspace{1em}
  \begin{subfigure}[t]{0.4\linewidth}
    \centering
    \includegraphics[width=\linewidth]{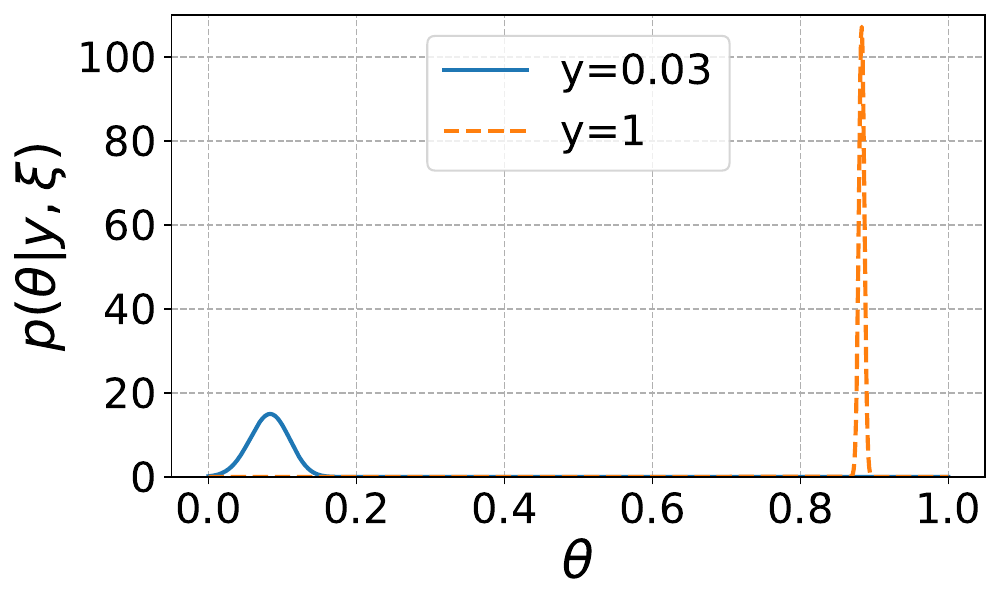}
    \caption{$\design=1$}
    \label{fig:roed_1Dnonlin_post_d_1}
  \end{subfigure}
  \caption{Case 2: Posterior densities $\pdf(\param|y,\design)$ for $y=0.03$ and $y=1$.}
  \label{fig:roed_1Dnonlin_post}
\end{figure}

This difference is further illustrated in \cref{fig:roed_1Dnonlin_UvsY}, which plots $u(\design,y)$ against $y$. At $\design=0.2$, the utility is relatively insensitive to the observation over a large portion of the observation range. At $\design=1$, in contrast, the utility changes much more strongly with $y$, producing the larger variance. \Cref{fig:roed_1Dnonlin_post} shows representative posterior distributions conditioned on $y=0.03$ and $y=1$. These two observations lead to similar posterior uncertainty at $\design=0.2$, but markedly different posterior uncertainty at $\design=1$, again illustrating the greater stability of $\design=0.2$.

\begin{figure}[htbp]
  \centering
  \includegraphics[width=0.65\linewidth]{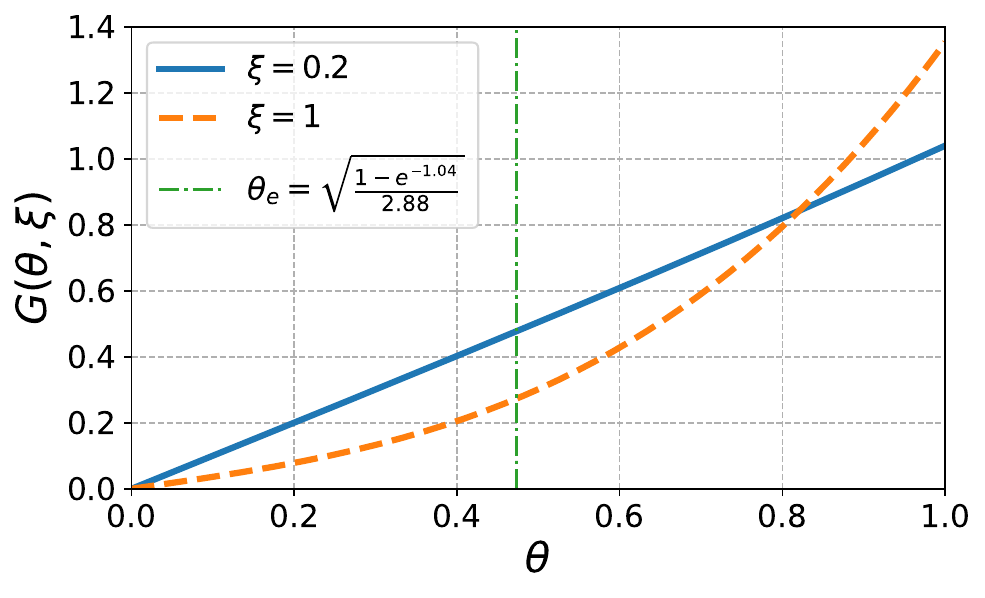}
  \caption{Case 2: Forward model $G(\param,\design)$ as a function of $\param$ for $\design=0.2$ and $\design=1$.}
  \label{fig:roed_1Dnonlin_model}
\end{figure}

This behavior can be understood from the forward model itself in     \cref{eq:roed_nonlin_model}. At $\design=0.2$, the model is dominated by the approximately linear term in $\param$, whereas at $\design=1$ the cubic term becomes dominant. As shown in \cref{fig:roed_1Dnonlin_model}, $G(\param,\design=0.2)$ is close to linear in $\param$, while $G(\param,\design=1)$ is strongly nonlinear. The slope of $G(\param,\design=0.2)$ is nearly constant across $\param$, whereas the slope of $G(\param,\design=1)$ varies substantially. Since information gain is intuitively connected to the sensitivity of the observation to the parameter, this variation in slope helps explain why $\design=1$ has a much larger utility variance.

\begin{figure}[htbp]
  \centering
  \begin{subfigure}[t]{0.4\linewidth}
    \centering
    \includegraphics[width=\linewidth]{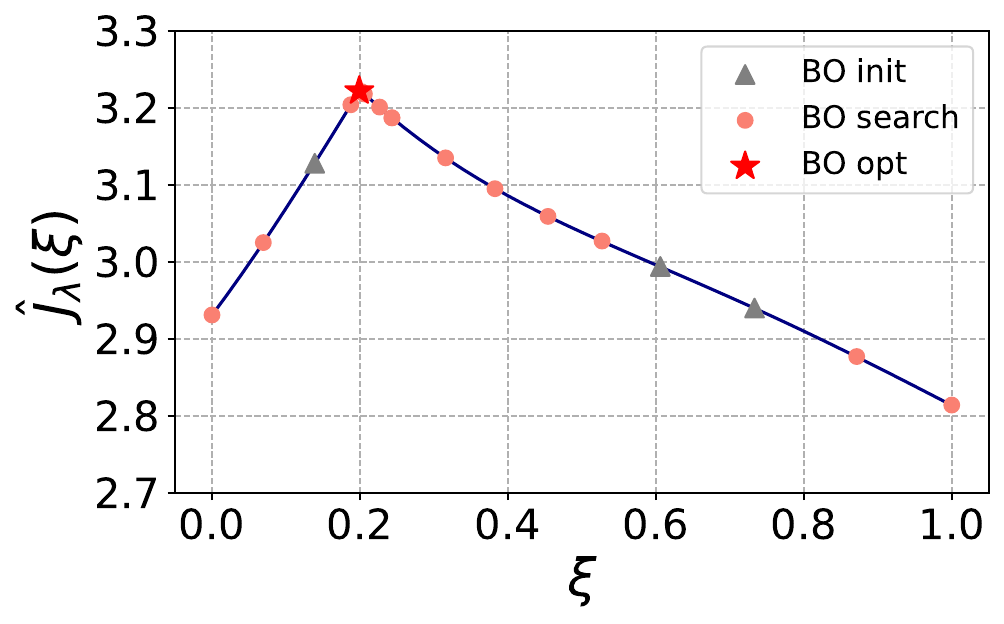}\\
    \includegraphics[width=\linewidth]{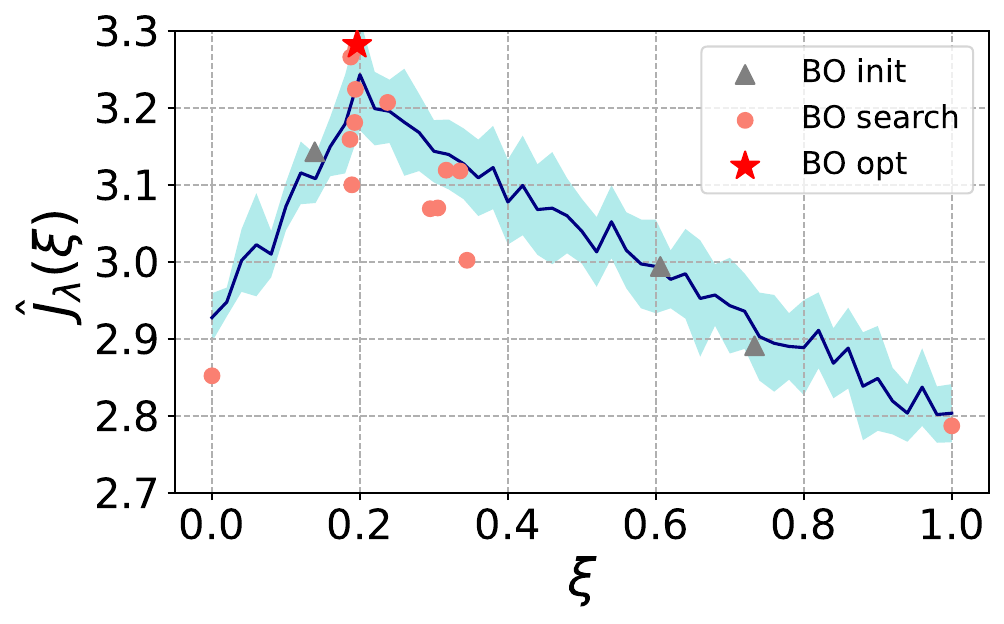}
    \caption{Evaluation locations overlaid on the objective estimate}
    \label{fig:roed_1Dnonlin_CRS_BO_obj}
  \end{subfigure}
  \hspace{1em}
  \begin{subfigure}[t]{0.4\linewidth}
    \centering
    \includegraphics[width=\linewidth]{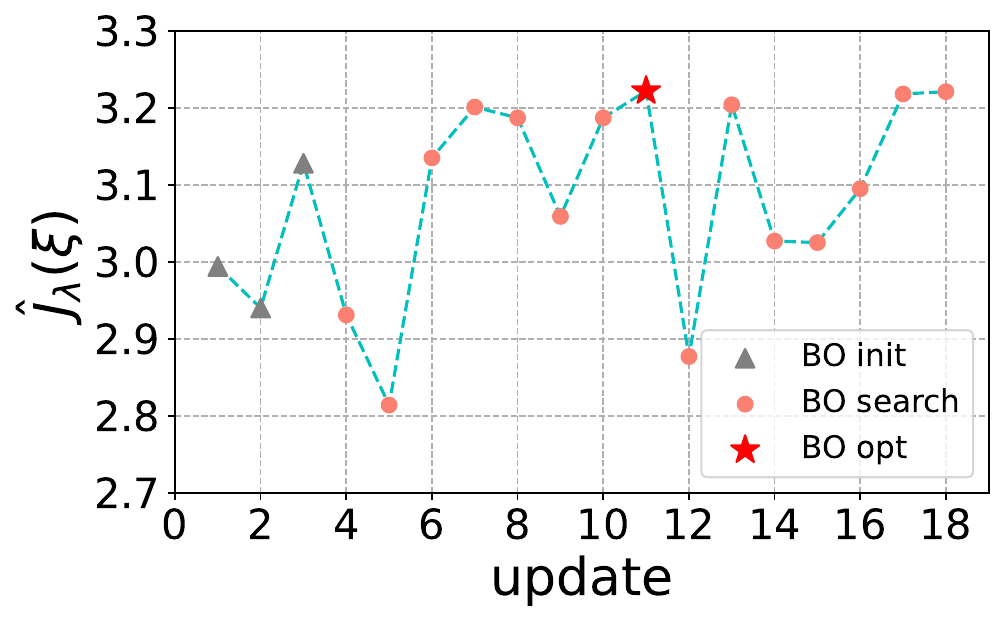}\\
    \includegraphics[width=\linewidth]{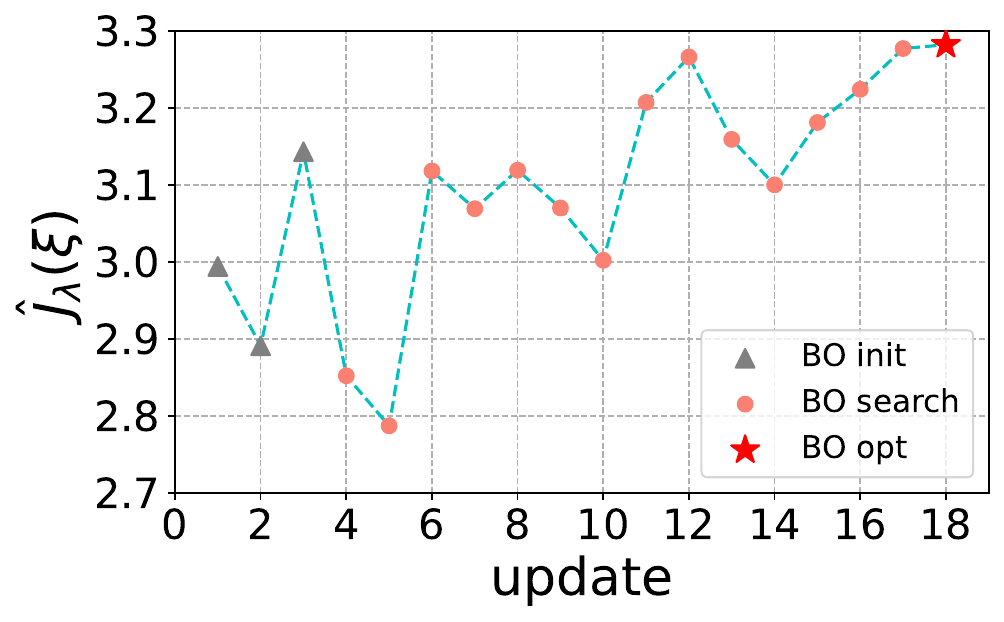}
    \caption{Optimization history}
    \label{fig:roed_1Dnonlin_CRS_BO_hist}
  \end{subfigure}
  \caption{Case 2: BO for $\lambda=1$ with (top row) and without (bottom row) CRS.}
  \label{fig:roed_1Dnonlin_CRS_noCRS_BO}
\end{figure}

We next examine the behavior of BO with $\lambda=1$. \Cref{fig:roed_1Dnonlin_CRS_noCRS_BO} (top row) shows the optimization history when CRS is used. BO rapidly identifies the optimal region near $\design=0.2$, while continuing to explore the design space in case better candidates exist. This behavior appears as occasional dips in the objective history after the optimum has already been found.
To assess BO under noisier objective evaluations, we repeat the optimization without CRS; the results are shown in \cref{fig:roed_1Dnonlin_CRS_noCRS_BO} (bottom row). BO still identifies the correct optimum, but the search becomes more locally concentrated around $\design=0.2$, with less exploration of the remainder of the design space. This behavior reflects increased uncertainty in the objective estimates: BO spends more evaluations refining the apparent optimum because the noise makes it less certain about the true landscape.

\begin{figure}[htbp]
  \centering
  \begin{subfigure}[t]{0.3\linewidth}
    \centering
    \includegraphics[width=\linewidth]{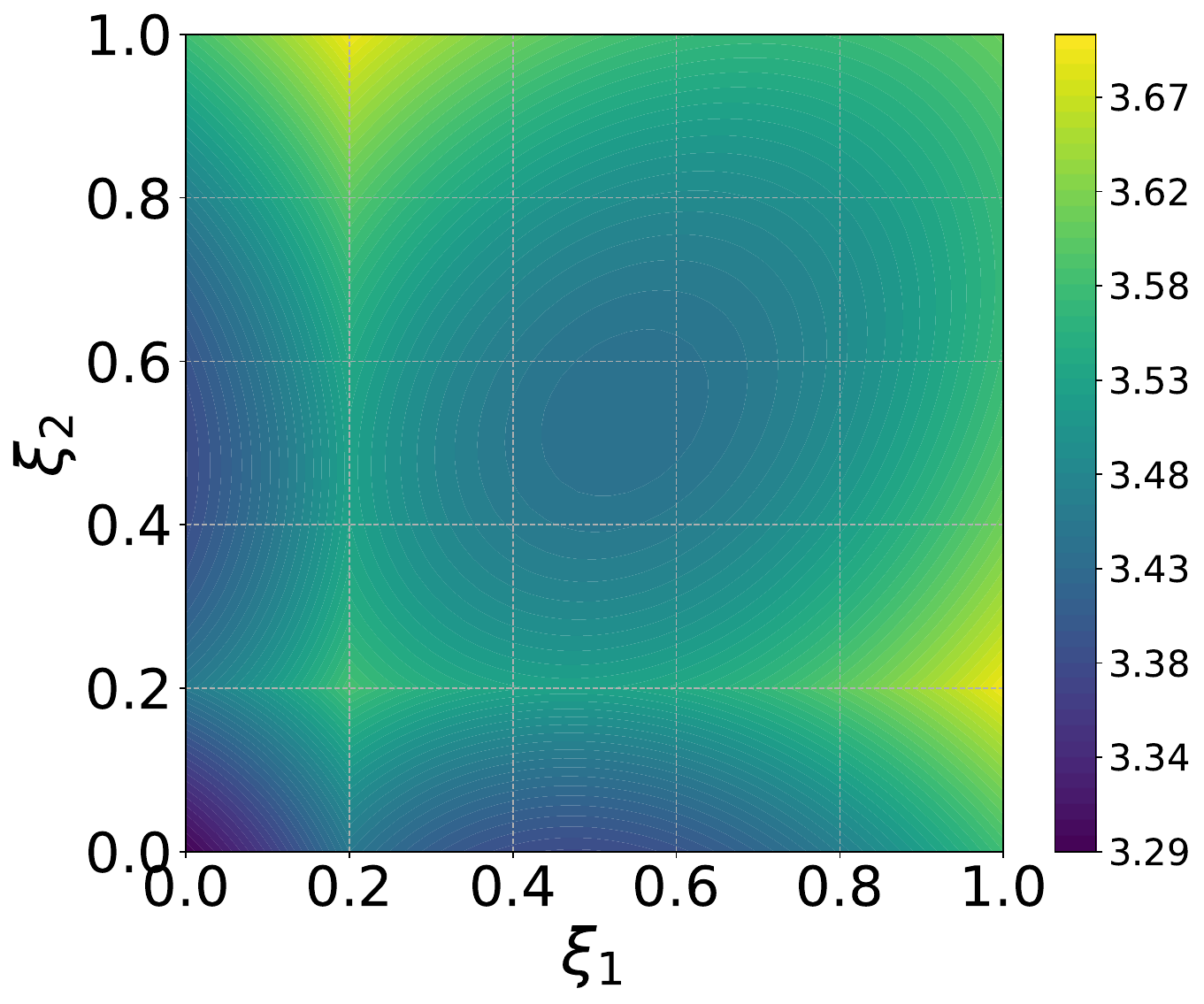}\\
    \includegraphics[width=\linewidth]{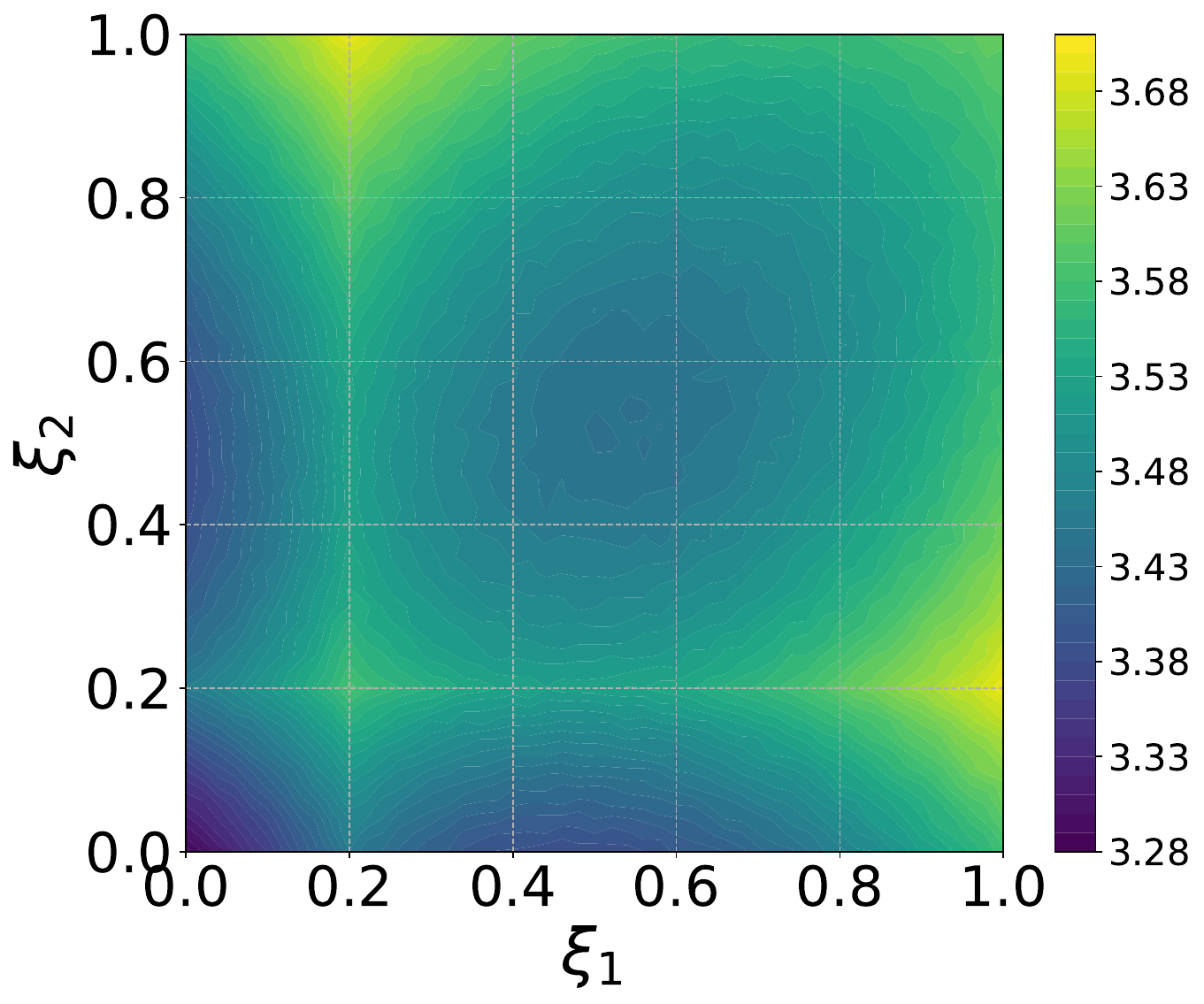}
    \caption{$\widehat{U}(\design)$}
    \label{fig:roed_2Dnonlin_CRS_util}
  \end{subfigure}
  \begin{subfigure}[t]{0.3\linewidth}
    \centering
    \includegraphics[width=\linewidth]{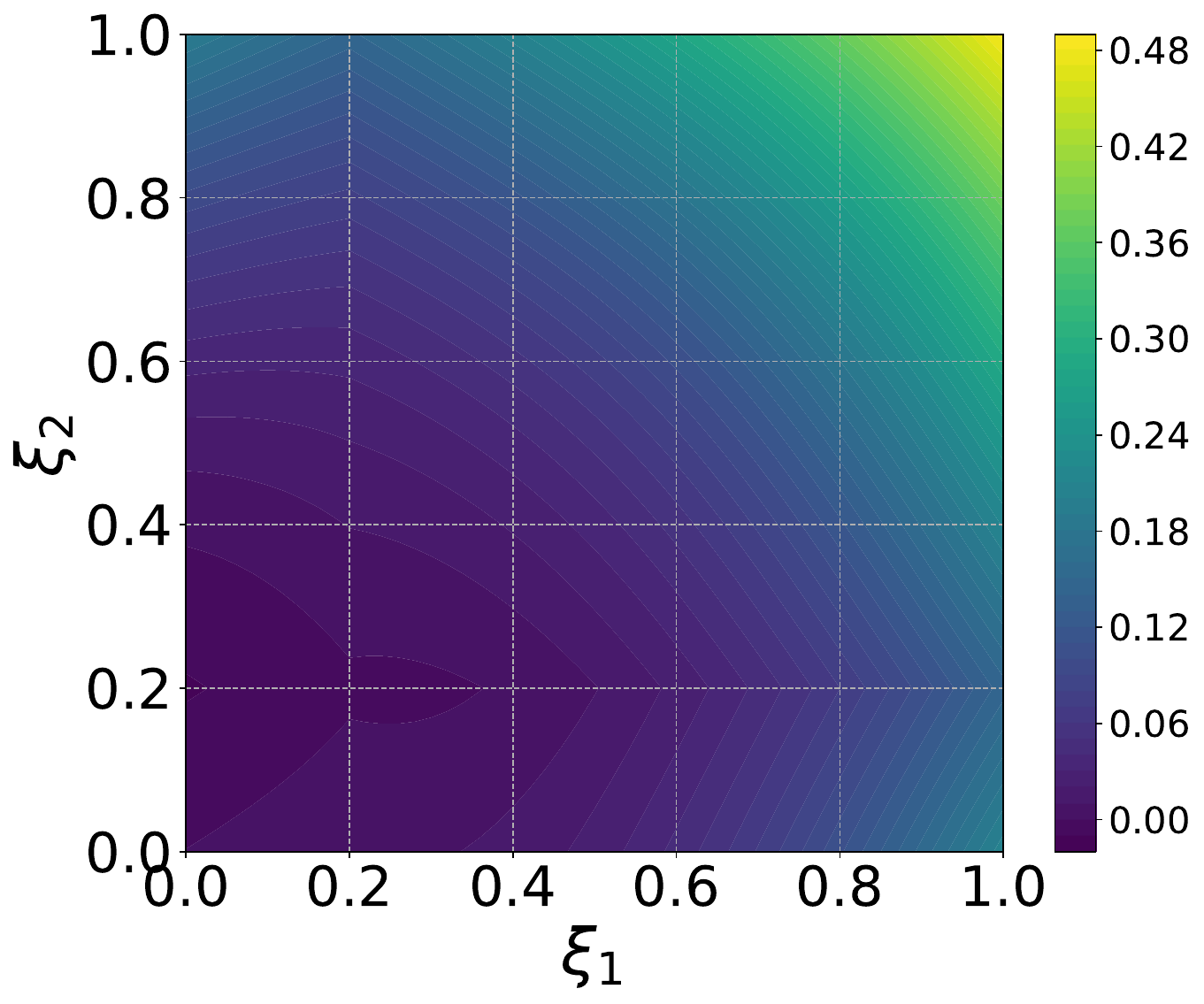}\\
    \includegraphics[width=\linewidth]{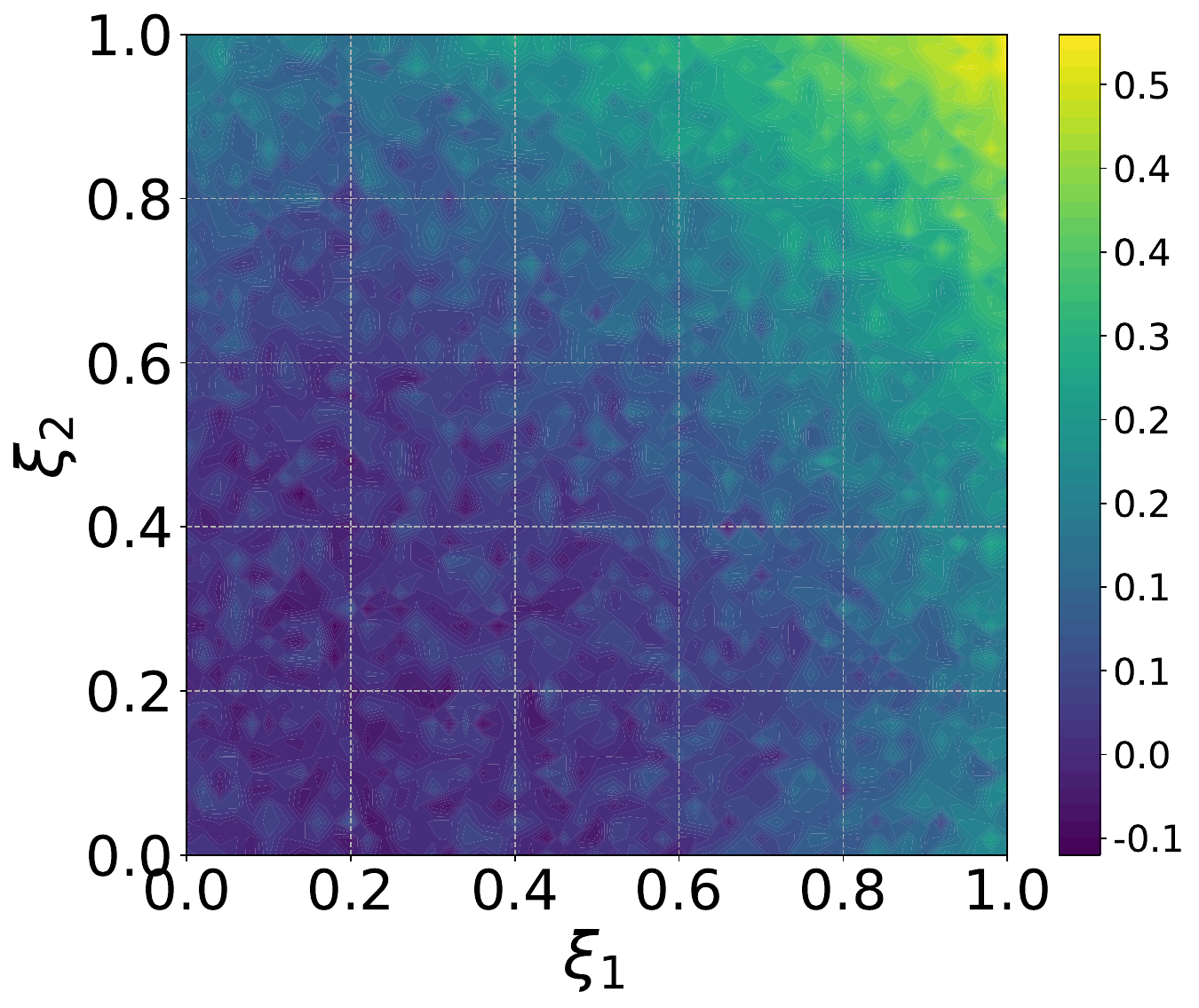}
    \caption{$\widehat{V}(\design)$}
    \label{fig:roed_2Dnonlin_CRS_utilval}
  \end{subfigure}
  \begin{subfigure}[t]{0.3\linewidth}
    \centering
    \includegraphics[width=\linewidth]{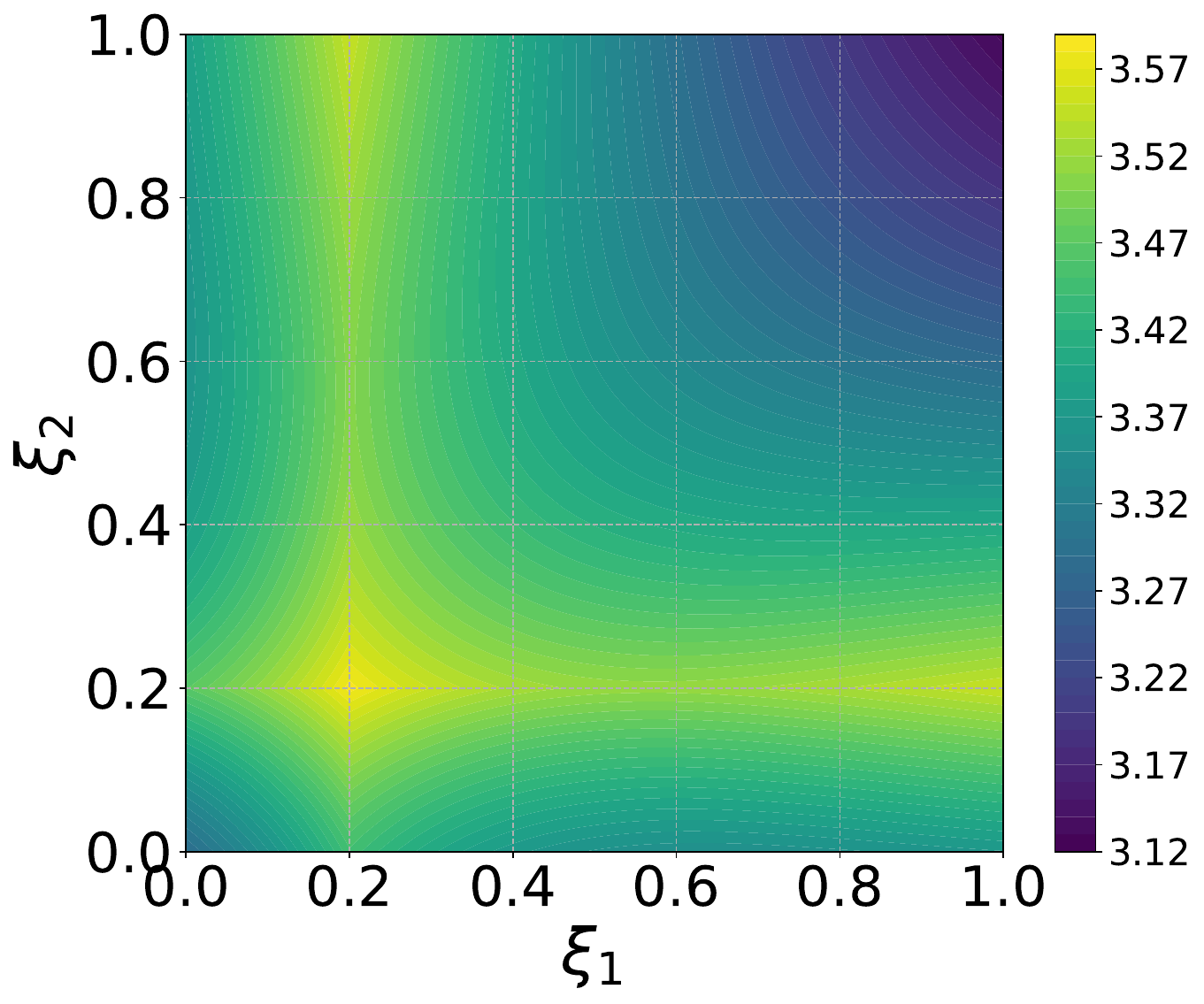}\\
    \includegraphics[width=\linewidth]{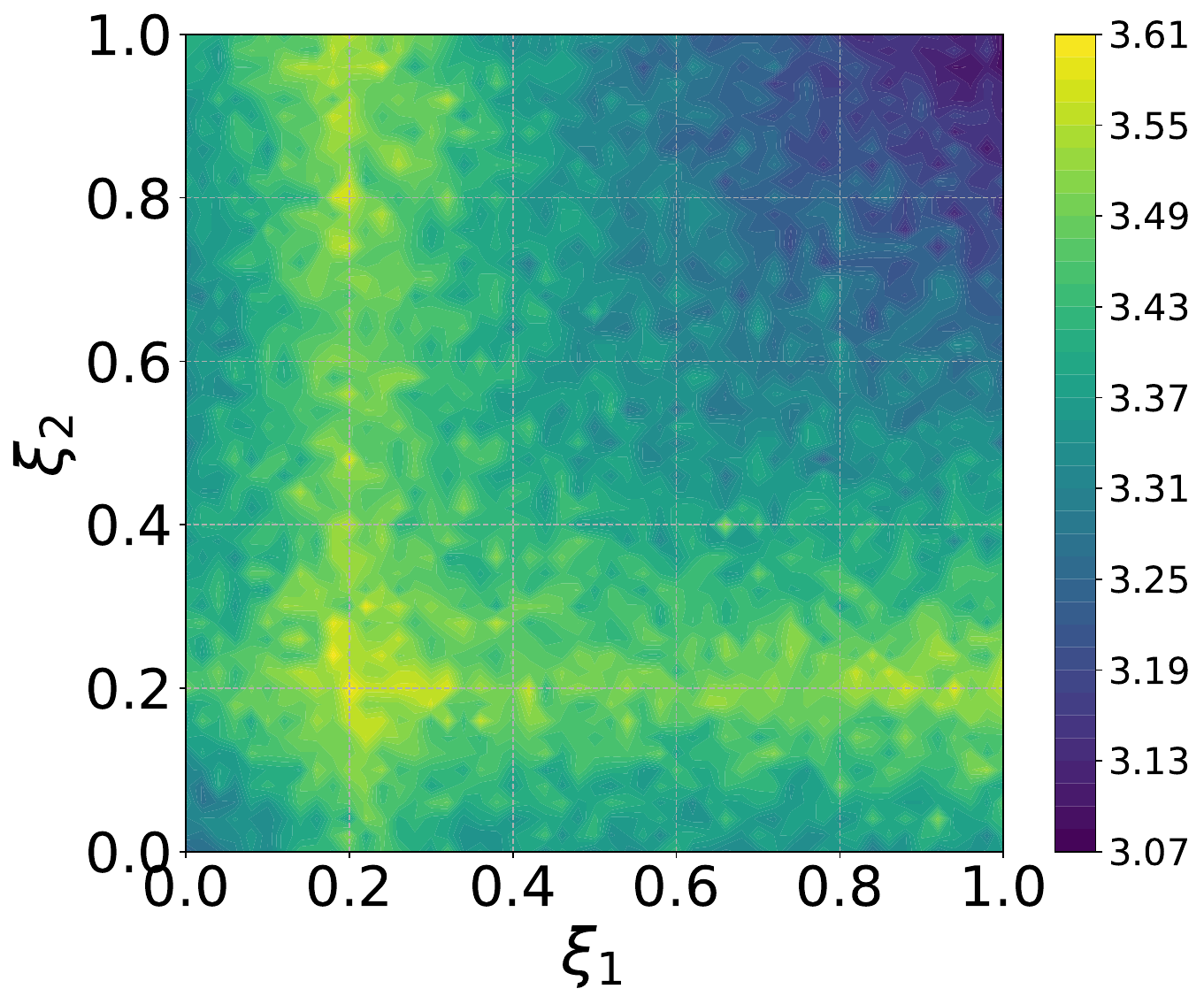}
    \caption{$\widehat{J}_{\lambda}(\design)$ with $\lambda=1$}
    \label{fig:roed_2Dnonlin_CRS_uvp}
  \end{subfigure}
  \caption{Case 2: Estimated expected utility, utility variance, and mean--variance objective with (top row) and without (bottom row) CRS.}
  \label{fig:roed_2Dnonlin_CRS}
\end{figure}

We now turn to the 2D design case. \Cref{fig:roed_2Dnonlin_CRS} shows contours of the estimated expected utility, utility variance, and mean--variance objective for $\lambda=1$ when CRS is used. The expected utility is maximized near $\design=[0.2,1]$ and $\design=[1,0.2]$, whereas the mean--variance criterion selects $\design=[0.2,0.2]$ as the optimal design. Thus, as in the 1D case, accounting for utility variance changes the preferred design substantially.
Using CRS introduces a small finite-sample shift, visible for example in the slight asymmetry of the lower-left region of the domain. Nevertheless, comparison shows that CRS dramatically smooth the objective landscape.

\begin{figure}[htbp]
  \centering
  \begin{subfigure}[t]{0.38\linewidth}
    \centering
    \includegraphics[width=\linewidth]{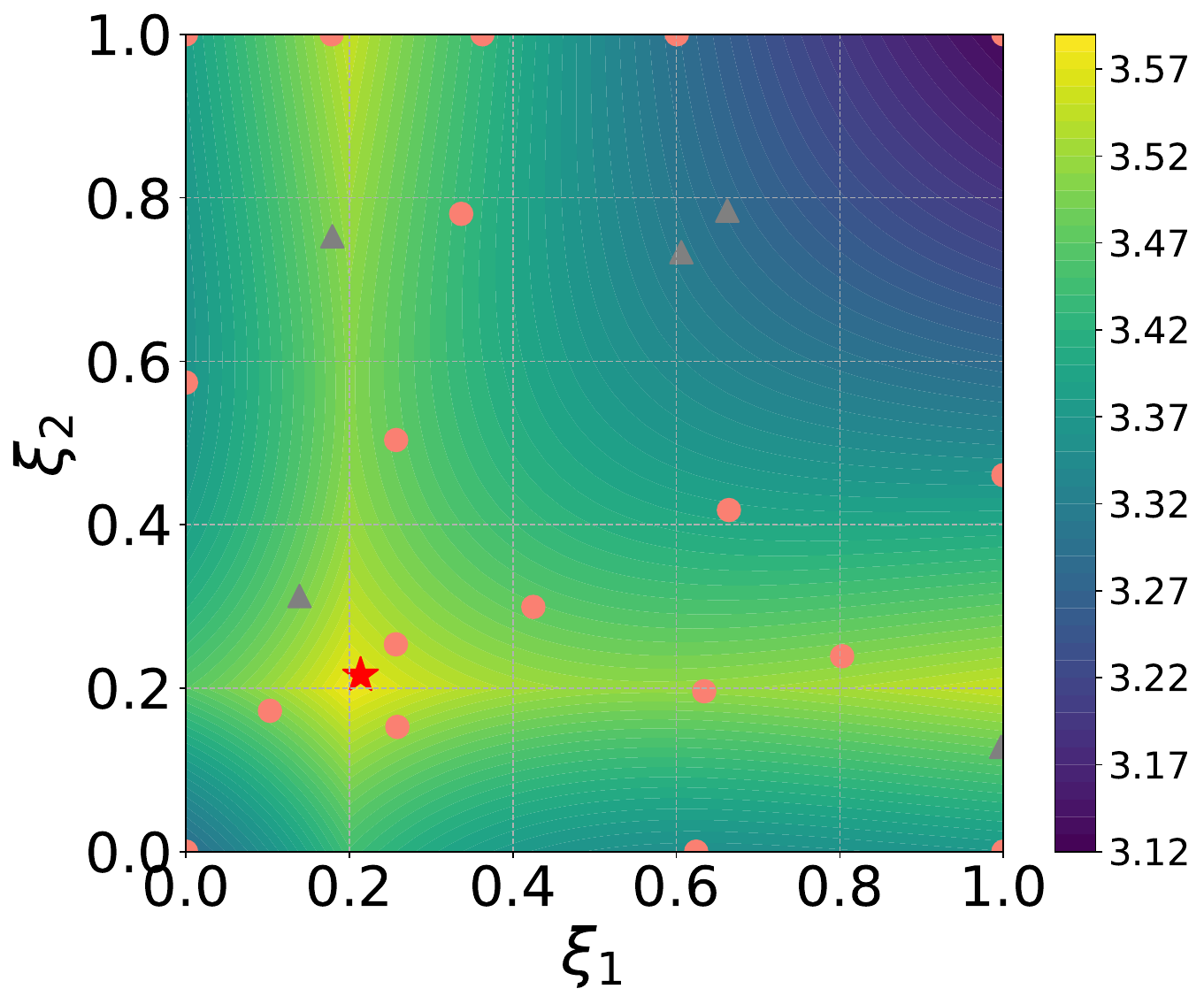}\\
    \includegraphics[width=\linewidth]{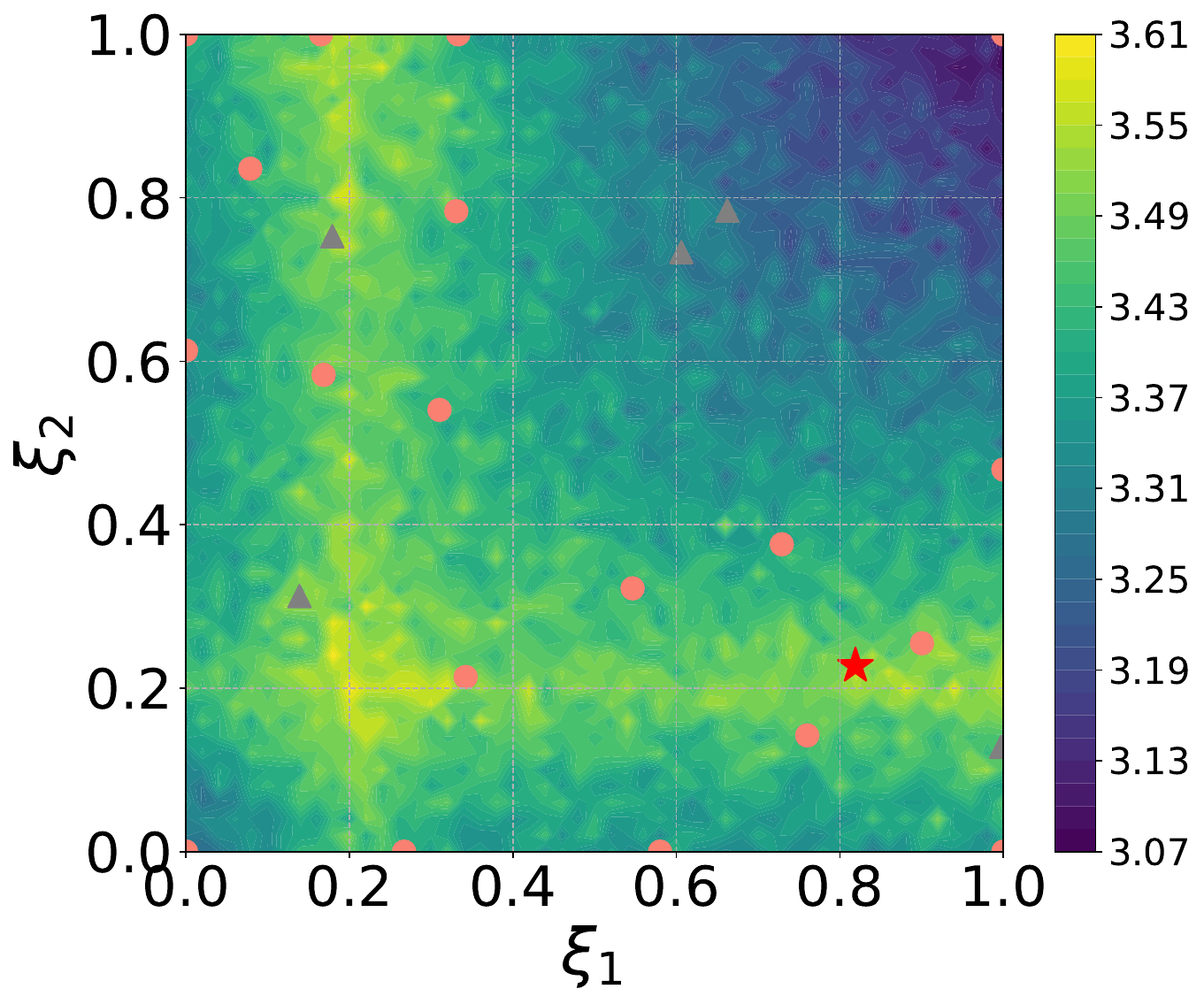}
    \caption{Evaluation locations overlaid on the objective contour}
    \label{fig:roed_2Dnonlin_CRS_BO_contour}
  \end{subfigure}
  \hspace{1em}
  \begin{subfigure}[t]{0.48\linewidth}
    \centering
    \includegraphics[width=\linewidth]{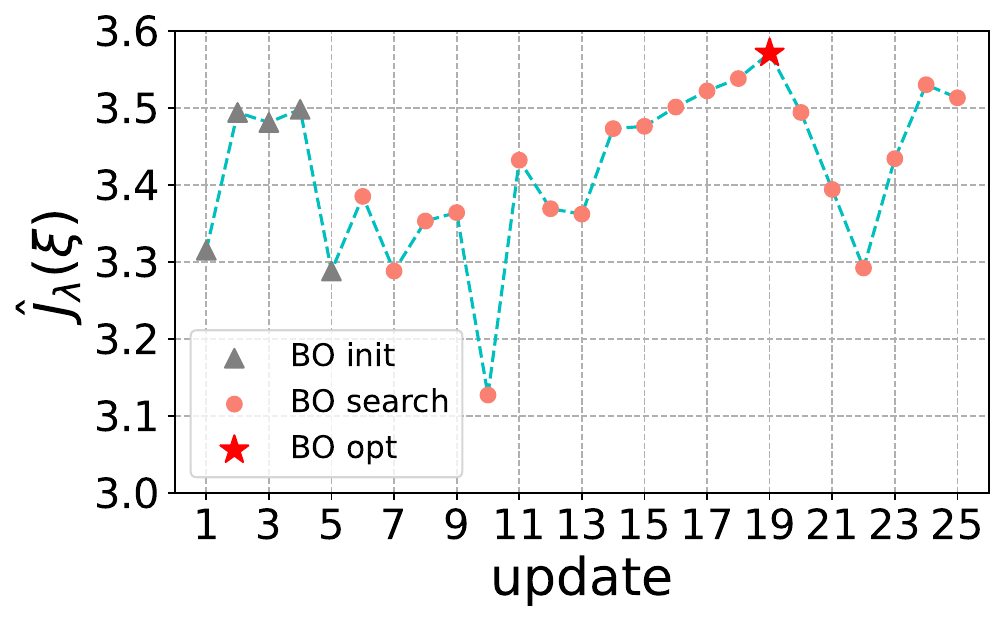}\\
    \includegraphics[width=\linewidth]{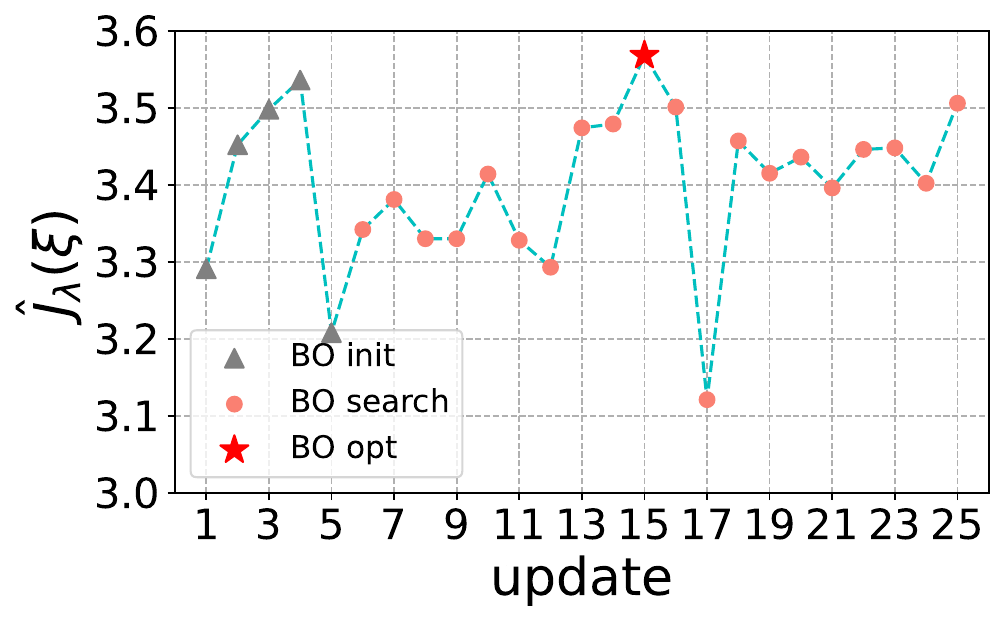}
    \caption{Optimization history}
    \label{fig:roed_2Dnonlin_CRS_BO_hist}
  \end{subfigure}
  \caption{Case 2: BO for $\lambda=1$ with (top row) and without (bottom row) CRS.}
  \label{fig:roed_2Dnonlin_CRS_noCRS_BO}
\end{figure}

The corresponding BO histories are shown in \cref{fig:roed_2Dnonlin_CRS_noCRS_BO}. With CRS (top row), BO successfully identifies the global optimum $\design^\ast_\lambda=[0.2,0.2]$, despite the presence of competing local optima near $[0.2,1]$ and $[1,0.2]$. Without CRS (bottom row), BO still detects the high-value cross-shaped region, but struggles to localize the true optimum. This supports again that smoothing the objective via CRS can substantially improve optimization performance.

\subsection{Case 3: Contaminant source inversion in a diffusion field}
\label{sec:roed_ex_source_wo_building}

\subsubsection{Problem setup}

We next consider a 2D contaminant source inversion problem. The contaminant concentration in the square domain $[0,1]^2$ is modeled by the scalar diffusion partial differential equation (PDE):
\begin{align}
    \frac{\partial G(z,t;\Param)}{\partial t}
    =
    \nabla^2 G + S(z,t;\Param),
    \qquad z\in [0,1]^2,\quad t>0,
    \label{eq:roed_PDE}
\end{align}
where $\Param=[\Param_x,\Param_y]\in \mathbb{R}^2$ denotes the unknown source location. We place a uniform prior on the source location, $\Param \sim \mathcal{U}[0,1]^2$. The source term is
\begin{align}
    S(z,t;\Param)
    =
    \frac{s}{2\pi h^2}
    \exp\left(
    -\frac{\|\Param-z\|^2}{2h^2}
    \right),
\end{align}
with source strength $s=2$ and source width $h=0.05$. The initial condition is $G(z,0;\Param)=0$, and homogeneous Neumann boundary conditions are imposed on all four sides of the domain.

The PDE is solved using a second-order finite-volume discretization on a uniform grid with $\Delta z_x=\Delta z_y=0.01$, together with a second-order fractional-step time integrator with $\Delta t=5\times 10^{-4}$.

The design variable is the sensor location used to measure the contaminant concentration. We consider a single measurement time, $t=0.16$. If $m$ sensors are used, then the design variable is
\begin{align}
    \design = [z^{(1)},\dots,z^{(m)}]\in [0,1]^{2m},
\end{align}
and the observation vector is
\begin{align}
    Y = [Y(z^{(1)}),\dots,Y(z^{(m)})]\in \mathbb{R}^m.
\end{align}
Thus the design dimension is $2m$, the observation dimension is $m$, and the parameter dimension remains $2$. The measurement model is
\begin{align}
    Y
    =
    G(z,t=0.16;\Param)+\mathcal{E},
\end{align}
where $\mathcal{E}\sim\mathcal{N}(0,0.05^2)$ for the one-sensor case, and for multiple sensors the measurement noise is assumed independent and identically distributed across sensors.

\subsubsection{Surrogate model}

Direct use of the PDE solver as the forward model is feasible but expensive. On a single 2.6 GHz CPU core, one forward solve requires approximately 1.2 seconds. As a result, estimating $J_\lambda(\design)$ with $N=10000$ MC samples would take roughly 3.3 hours for a single design.

To accelerate the computation, we replace the PDE solver with a deep neural network (DNN) surrogate for $G(z,t=0.16;\param)$. The surrogate takes the four-dimensional input $(\param,z)$ and outputs the scalar concentration value $G$. The network has five hidden layers with 100, 200, 100, 50, and 20 neurons, respectively, and uses ReLU activations. Training data are generated by sampling 1000 source locations uniformly from the parameter domain and evaluating the PDE solution on a uniform spatial grid. The resulting dataset is split into 80\% training data and 20\% testing data. After training, the test mean squared error is on the order of $10^{-6}$.

\begin{figure}[htbp]
  \centering
  \begin{subfigure}[t]{0.48\linewidth}
    \centering
    \includegraphics[width=\linewidth]{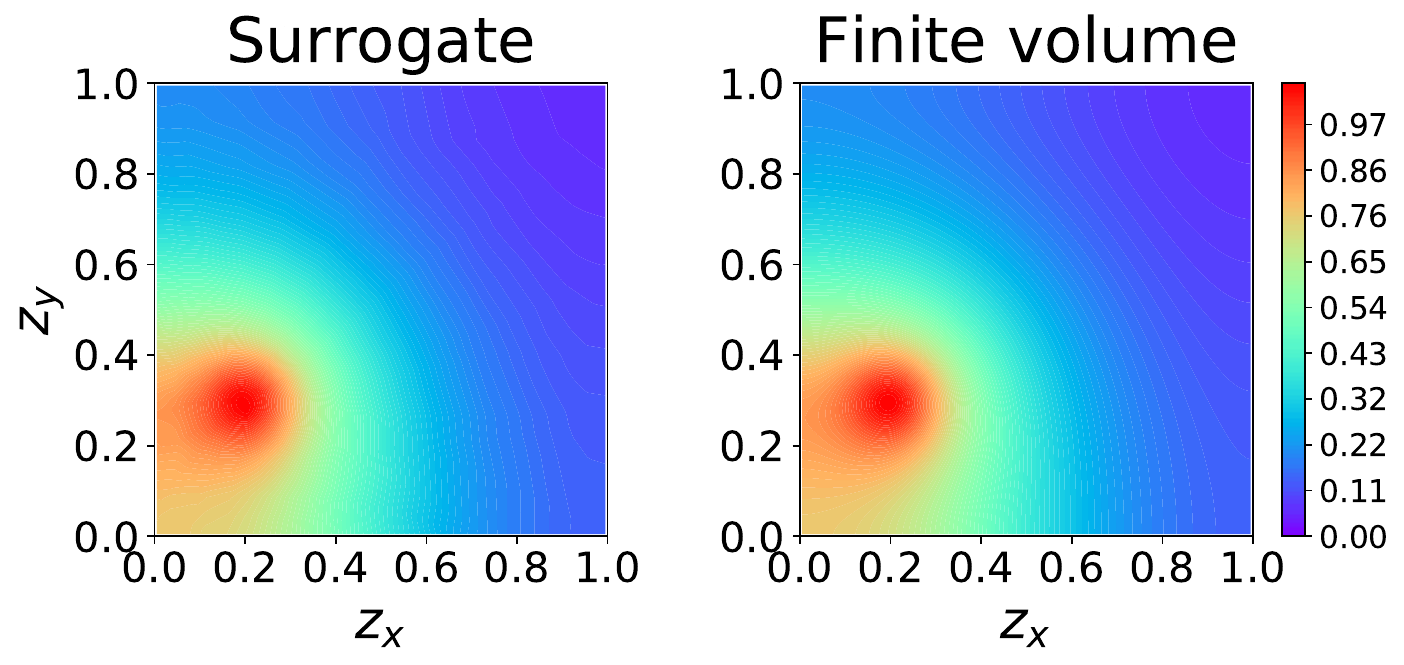}
    \caption{Example 1}
    \label{fig:roed_Source_surrogate_1_noBuilding}
  \end{subfigure}\hspace{1em}
  \begin{subfigure}[t]{0.48\linewidth}
    \centering
    \includegraphics[width=\linewidth]{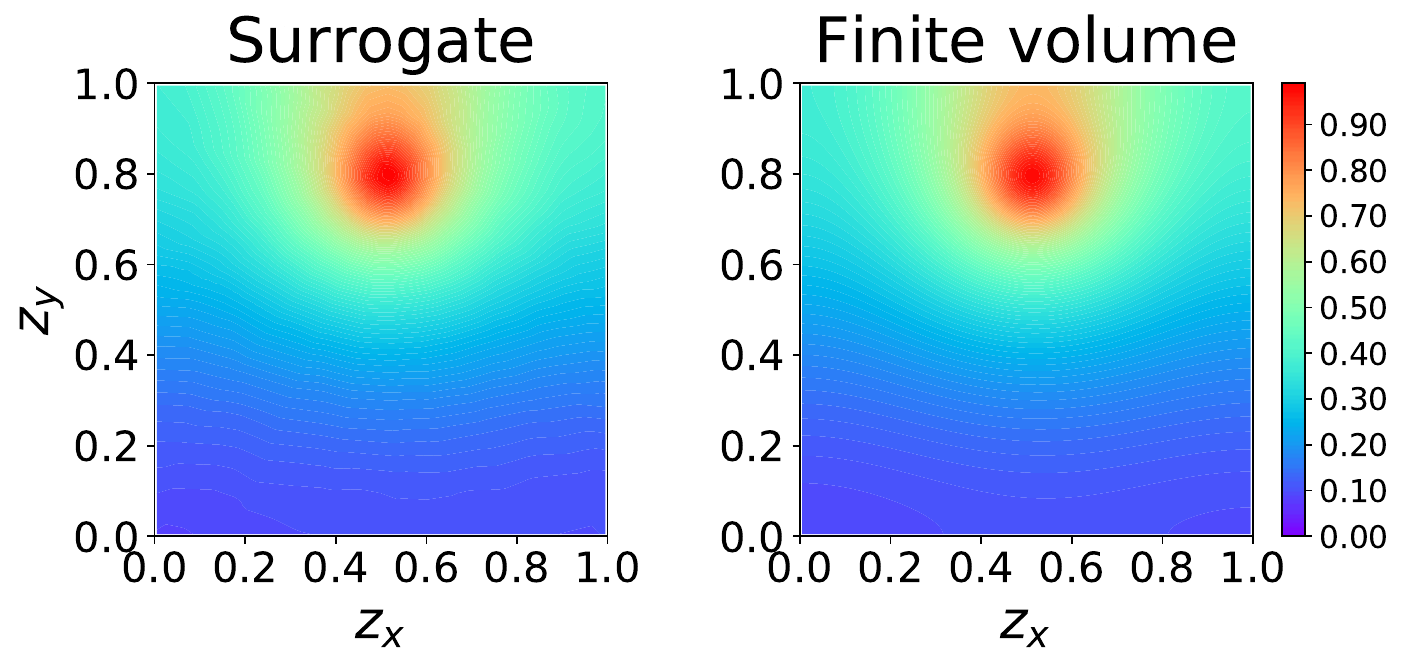}
    \caption{Example 2}
    \label{fig:roed_Source_surrogate_2_noBuilding}
  \end{subfigure}
  \caption{Case 3: Comparison of the concentration field $G$ computed by the DNN surrogate and the finite-volume solver.}
  \label{fig:roed_Source_surrogate_noBuilding}
\end{figure}

A representative comparison between the surrogate and the finite-volume solver is shown in \cref{fig:roed_Source_surrogate_noBuilding}. The two are visually indistinguishable. More importantly, the surrogate provides an approximate speedup of $10^5\times$.

\subsubsection{Results: One sensor}

We first consider the one-sensor case. All objective estimates use $N=30000$ MC samples together with CRS across different designs.

\begin{figure}[htbp]
  \centering
  \begin{subfigure}[t]{0.3\linewidth}
    \centering
    \includegraphics[width=\linewidth]{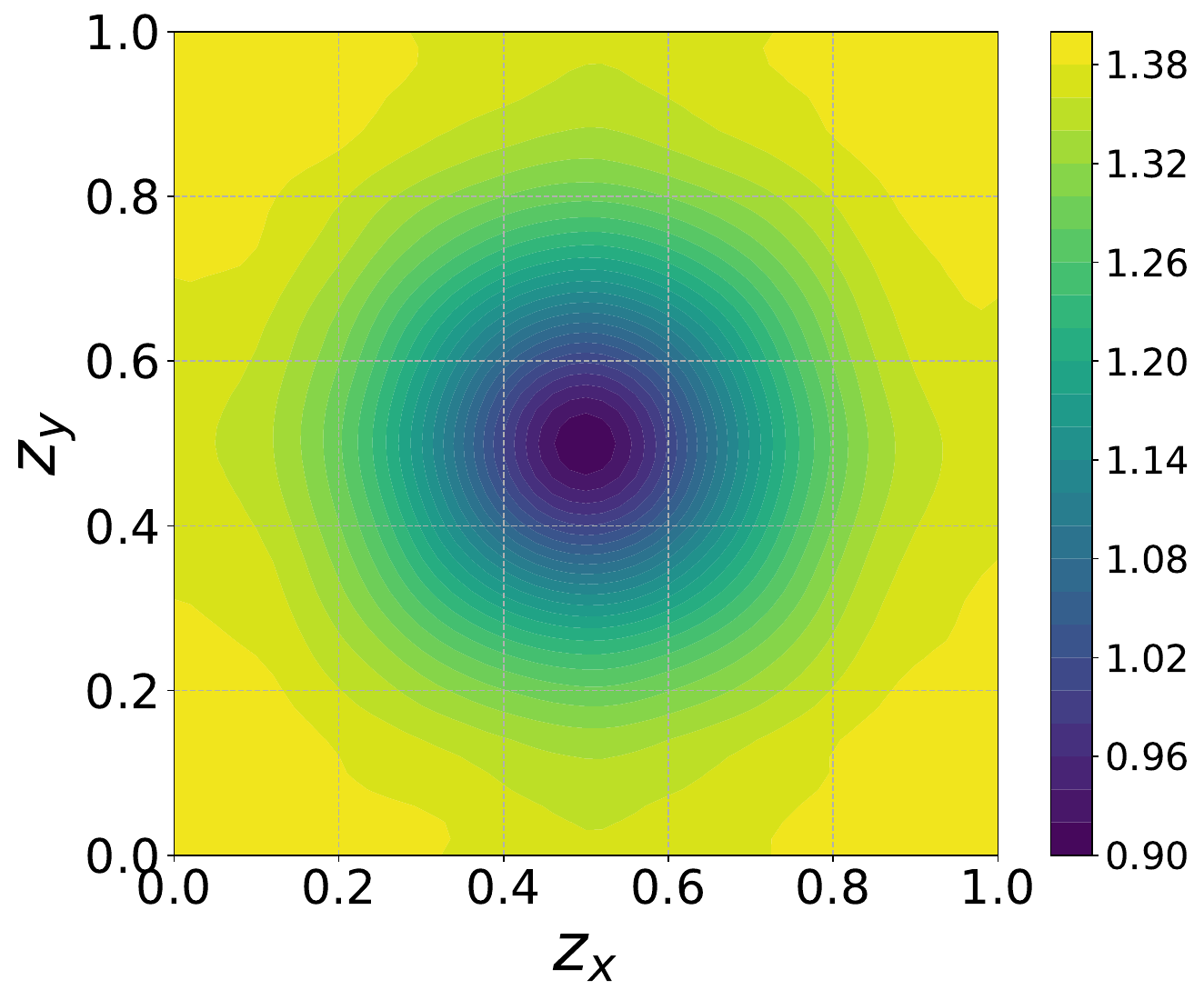}
    \caption{$\widehat{U}(\design)$}
    \label{fig:roed_2Dsource_1dsgn_b0_util}
  \end{subfigure}
  \begin{subfigure}[t]{0.3\linewidth}
    \centering
    \includegraphics[width=\linewidth]{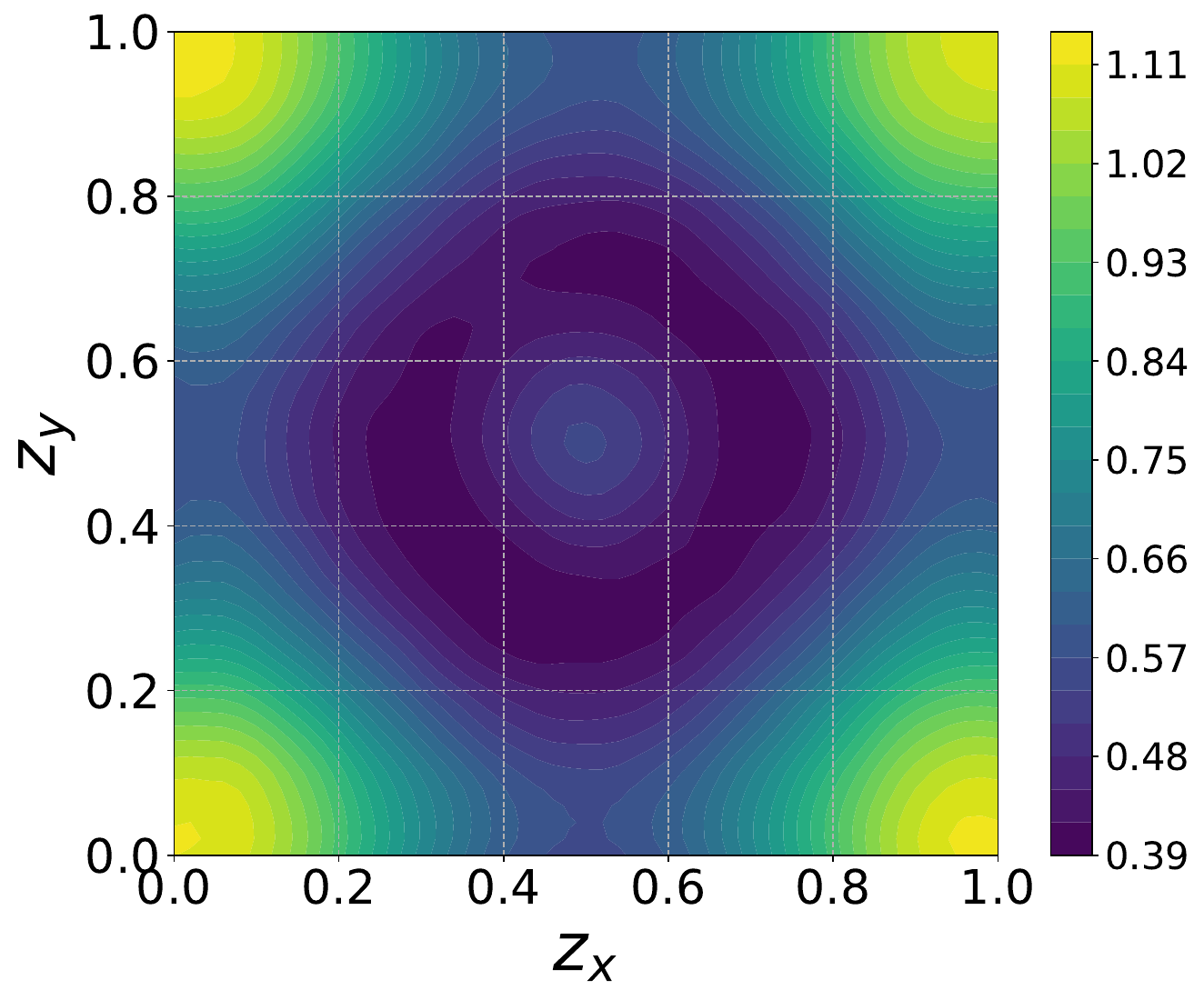}
    \caption{$\widehat{V}(\design)$}
    \label{fig:roed_2Dsource_1dsgn_b0_utilvar}
  \end{subfigure}
  \begin{subfigure}[t]{0.3\linewidth}
    \centering
    \includegraphics[width=\linewidth]{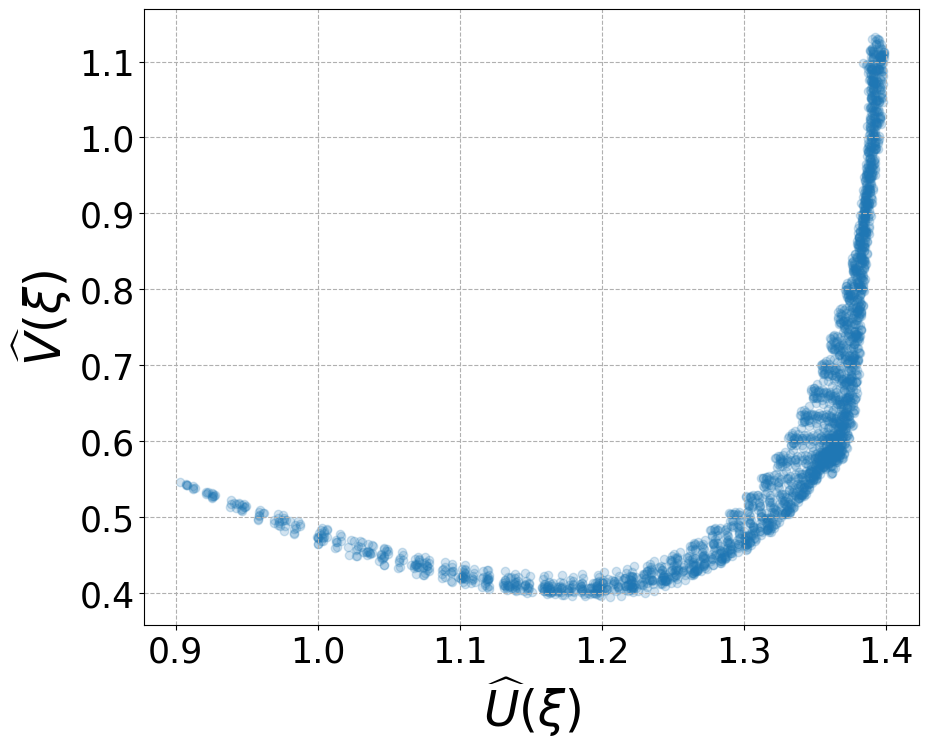}
    \caption{$\widehat{V}(\design)$ versus $\widehat{U}(\design)$}
    \label{fig:roed_2Dsource_1dsgn_b0_scatter}
  \end{subfigure}
  \caption{Case 3: Estimated expected utility, utility variance, and their trade-off for the one-sensor problem.}
  \label{fig:roed_2Dsource_1dsgn_b0}
\end{figure}

\Cref{fig:roed_2Dsource_1dsgn_b0} shows the estimated expected utility, utility variance, and the corresponding scatter plot of variance against expected utility. The expected utility in \cref{fig:roed_2Dsource_1dsgn_b0_util} is largest near the four corners of the domain, while the domain center yields the lowest. This behavior is consistent with the geometry of isotropic diffusion: a single concentration measurement provides distance information to the source and not direction, and corner sensors benefit from the truncation imposed by the square domain boundaries in this case.

However, the corners also exhibit substantially larger utility variance, as shown in \cref{fig:roed_2Dsource_1dsgn_b0_utilvar}. Intuitively, the informativeness of a corner sensor depends strongly on the distance between the source and the sensor, which leads to large variability in utility across realizations. The scatter plot in \cref{fig:roed_2Dsource_1dsgn_b0_scatter} further reveals a steep trade-off in the region of high expected utility: many designs have similar expected utility but markedly different utility variances. This is the setting in which the mean--variance formulation becomes useful.

\begin{figure}[htbp]
  \centering
  \includegraphics[width=0.95\linewidth]{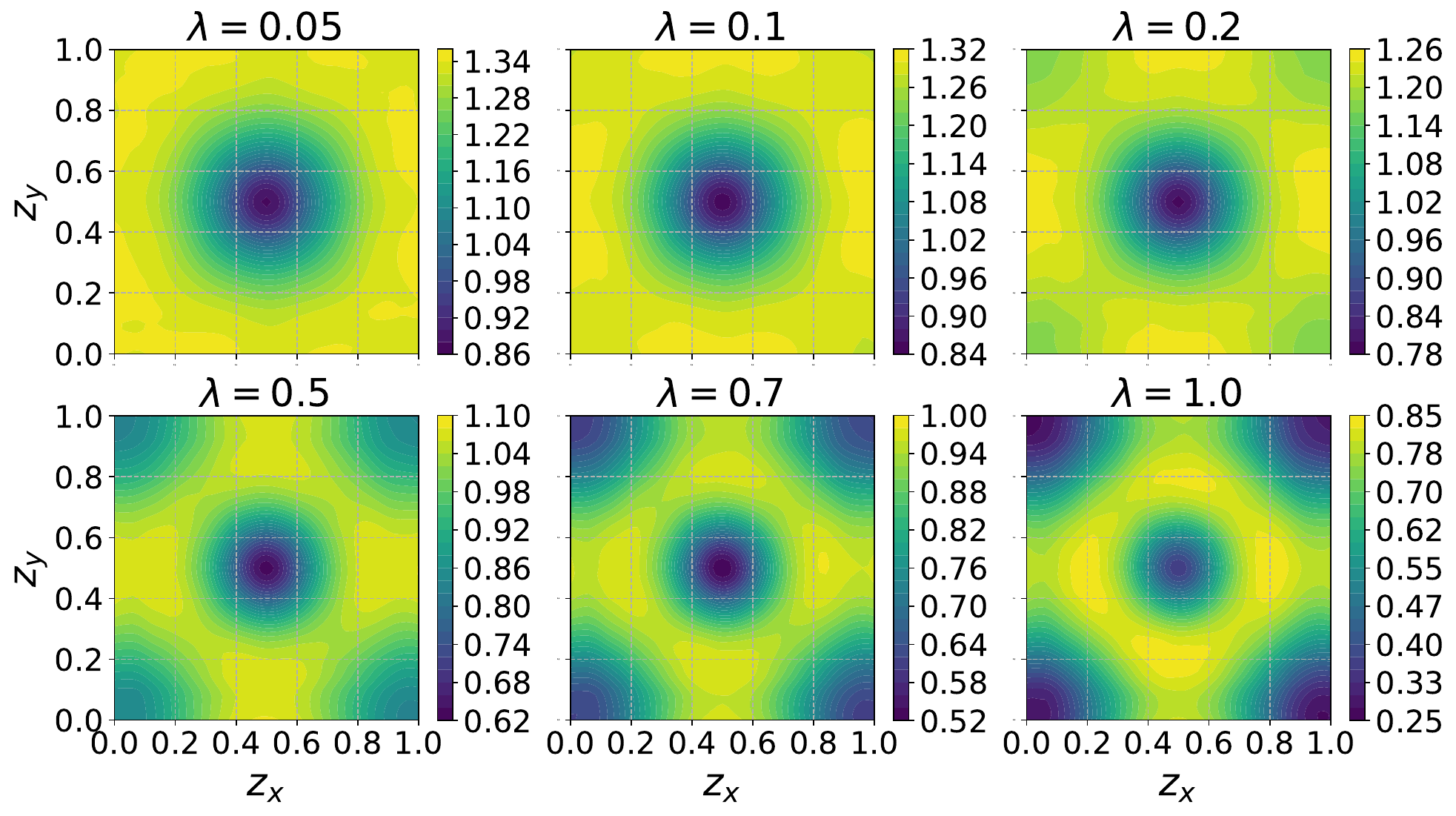}
  \caption{Case 3: Estimated mean--variance objective with different values of $\lambda$  for the one-sensor problem.}
  \label{fig:roed_2Dsource_1dsgn_b0_uvp}
\end{figure}

\begin{figure}[htbp]
  \centering
  \includegraphics[width=0.95\linewidth]{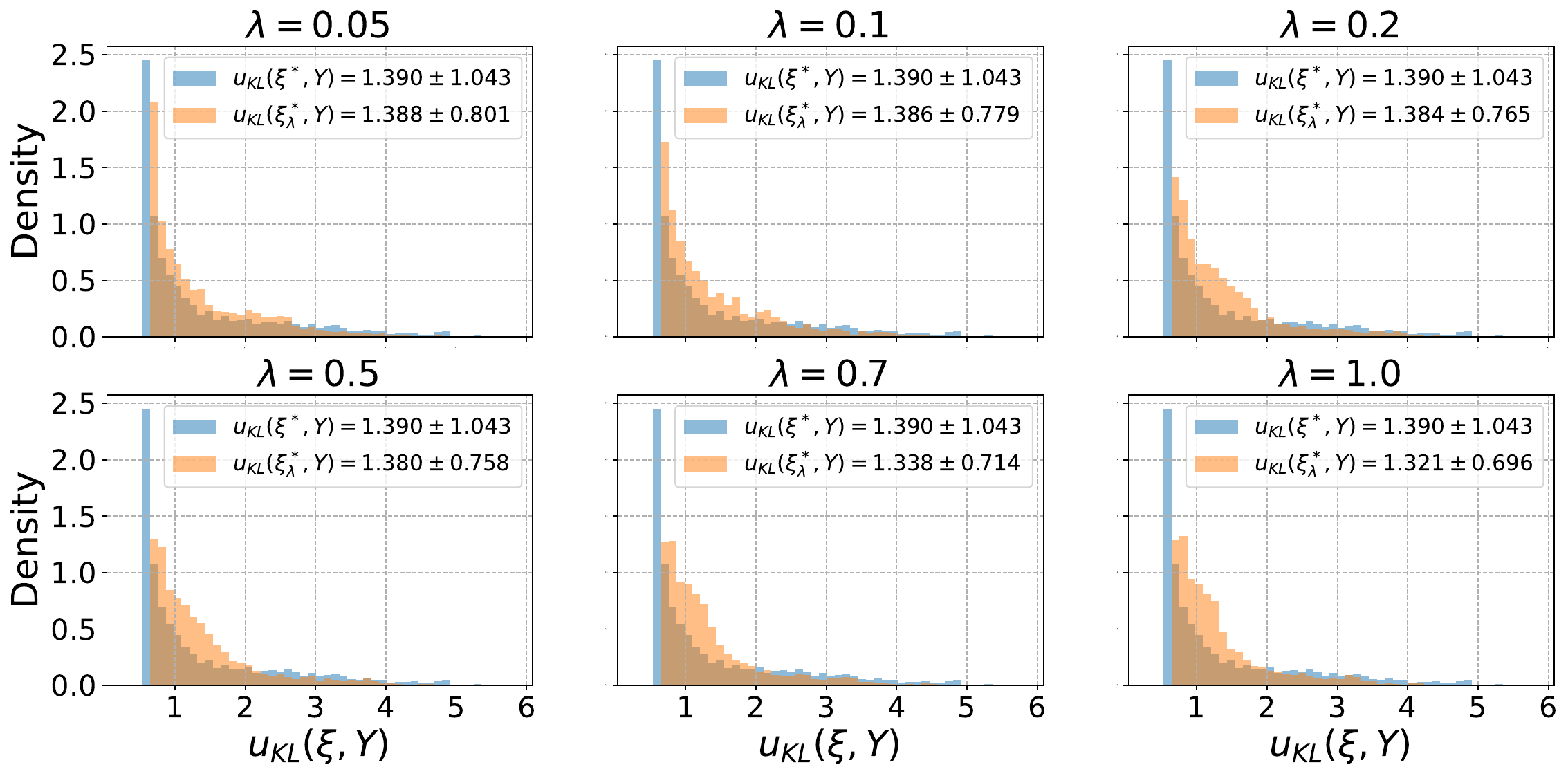}
  \caption{Case 3: Histogram distributions of $u_{\mathrm{KL}}(\design^\ast,Y)$ and $u_{\mathrm{KL}}(\design^\ast_\lambda,Y)$ for the one-sensor problem. The values before and after $\pm$ are the mean and standard deviation, respectively.}
  \label{fig:roed_2Dsource_1dsgn_b0_hist}
\end{figure}

\Cref{fig:roed_2Dsource_1dsgn_b0_uvp} shows contours of the estimated mean--variance objective for several values of $\lambda$, and \cref{fig:roed_2Dsource_1dsgn_b0_hist} compares the histogram distributions of $u_{\mathrm{KL}}(\design,Y)$ at the designs maximizing $\widehat{U}(\design)$ and $\widehat{J}_\lambda(\design)$. As $\lambda$ increases, the optimal sensor location shifts from a corner toward the middle of the boundary and then toward the interior. The utility variance decreases substantially, while the loss in expected utility remains modest. This trade-off is visible directly in the reported means and standard deviations in \cref{fig:roed_2Dsource_1dsgn_b0_hist}.

\begin{figure}[htbp]
  \centering
  \begin{subfigure}[t]{0.38\linewidth}
    \centering
    \includegraphics[width=\linewidth]{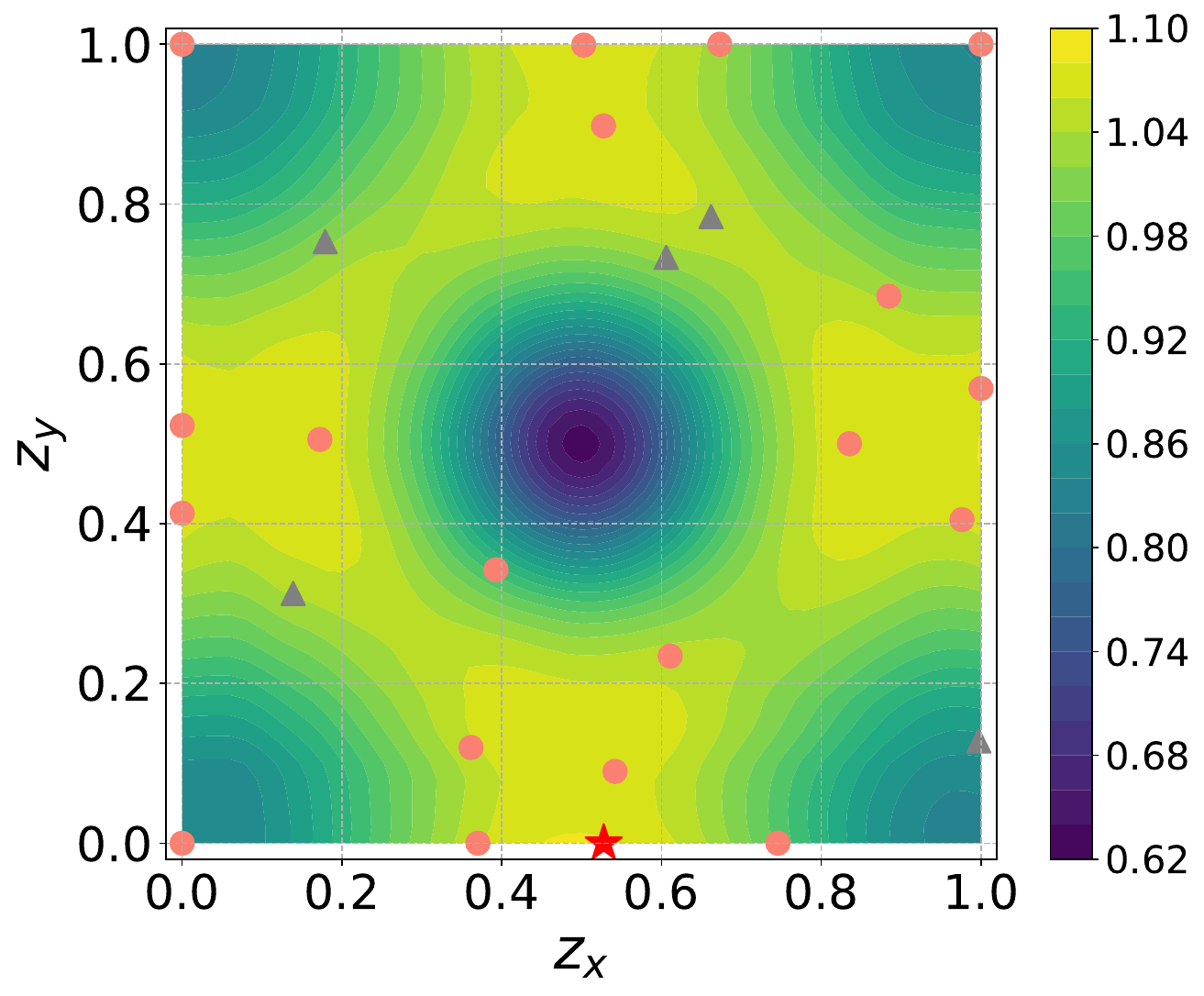}
    \caption{Evaluation locations overlaid on the objective contour}
    \label{fig:roed_2Dsource_1dsgn_b0_BO_contour}
  \end{subfigure}\hspace{1em}
  \begin{subfigure}[t]{0.48\linewidth}
    \centering
    \includegraphics[width=\linewidth]{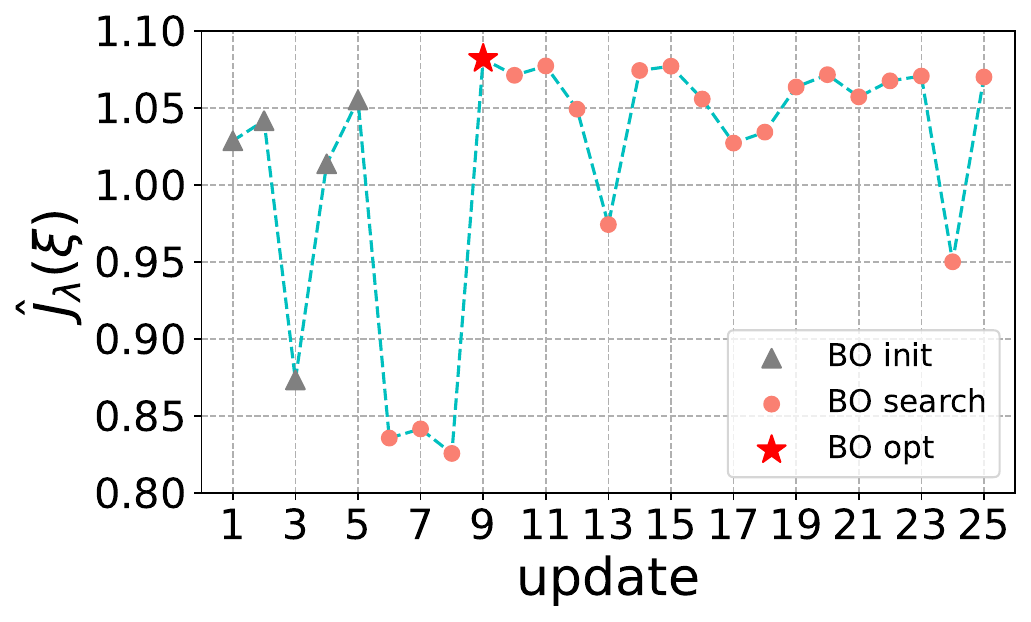}
    \caption{Optimization history}
    \label{fig:roed_2Dsource_1dsgn_b0_BO_hist}
  \end{subfigure}
  \caption{Case 3: BO with $\lambda=0.5$ for the one-sensor problem.}
  \label{fig:roed_2Dsource_1dsgn_b0_BO}
\end{figure}

To illustrate the optimization process, we fix $\lambda=0.5$ and apply BO. \Cref{fig:roed_2Dsource_1dsgn_b0_BO} shows the BO trajectory and objective history. BO rapidly converges to a design near the midpoint of one of the domain boundaries and explores all four symmetric high-value regions, indicating effective global search behavior.

\begin{figure}[htbp]
  \centering
  \includegraphics[width=\linewidth]{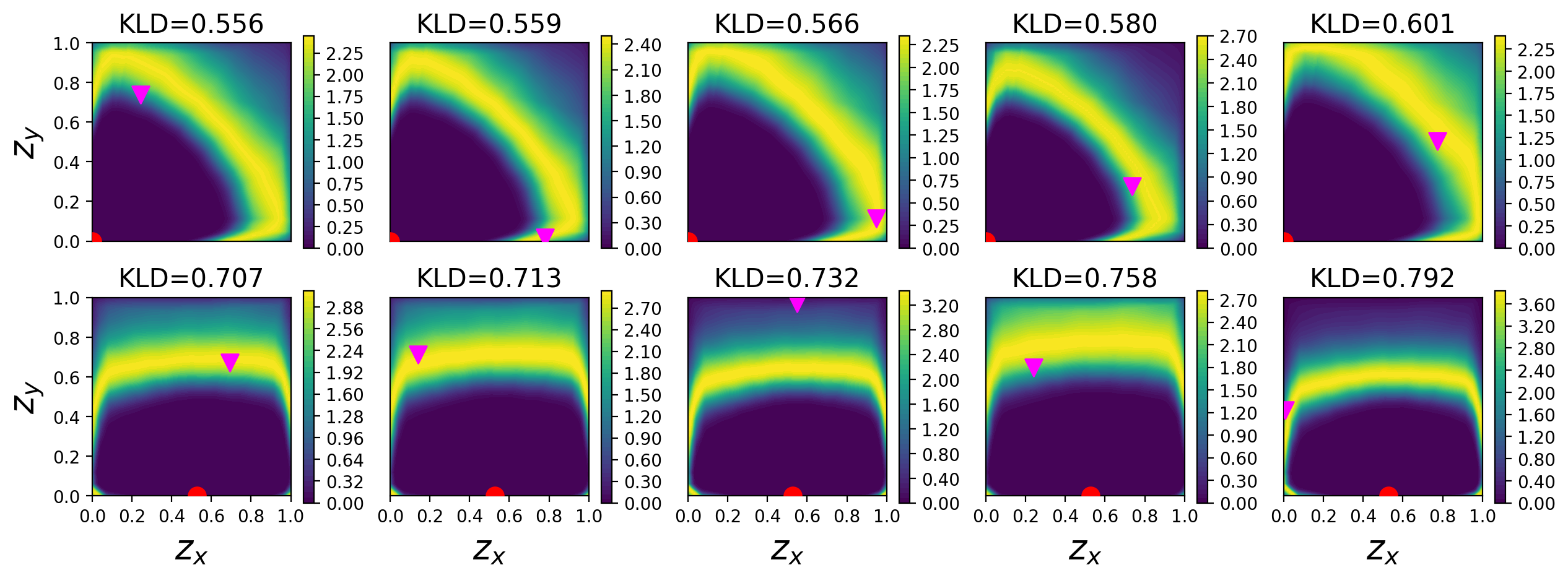}
  \caption{Case 3: Representative lowest-utility posterior distributions for the one-sensor problem. The top row corresponds to the formulation with $\widehat{U}(\design)$ and the bottom row with  $\widehat{J}_\lambda(\design)$ ($\lambda=0.5$). The red star denotes the sensor location and the magenta triangle denotes the true source location.}
  \label{fig:roed_2Dsource_1dsgn_b0_post}
\end{figure}

To further contrast the mean-optimal and mean--variance-optimal designs, we draw 3000 prior samples of $\Param$, generate the corresponding observations using the surrogate model, compute the resulting KL divergences, and then select the five lowest-utility cases for each design. The corresponding posterior distributions are shown in \cref{fig:roed_2Dsource_1dsgn_b0_post}. The worst cases under the mean-optimal design (top row) have lower KL divergence and flatter posteriors than those under the mean--variance-optimal design (bottom row). This reflects the fact that corner sensors can produce weakly informative observations when the source lies near the diagonal opposite the sensor, whereas boundary-midpoint sensors yield more consistently informative measurements.

\subsubsection{Results: Two sensors}

We next consider two sensors, so that $\design \in [0,1]^4$. For visualization, we randomly sample 1000 candidate sensor pairs and estimate the objective at each pair using $N=30000$ MC samples.

\begin{figure}[htbp]
  \centering
  \begin{subfigure}[t]{0.3\linewidth}
    \centering
    \includegraphics[width=\linewidth]{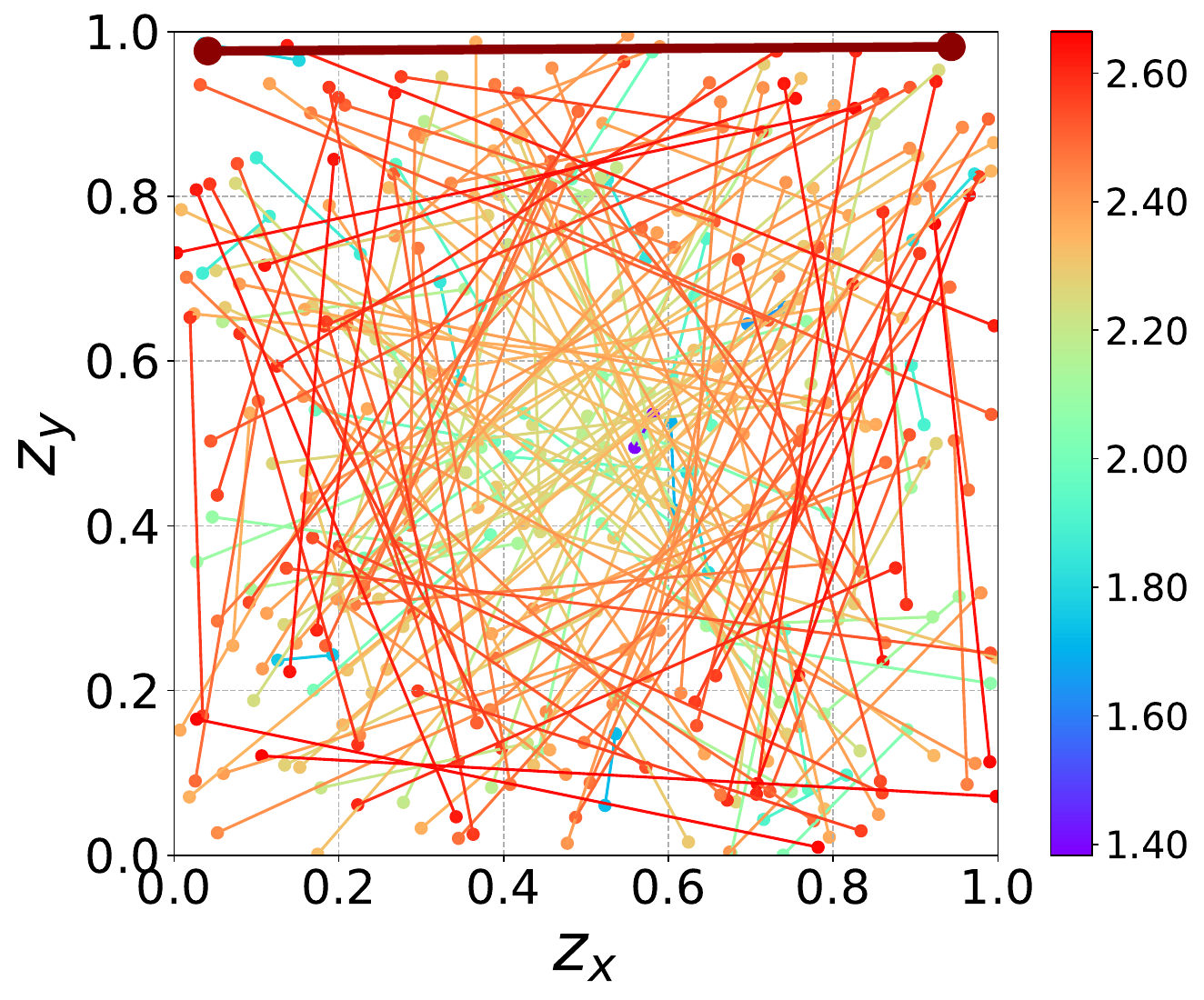}
    \caption{$\widehat{U}(\design)$}
    \label{fig:roed_2Dsource_2dsgn_b0_util}
  \end{subfigure}
  \begin{subfigure}[t]{0.3\linewidth}
    \centering
    \includegraphics[width=\linewidth]{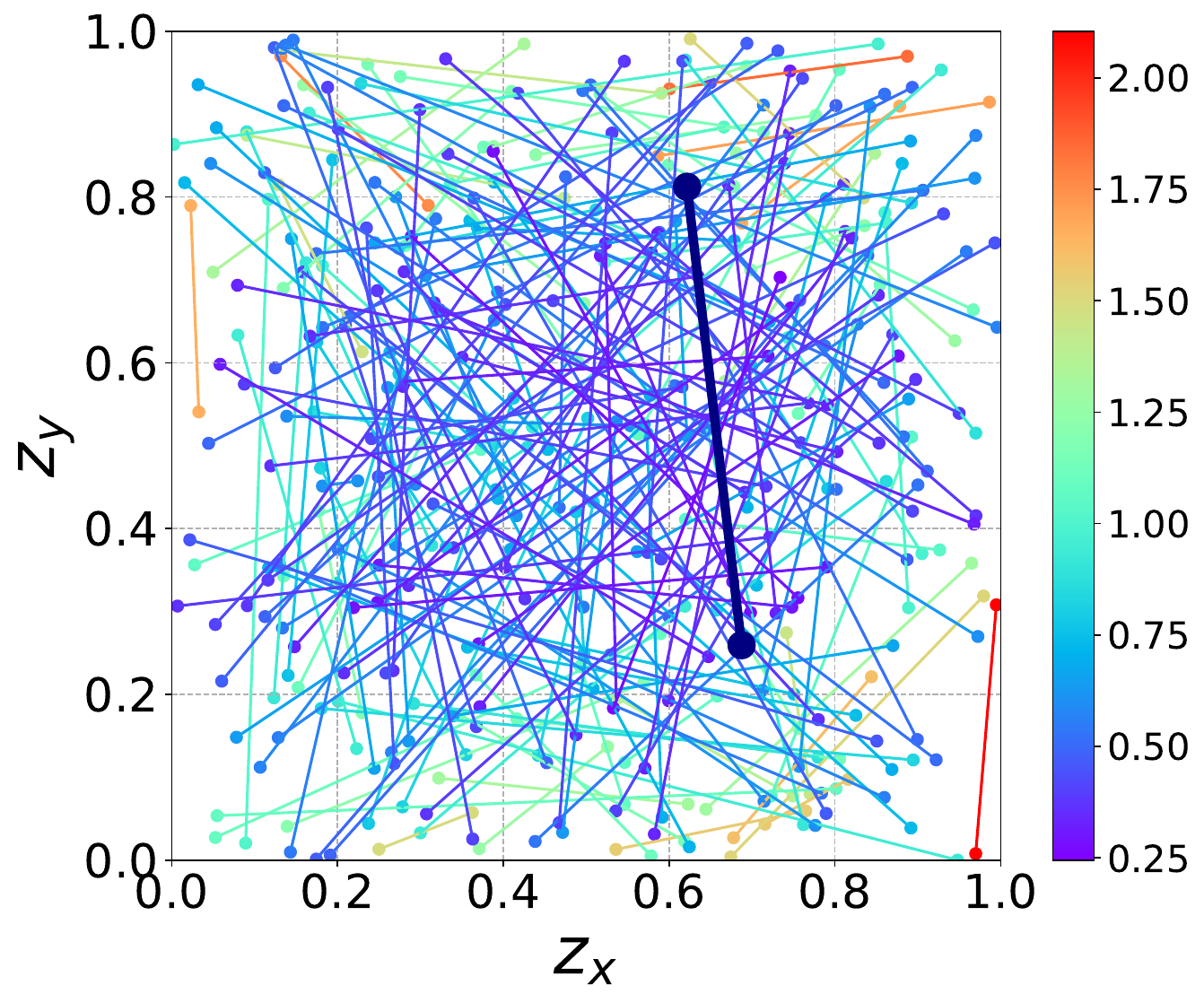}
    \caption{$\widehat{V}(\design)$}
    \label{fig:roed_2Dsource_2dsgn_b0_utilvar}
  \end{subfigure}
  \begin{subfigure}[t]{0.3\linewidth}
    \centering
    \includegraphics[width=\linewidth]{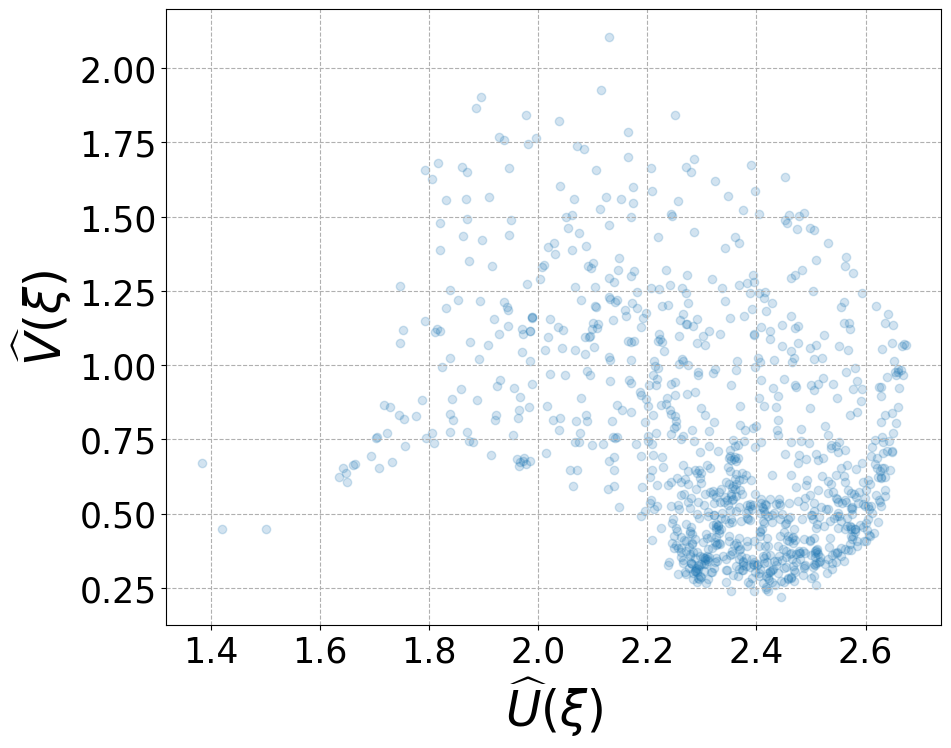}
    \caption{$\widehat{V}(\design)$ versus $\widehat{U}(\design)$}
    \label{fig:roed_2Dsource_2dsgn_b0_scatter}
  \end{subfigure}
  \caption{Case 3: Estimated expected utility, utility variance, and their trade-off for the two-sensor problem.}
  \label{fig:roed_2Dsource_2dsgn_b0}
\end{figure}

\Cref{fig:roed_2Dsource_2dsgn_b0} shows the estimated expected utility, utility variance, and the corresponding scatter plot of variance versus expected utility. The highest expected utility is obtained by placing the two sensors near adjacent corners, whereas lower-variance designs tend to move the sensors closer to the center of the domain. As in the one-sensor problem, the scatter plot exhibits a steep trade-off in the region with high expected utility.

\begin{figure}[htbp]
  \centering
  \includegraphics[width=0.95\linewidth]{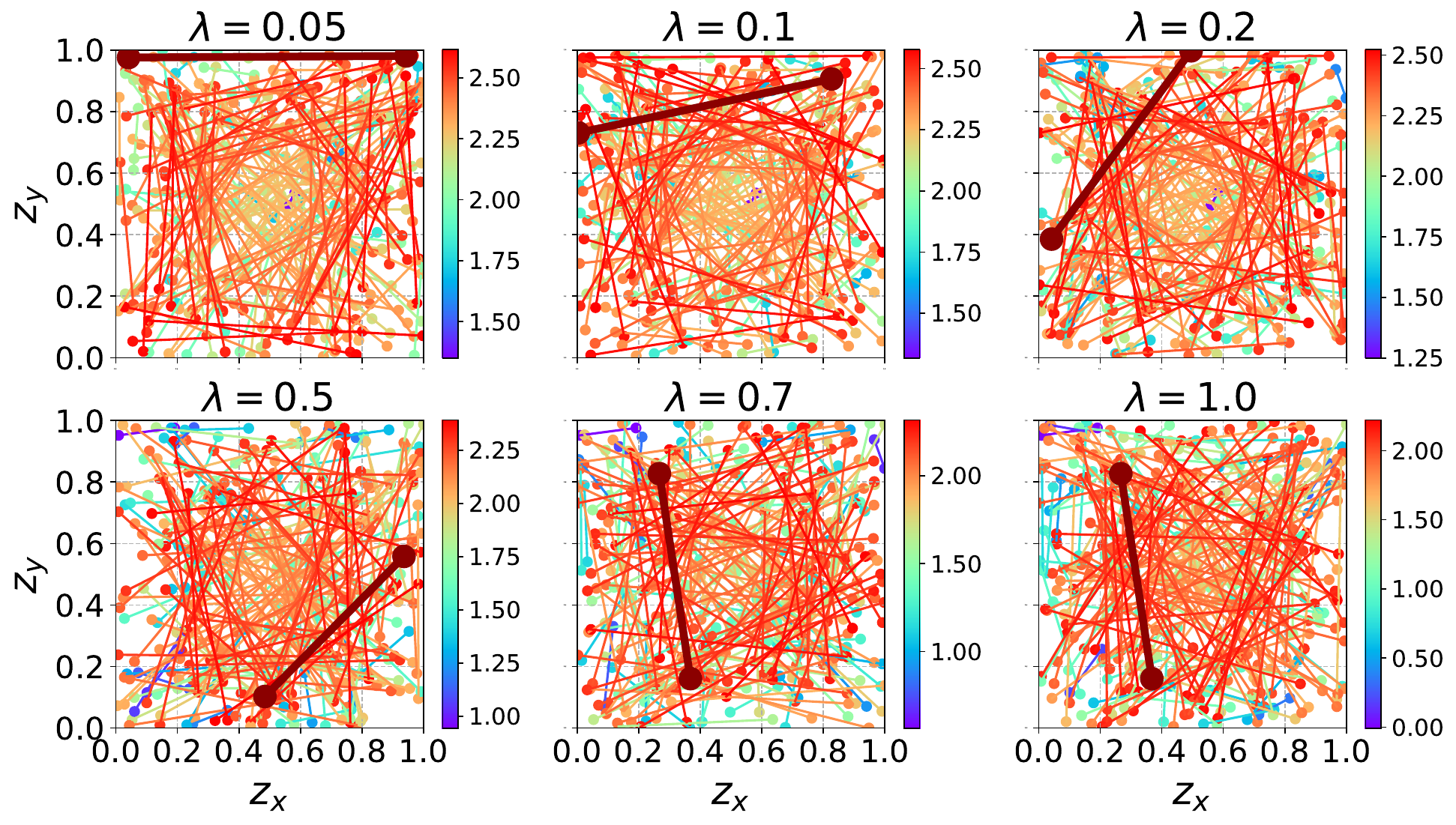}
  \caption{Case 3: Estimated mean--variance objective with different values of $\lambda$ for the two-sensor problem.}
  \label{fig:roed_2Dsource_2dsgn_b0_uvp}
\end{figure}

\begin{figure}[htbp]
  \centering
  \includegraphics[width=0.95\linewidth]{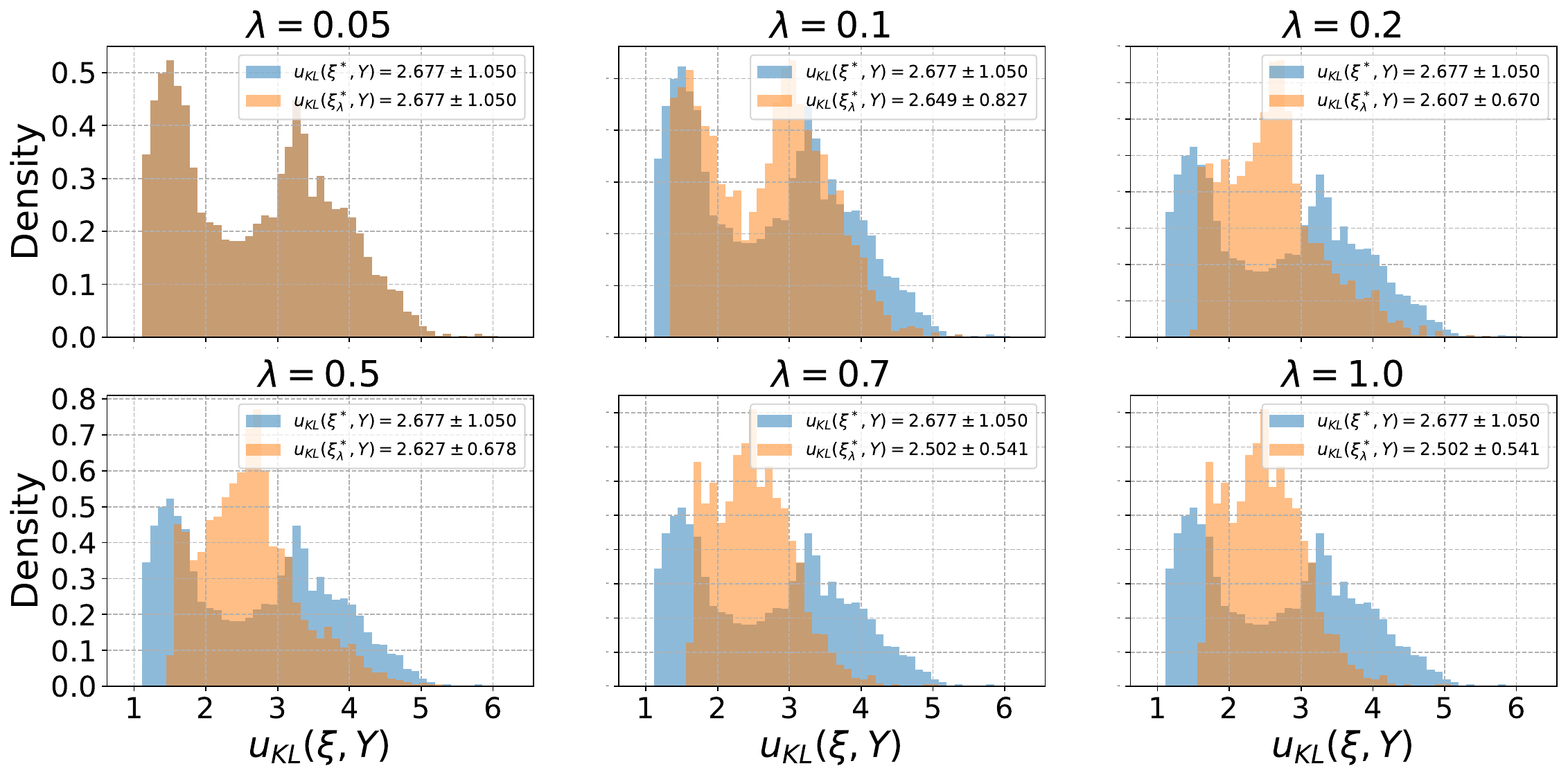}
  \caption{Case 3: Histogram distributions of $u_{\mathrm{KL}}(\design^\ast,Y)$ and $u_{\mathrm{KL}}(\design^\ast_\lambda,Y)$ for the two-sensor problem. The values before and after $\pm$ are the mean and standard deviation, respectively.}
  \label{fig:roed_2Dsource_2dsgn_b0_hist}
\end{figure}

\Cref{fig:roed_2Dsource_2dsgn_b0_uvp} shows the estimated mean--variance objective for several values of $\lambda$, and \cref{fig:roed_2Dsource_2dsgn_b0_hist} compares the corresponding utility histogram distributions. As $\lambda$ increases, the optimal design moves away from the boundary toward the interior, again achieving a substantial reduction in variance for only a modest reduction in mean utility. The distribution for the mean-optimal design is visibly more multimodal, whereas the mean--variance-optimal design produces a more concentrated and stable utility distribution.

\begin{figure}[htbp]
  \centering
  \begin{subfigure}[t]{0.48\linewidth}
    \centering
    \includegraphics[width=\linewidth]{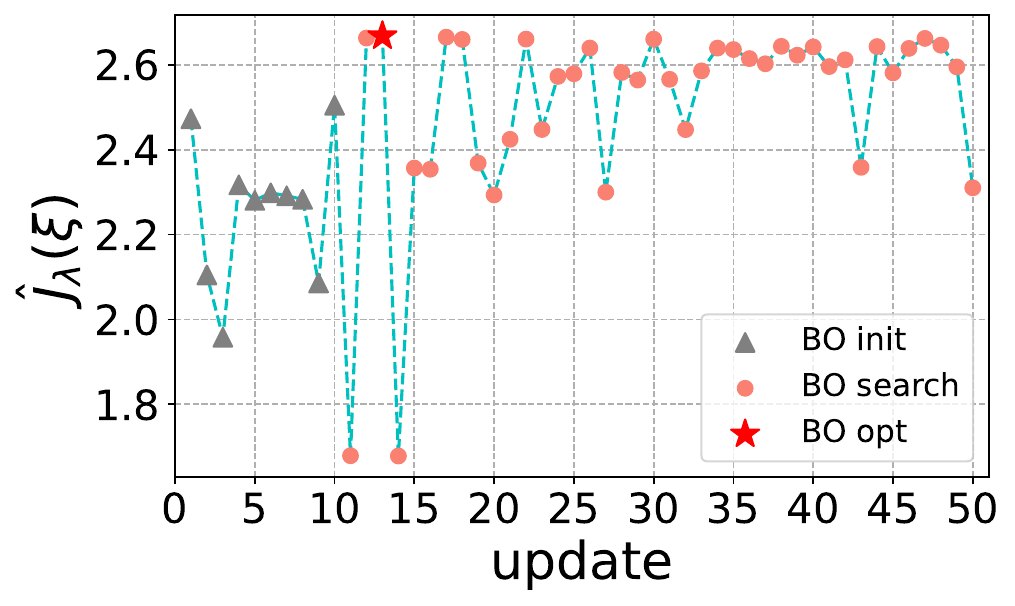}
    \caption{$\lambda=0$}
    \label{fig:roed_2Dsource_2dsgn_b0_lambda0_BO_hist}
  \end{subfigure}\hspace{1em}
  \begin{subfigure}[t]{0.48\linewidth}
    \centering
    \includegraphics[width=\linewidth]{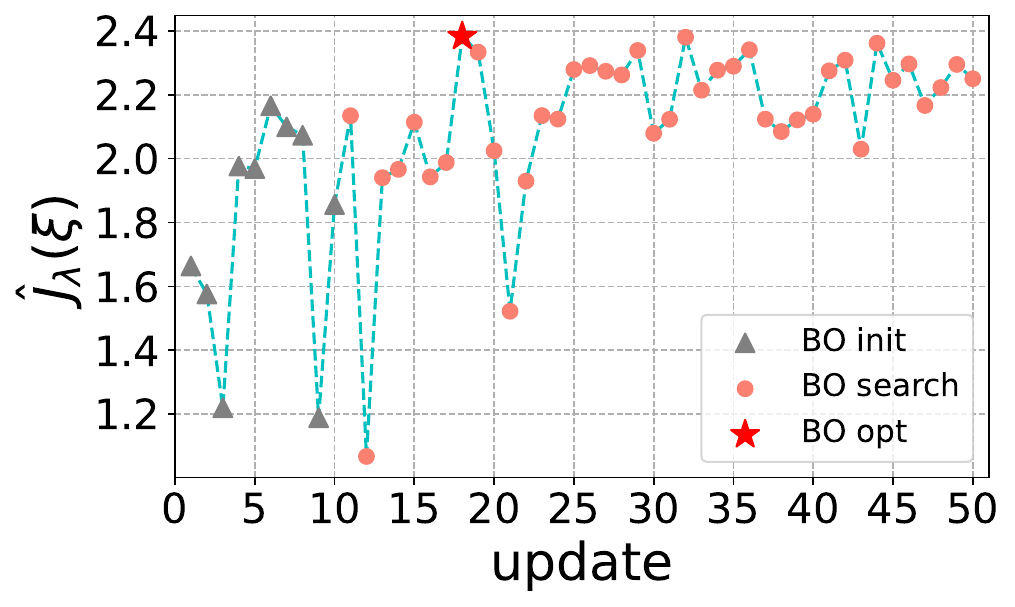}
    \caption{$\lambda=0.5$}
    \label{fig:roed_2Dsource_2dsgn_b0_lambda05_BO_hist}
  \end{subfigure}
  \caption{Case 3: BO histories for the two-sensor problem.}
  \label{fig:roed_2Dsource_2dsgn_b0_BO}
\end{figure}

The BO results for $\lambda=0$ and $\lambda=0.5$ are shown in \cref{fig:roed_2Dsource_2dsgn_b0_BO}. In both cases, BO finds designs whose objective values are close to the best among the 1000 randomly sampled sensor pairs, but with substantially fewer evaluations. In particular, BO identifies high-quality designs in fewer than 20 objective evaluations.

\begin{figure}[htbp]
  \centering
  \includegraphics[width=0.95\linewidth]{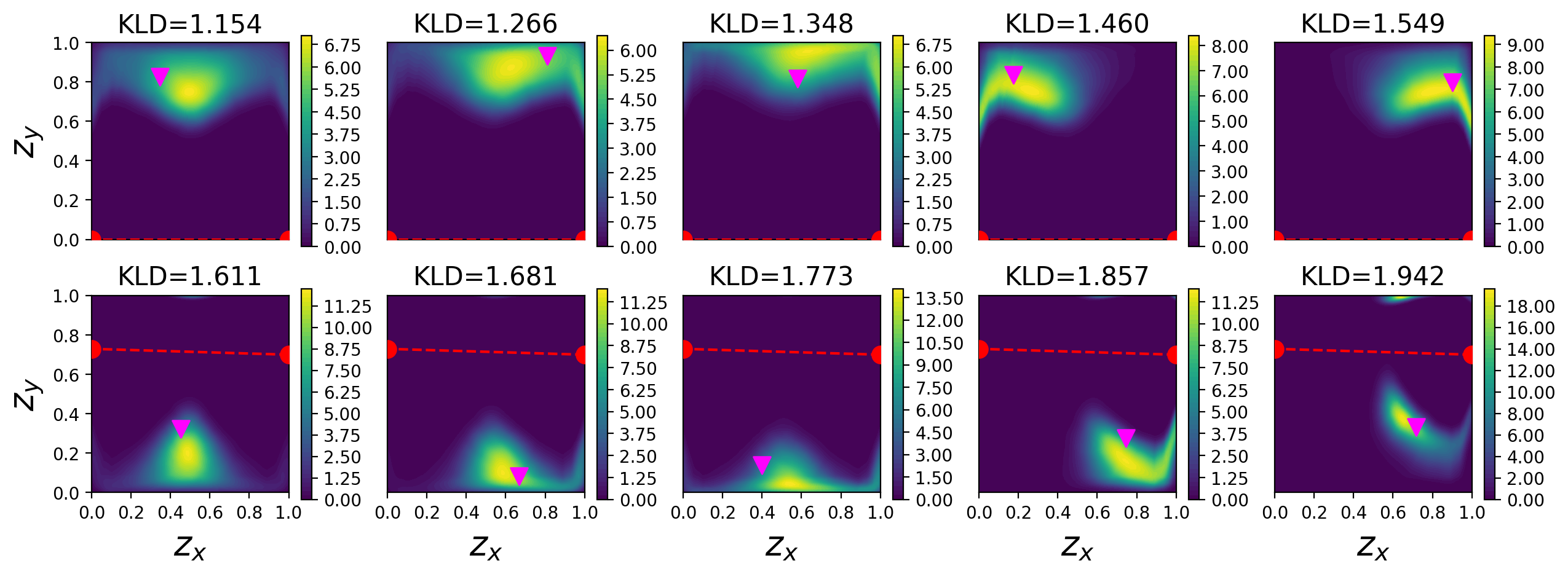}
  \caption{Case 3: Representative lowest-utility posterior distributions for the two-sensor problem. The top row corresponds to the formulation with $\widehat{U}(\design)$ and the bottom row with $\widehat{J}_\lambda(\design)$ ($\lambda=0.5$). The red stars denote the sensor locations and the magenta triangle denotes the true source location.}
  \label{fig:roed_2Dsource_2dsgn_b0_post}
\end{figure}

Finally, \cref{fig:roed_2Dsource_2dsgn_b0_post} shows representative lowest-utility posterior distributions for the two-sensor problem. As in the one-sensor setting, the worst cases under the mean-optimal design have substantially lower utility and flatter posterior distributions than the worst cases under the mean--variance-optimal design. This again illustrates that the proposed formulation sacrifices a small amount of expected utility in order to reduce the risk of poor experimental outcomes.

\subsection{Case 4: Contaminant source inversion with building obstacles}
\label{sec:roed_ex_source_w_building}

We now introduce building obstacles into the diffusion domain to obtain a more realistic sensor-placement problem. The prior on the source location remains uniform over the accessible region, with zero density inside the obstacles.

\begin{figure}[htbp]
  \centering
  \begin{subfigure}[t]{\linewidth}
    \centering
    \includegraphics[width=\linewidth]{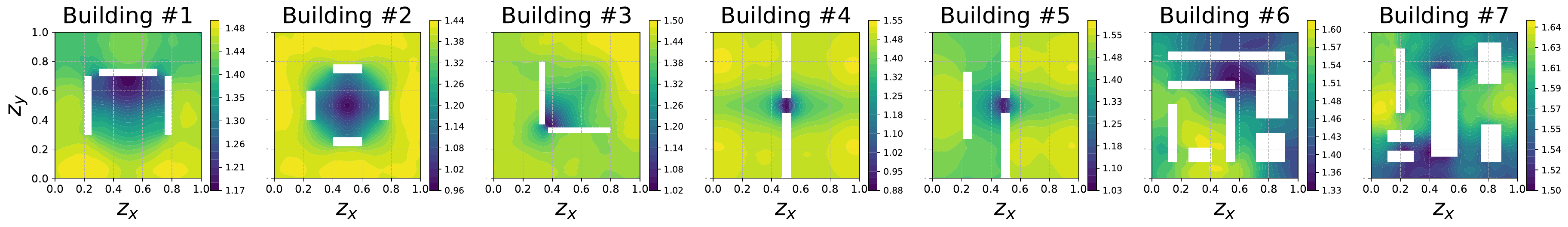}
    \caption{$\widehat{U}(\design)$}
    \label{fig:roed_2Dsource_1dsgn_bs_utils}
  \end{subfigure}

  \begin{subfigure}[t]{\linewidth}
    \centering
    \includegraphics[width=\linewidth]{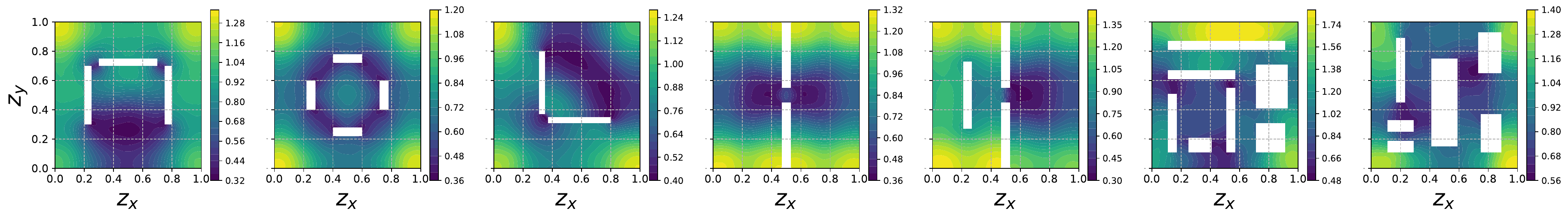}
    \caption{$\widehat{V}(\design)$}
    \label{fig:roed_2Dsource_1dsgn_bs_utilvars}
  \end{subfigure}

  \begin{subfigure}[t]{\linewidth}
    \centering
    \includegraphics[width=\linewidth]{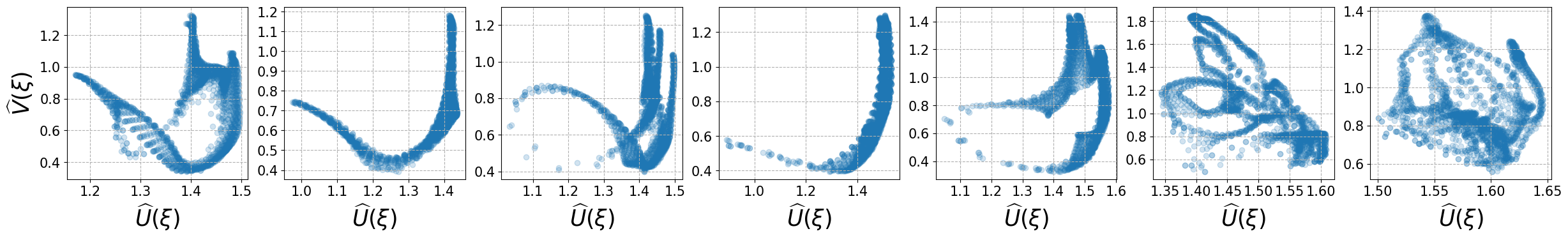}
    \caption{$\widehat{V}(\design)$ versus $\widehat{U}(\design)$}
    \label{fig:roed_2Dsource_1dsgn_bs_scatters}
  \end{subfigure}
  \caption{Case 4: Estimated expected utility, utility variance, and their trade-off under seven obstacle configurations for the one-sensor problem.}
  \label{fig:roed_2Dsource_1dsgn_bs}
\end{figure}

\Cref{fig:roed_2Dsource_1dsgn_bs} shows the estimated expected utility, utility variance, and the corresponding scatter plot trade-off for seven different obstacle configurations in the one-sensor setting. Each column corresponds to a different building layout. Across all cases, the same qualitative pattern is observed: there is generally a steep trade-off in the region of high expected utility, indicating that designs with similar expected utility can have substantially different utility variances. Thus, even in the presence of obstacles, the mean--variance criterion can identify designs that sacrifice little expected utility while substantially reducing variability.

We next focus on two representative examples: building~\#4 with one sensor, and building~\#5 with two sensors.

\subsubsection{Building \#4: One sensor}

\begin{figure}[htbp]
  \centering
  \includegraphics[width=0.95\linewidth]{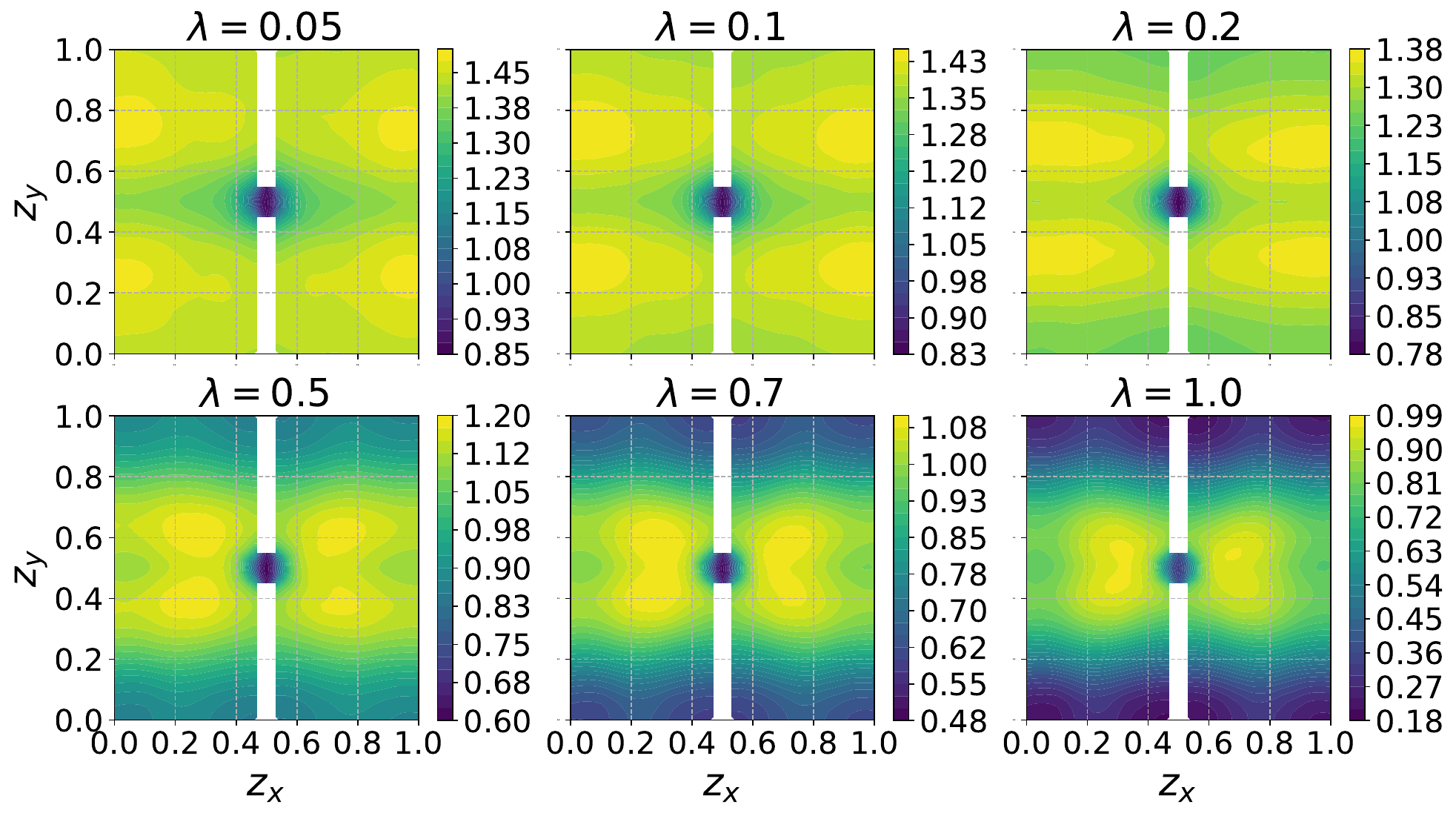}
  \caption{Case 4: Estimated mean--variance objective with different values of $\lambda$ for the one-sensor problem with building~\#4.}
  \label{fig:roed_2Dsource_1dsgn_b19_uvp}
\end{figure}

\begin{figure}[htbp]
  \centering
  \includegraphics[width=0.95\linewidth]{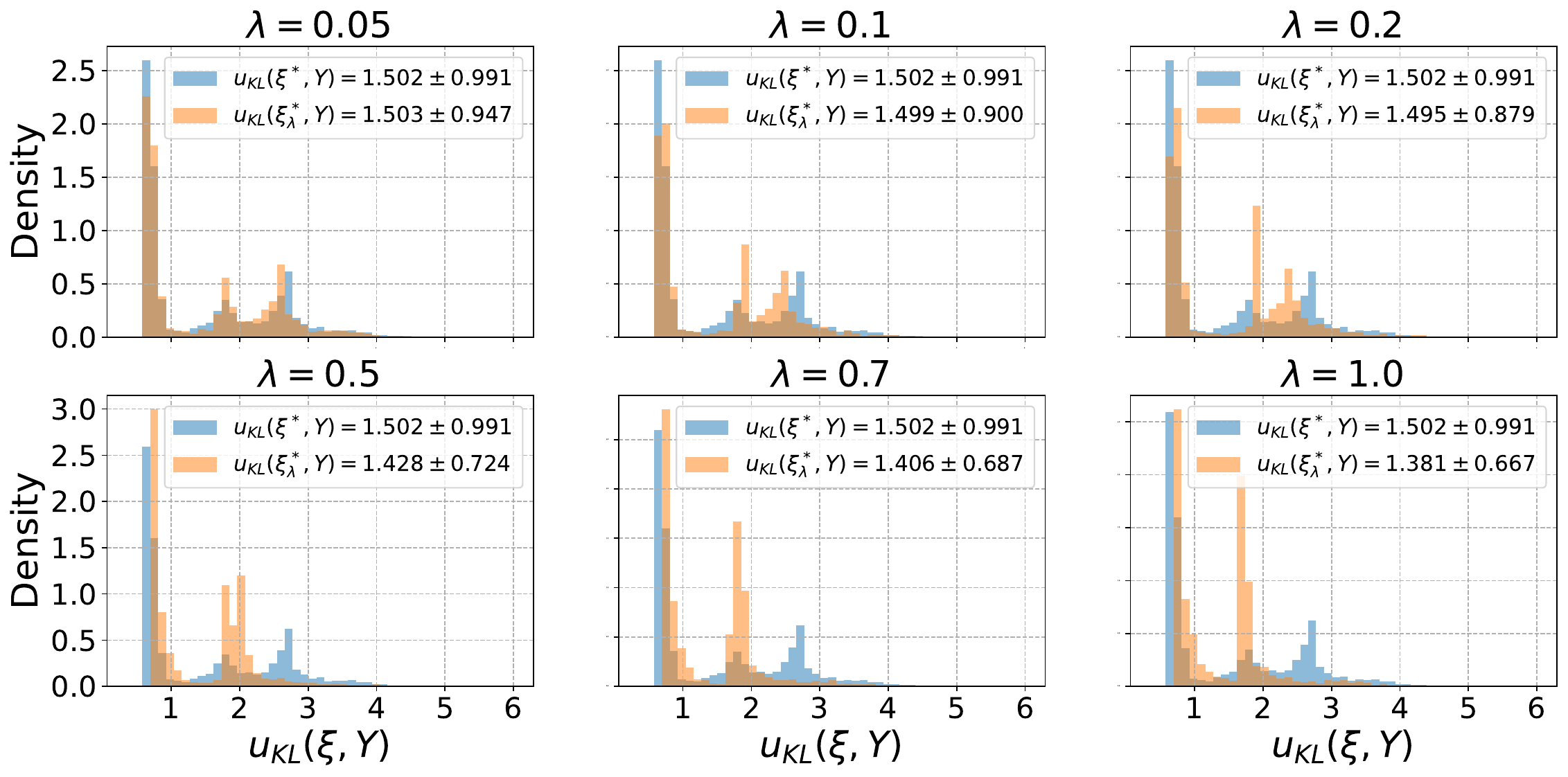}
  \caption{Case 4: Histogram distributions of $u_{\mathrm{KL}}(\design^\ast,Y)$ and $u_{\mathrm{KL}}(\design^\ast_\lambda,Y)$ for the one-sensor problem with building~\#4. The values before and after $\pm$ are the mean and standard deviation, respectively.}
  \label{fig:roed_2Dsource_1dsgn_b19_hist}
\end{figure}

For building~\#4, \cref{fig:roed_2Dsource_1dsgn_b19_uvp} shows contours of the estimated mean--variance objective for different values of $\lambda$, and \cref{fig:roed_2Dsource_1dsgn_b19_hist} compares the utility histogram distributions at the designs maximizing $\widehat{U}(\design)$ and $\widehat{J}_\lambda(\design)$. Here the reference designs are obtained from a dense grid search. As $\lambda$ increases, the optimal sensor location shifts away from the most aggressive high-utility regions and toward the center of the accessible domain, yielding a substantially more stable utility distribution.

\begin{figure}[htbp]
  \centering
  \begin{subfigure}[t]{0.38\linewidth}
    \centering
    \includegraphics[width=\linewidth]{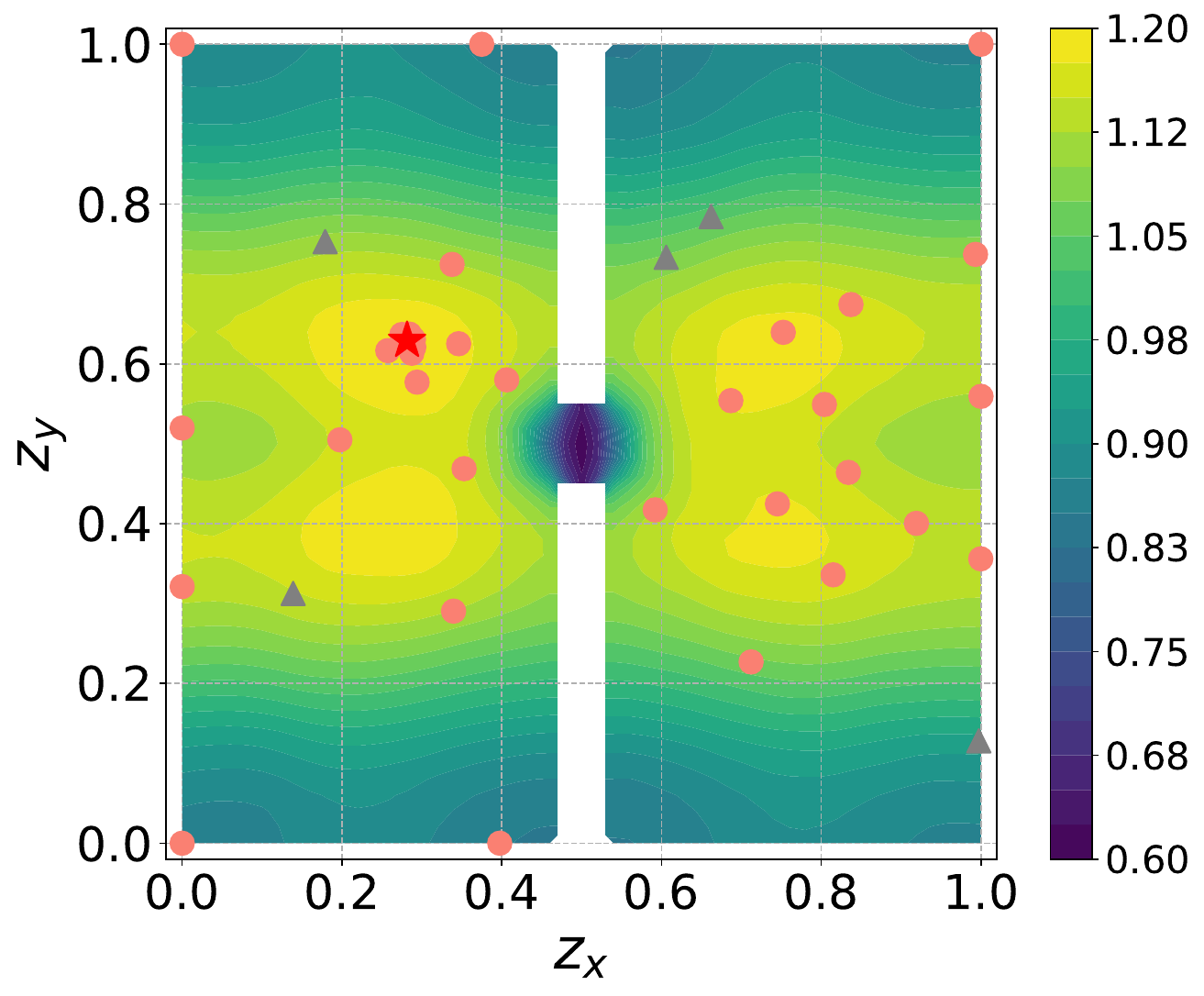}
    \caption{Evaluation locations overlaid on the objective contour}
    \label{fig:roed_2Dsource_1dsgn_b19_BO_contour}
  \end{subfigure}\hspace{1em}
  \begin{subfigure}[t]{0.48\linewidth}
    \centering
    \includegraphics[width=\linewidth]{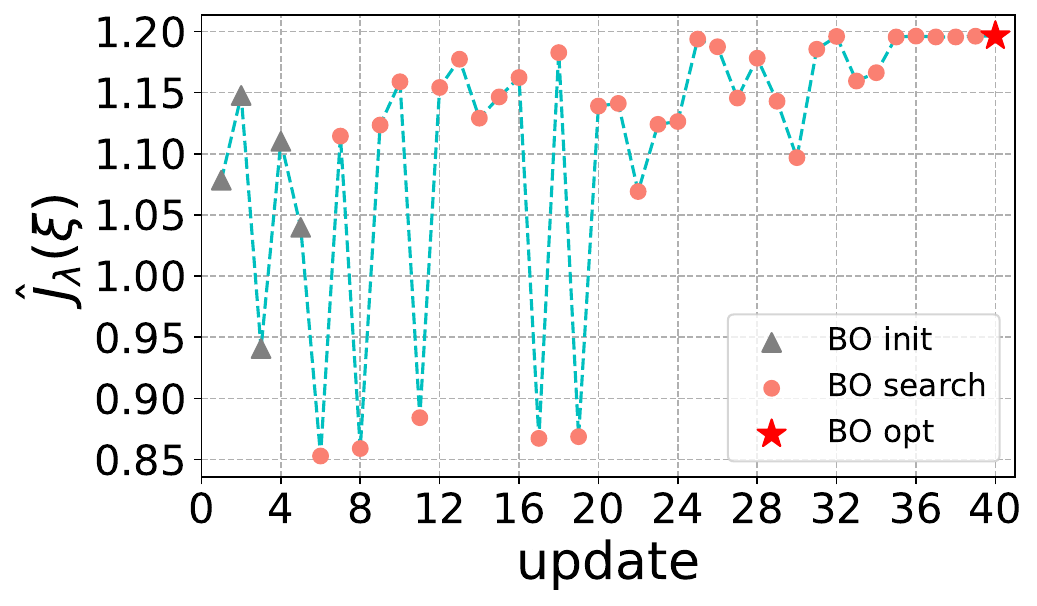}
    \caption{Optimization history}
    \label{fig:roed_2Dsource_1dsgn_b19_BO_hist}
  \end{subfigure}
  \caption{Case 4: BO with $\lambda=0.5$ for the one-sensor problem with building~\#4.}
  \label{fig:roed_2Dsource_1dsgn_b19_BO}
\end{figure}

We then apply BO to solve for the optimal design with $\lambda=0.5$. The results are shown in \cref{fig:roed_2Dsource_1dsgn_b19_BO}. BO identifies three of the four symmetric local optima within a relatively small number of iterations. During optimization, the obstacle constraints are enforced so that sensors cannot be placed inside buildings.

\begin{figure}[htbp]
  \centering
  \includegraphics[width=\linewidth]{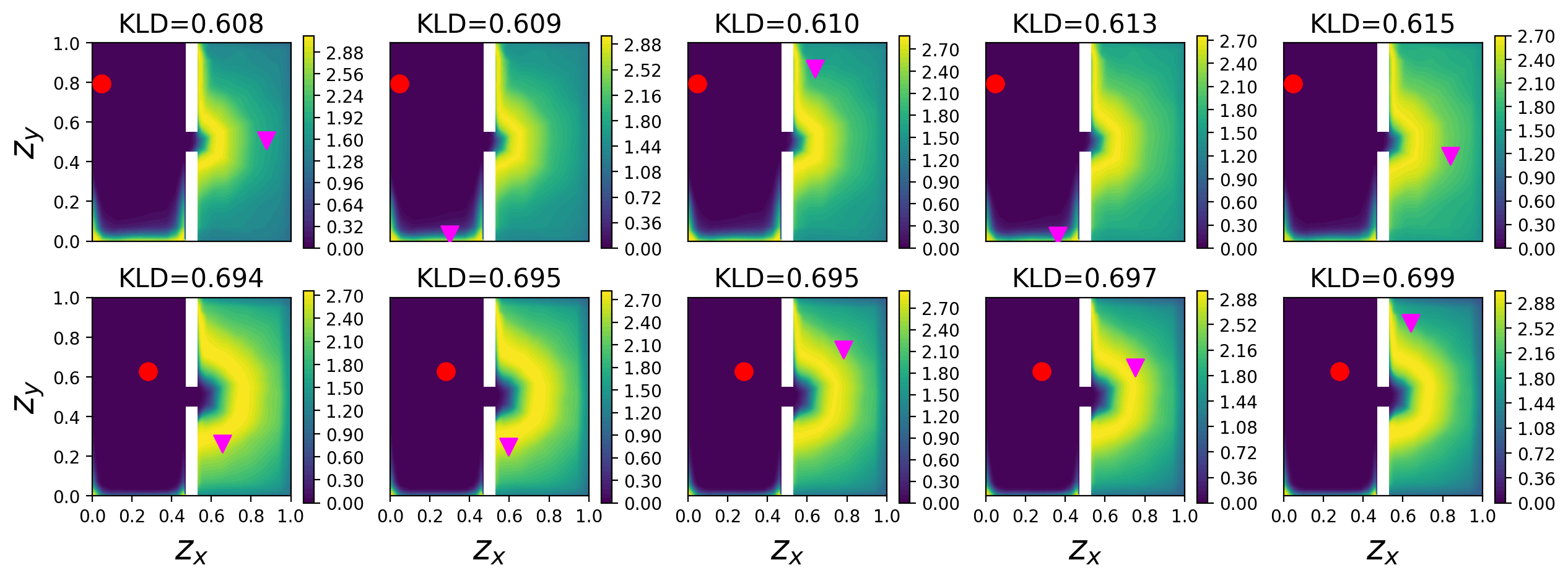}
  \caption{Case 4: Representative lowest-utility posterior distributions for the one-sensor problem with building~\#4. The top row corresponds to the formulation with $\widehat{U}(\design)$ and the bottom row with $\widehat{J}_\lambda(\design)$ ($\lambda=0.5$). The red star denotes the sensor location and the magenta triangle denotes the true source location.}
  \label{fig:roed_2Dsource_1dsgn_b19_post}
\end{figure}

To better understand the difference between the mean-optimal and mean--variance-optimal designs, \cref{fig:roed_2Dsource_1dsgn_b19_post} shows representative lowest-utility posterior distributions obtained from BO solutions. The worst cases under the mean-optimal design often exhibit ambiguity between the left and right sides of the obstacle and may place substantial posterior mass on the wrong side altogether. In contrast, the worst cases under the mean--variance-optimal design still identify the correct side of the source more reliably, leading to higher utility in adverse scenarios.

\subsubsection{Building \#5: Two sensors}

\begin{figure}[htbp]
  \centering
  \includegraphics[width=0.65\linewidth]{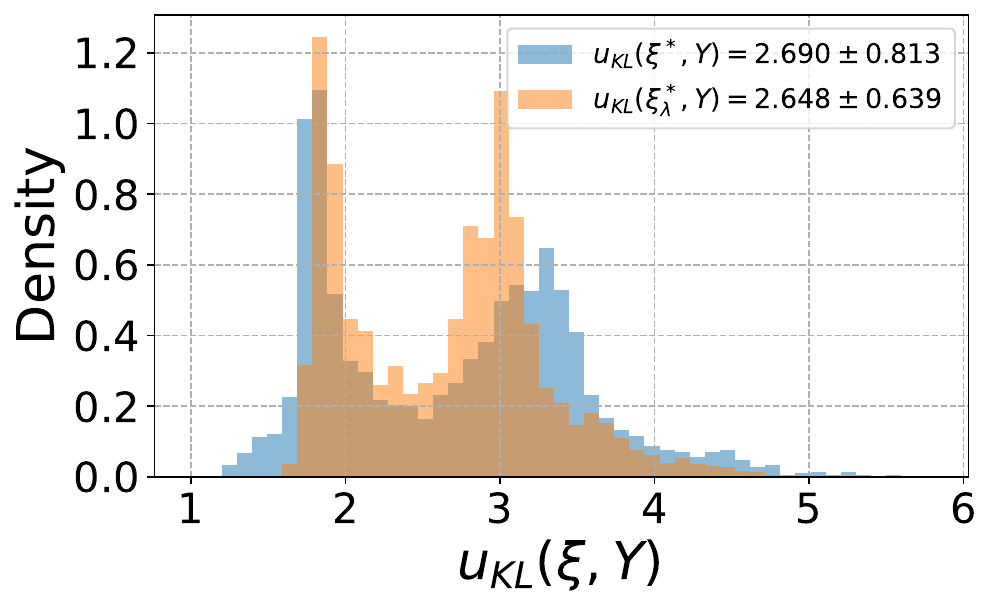}
  \caption{Case 4: Histogram distributions of $u(\design^\ast,Y)$ and $u(\design^\ast_\lambda,Y)$ for the two-sensor problem with building~\#5. The values before and after $\pm$ are the mean and standard deviation, respectively.}
  \label{fig:roed_2Dsource_2dsgn_b23_hist}
\end{figure}

\begin{figure}[htbp]
  \centering
  \includegraphics[width=\linewidth]{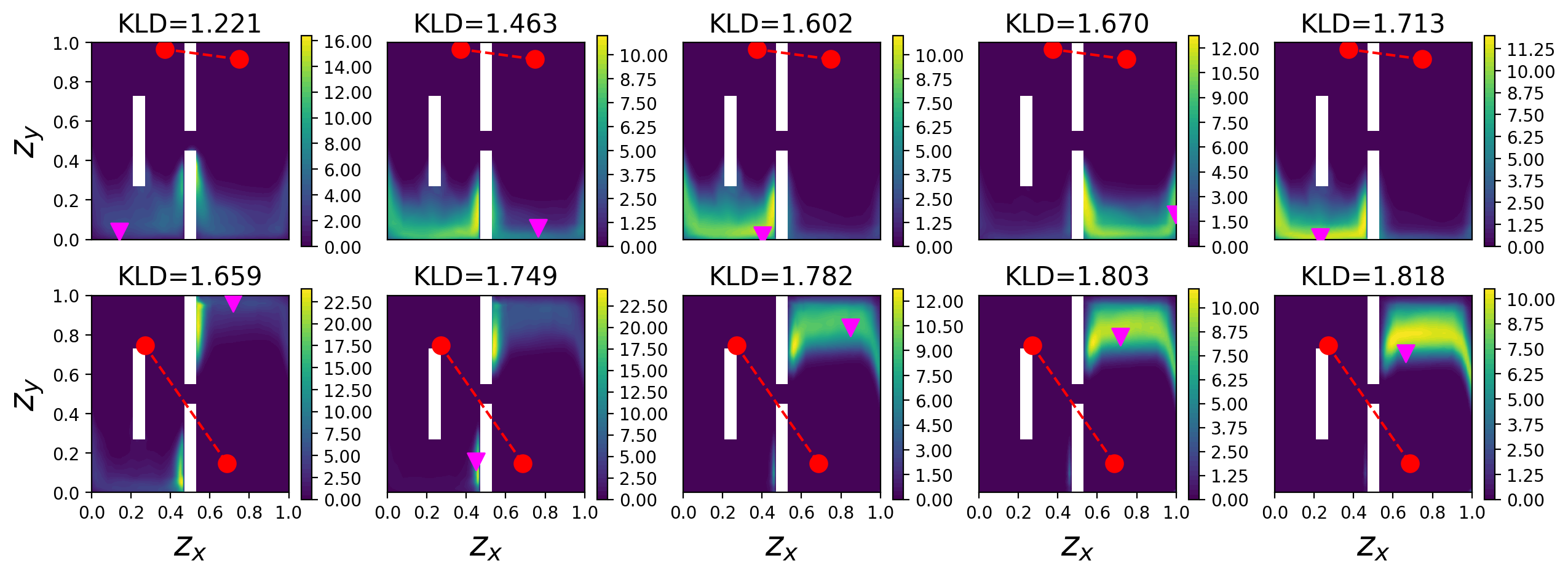}
  \caption{Case 4: Representative lowest-utility posterior distributions for the two-sensor problem with building~\#5. The top row corresponds to the formulation with $\widehat{U}(\design)$ and the bottom row with $\widehat{J}_\lambda(\design)$ ($\lambda=0.5$). The red stars denote the sensor locations and the magenta triangle denotes the true source location.}
  \label{fig:roed_2Dsource_2dsgn_b23_post}
\end{figure}

We next consider building~\#5 with two sensors. \Cref{fig:roed_2Dsource_2dsgn_b23_hist} compares the utility histogram distributions under the mean-optimal and mean--variance-optimal designs, while \cref{fig:roed_2Dsource_2dsgn_b23_post} shows representative lowest-utility posterior distributions.

In this case, both designs place one sensor on each side of the obstacle, which is intuitively reasonable for resolving source locations on either side. The mean--variance-optimal design further spreads the two sensors vertically, placing one near the bottom and one near the top. This produces a more space-filling layout and reduces the risk of both sensors being simultaneously far from the source, which helps mitigate the poor-outcome cases that arise under the mean-optimal design.

%% file: sections/5_conclusions.tex
\section{Conclusions}
\label{sec:roed_conclusions}

This work introduced a mean--variance risk-aware formulation of Bayesian OED that augments the classical expected-utility objective with a penalty on the variance of utility. The resulting criterion explicitly balances expected performance against variability in experimental outcomes. In the information-gain setting considered here, this leads to designs that may sacrifice a small amount of EIG in exchange for substantially more stable utility realizations.

On the methodological side, we developed MC estimators for the expected utility, the utility second moment, the utility variance, and the resulting mean--variance objective. The construction is based on prior sampling and avoids direct posterior sampling, which would otherwise be prohibitively expensive in the second-moment calculation. We also analyzed the leading-order bias and variance of these estimators. In addition, we considered practical strategies for improving efficiency, including sample reuse across nested loops and CRS across objective evaluations.

The numerical experiments illustrate both the accuracy of the estimators and the value of accounting for utility variability. In the linear-Gaussian benchmark, the proposed estimators exhibit the predicted convergence behavior and agree closely with the analytical solution. In the nonlinear test problem, the design favored by the mean--variance criterion differs substantially from the design that maximizes expected utility alone, demonstrating that high expected utility can be accompanied by considerable risk. In the contaminant source inversion examples, both with and without building obstacles, the mean--variance formulation consistently identifies designs with much lower utility variance while retaining competitive expected utility, and it remains effective in more realistic, constrained physical settings.

For optimization, we emphasized that the proposed objective can in principle be coupled with any suitable optimizer, including gradient-based methods when gradient information is available. In this work, we used BO as a practical choice for handling expensive and noisy objective evaluations. The numerical results also show that CRS can substantially smooth the optimization landscape and improve BO performance.

Overall, the results show that variance-penalized Bayesian OED can provide a useful alternative to purely expectation-based design, particularly in problems where experimental outcomes can vary widely across realizations. The proposed formulation is general and can be applied beyond information gain to other utility choices within the same decision-theoretic framework.

Several directions remain open for future work. One is the principled selection of the penalty parameter $\lambda$, which is treated here as a user-specified preference parameter. Another is the development of more accurate and efficient estimators, for example through importance sampling, Laplace-based approximations, or multifidelity MC methods. 
A particularly important direction is the extension of the proposed framework to alternative risk functionals beyond the mean--variance criterion, including worst-case risk, entropic risk, and coherent risk measures such as CVaR. While such measures offer stronger theoretical guarantees for risk control and greater flexibility in capturing risk preferences, they introduce additional computational challenges due to nonsmooth objectives and more complex nested expectations. Developing efficient and scalable estimators for these formulations remains an open problem.

%% file: sections/6_appendix.tex
\appendix
\renewcommand{\thesection}{Appendix \Alph{section}}
\renewcommand{\thesubsection}{\thesection.\arabic{subsection}}
\renewcommand{\theequation}{\Alph{section}.\arabic{equation}}
\setcounter{equation}{0}

\section{Bias and variance derivations}

Throughout this appendix, all expectations and variances are understood to be
conditioned on the design $\design$; this conditioning is omitted for brevity.
We derive leading-order asymptotic expansions for the bias and variance of the
MC estimators introduced in the main text.

All results are obtained using conditional delta-method expansions together with
the law of total variance. We assume that the relevant moments exist and that
$\pdf(y|\design)>0$ almost everywhere. Remainder terms are expressed using
$o(\cdot)$ notation as $N,M_1,M_2 \to \infty$. The constants
$A_l(\design)$, $B_l(\design)$, $C_l(\design)$, $D_l(\design)$,
$E_l(\design)$, and $F_l(\design)$ are problem-dependent and correspond to
those introduced in the main text.

\subsection{A conditional delta-method lemma}
\label{app:delta_method}

Let $\widehat{X}$ be an estimator of $\mu \in \mathbb{R}^q$ based on $m$
independent samples, such that
\[
\mathbb{E}[\widehat{X}] = \mu + \delta + o(\epsilon_1),
\qquad
\Var(\widehat{X}) = \Sigma + o(\epsilon_2),
\]
where $\delta \in \mathbb{R}^q$ is the leading-order bias, $\Sigma \in
\mathbb{R}^{q \times q}$ is the leading-order variance, and $\epsilon_1,
\epsilon_2 \to 0$ are the corresponding asymptotic rates. Let
$\phi:\mathbb{R}^q\to\mathbb{R}$ be twice continuously differentiable near
$\mu$. Then
\begin{align}
\mathbb{E}[\phi(\widehat{X})]
&=
\phi(\mu)
+
\nabla\phi(\mu)^\top \delta
+
\frac{1}{2}\tr\!\big(\nabla^2 \phi(\mu)\Sigma\big)
+
o(\epsilon_1) + o(\epsilon_2),
\label{eq:delta_bias}
\\
\Var[\phi(\widehat{X})]
&=
\nabla \phi(\mathbb{E}[\widehat{X}])^\top \Sigma \nabla \phi(\mathbb{E}[\widehat{X}])
+
o(\epsilon_2),
\label{eq:delta_var}
\end{align}
where in \cref{eq:delta_var} the gradient is evaluated at
$\mathbb{E}[\widehat{X}] = \mu + \delta$ for greater accuracy, since the
variance expansion is centered at the true mean of $\widehat{X}$ rather than
at $\mu$. These formulas also hold conditionally, replacing moments by their
conditional counterparts.

\begin{remark}
The lemma is applied with different choices of $\delta$, $\Sigma$,
$\epsilon_1$, and $\epsilon_2$ throughout this appendix. When $\phi$ is applied
to $\widehat{U}(\design)$, the bias $\delta$ is of order $M_1^{-1}$ and the
variance $\Sigma$ decomposes into terms of order $N^{-1}$ and $(NM_1)^{-1}$,
arising from the outer and inner levels of sampling respectively. In such cases
the lemma is applied iteratively---first conditioning on the outer samples,
then on the inner---and the contributions at each level are tracked separately
using the law of total variance.
\end{remark}

\subsection{Bias and variance of $\widehat{U}(\design)^2$}
\label{app:esti_util_squared}

Recall from \cref{e:Uhat_bias} and \cref{e:Uhat_var} that $\widehat{U}(\design)$ satisfies
\begin{align}
\mathbb{E}[\widehat{U}(\design)]
&=
U(\design) + \frac{E_1(\design)}{M_1} + o(M_1^{-1}),
\label{eq:Uhat_mean}
\\
\Var[\widehat{U}(\design)]
&=
\frac{A_1(\design)}{N}
+
\frac{B_1(\design)}{NM_1}
+
o(N^{-1})+o((NM_1)^{-1}).
\label{eq:Uhat_var}
\end{align}
We identify $\delta_1(\design) = E_1(\design)/M_1$,
$\Sigma_1(\design) = A_1(\design)/N + B_1(\design)/(NM_1)$,
with rates $o(\epsilon_1) = o(M_1^{-1})$ and $o(\epsilon_2) = o(N^{-1})+o((NM_1)^{-1})$.

Applying \cref{eq:delta_var} with $\phi(x)=x^2$, so that $\nabla\phi(x)=2x$,
and evaluating the gradient at $\mathbb{E}[\widehat{U}(\design)] =
U(\design)+E_1(\design)/M_1$,
\begin{align}
\Var[\widehat{U}(\design)^2]
&=
4\left[U(\design)+\frac{E_1(\design)}{M_1}\right]^2
\left[
\frac{A_1(\design)}{N}
+
\frac{B_1(\design)}{NM_1}
\right]
+
o(N^{-1})+o((NM_1)^{-1})
\nonumber\\
&=
4U(\design)^2
\left[
\frac{A_1(\design)}{N}
+
\frac{B_1(\design)}{NM_1}
\right]
+
\frac{8U(\design)E_1(\design)A_1(\design)}{NM_1}
+
o(N^{-1})+o((NM_1)^{-1})
\nonumber\\
&=
\frac{A_3(\design)}{N}
+
\frac{B_3(\design)}{NM_1}
+
o(N^{-1})+o((NM_1)^{-1}),
\end{align}
where $A_3(\design)=4U(\design)^2 A_1(\design)$ and
$B_3(\design)=4U(\design)^2 B_1(\design)+8U(\design)E_1(\design)A_1(\design)$.

For the bias, applying \cref{eq:delta_bias} with $\nabla\phi(x)=2x$ and
$\nabla^2\phi(x)=2$, evaluated at $\mu=U(\design)$,
\begin{align}
\mathbb{E}[\widehat{U}(\design)^2 - U(\design)^2]
&=
2U(\design)\cdot\frac{E_1(\design)}{M_1}
+
\frac{A_1(\design)}{N}
+
\frac{B_1(\design)}{NM_1}
+
o(N^{-1})+o(M_1^{-1})
\nonumber\\
&=
\frac{D_3(\design)}{N}
+
\frac{E_3(\design)}{M_1}
+
o(N^{-1})+o(M_1^{-1}),
\end{align}
where $D_3(\design)=A_1(\design)$ and
$E_3(\design)=2U(\design)E_1(\design)$, and the $(NM_1)^{-1}$ cross
term is absorbed into the $o(M_1^{-1})$ remainder.

\subsection{Bias and variance of $\widehat{M}_{2,a}(\design)$}
\label{app:esti_util_mu2_a}

Recall from \cref{e:M2a_hat} that
\[
\widehat{M}_{2,a}(\design)
=
\frac{1}{N}\sum_{i=1}^{N}
\left[\log\widehat{\pdf}(y^{(i)}|\design)\right]^2.
\]
Define
\[
\mu(y)=\pdf(y|\design),
\qquad
\sigma^2(y)=\Var[\pdf(y|\Param^\ast,\design)\mid y],
\]
so that $\widehat{\pdf}(y|\design)$ defined in \cref{e:marginal_likelihood_est}
satisfies
\[
\mathbb{E}[\widehat{\pdf}(y|\design)\mid y]=\mu(y),
\qquad
\Var[\widehat{\pdf}(y|\design)\mid y]=\frac{\sigma^2(y)}{M_1}.
\]
Here $\delta=0$, $\epsilon_1=0$, and $\epsilon_2=M_1^{-1}$, so
$\mathbb{E}[\widehat{\pdf}(y|\design)\mid y] = \mu(y)$ exactly and the gradient
is evaluated at $\mu(y)$ in both \cref{eq:delta_bias} and
\cref{eq:delta_var}.

\paragraph{Step 1: Apply the lemma to $\log\widehat{\pdf}(y|\design)$.}
With $\phi(x)=\log x$, so that $\nabla\phi(x)=x^{-1}$ and
$\nabla^2\phi(x)=-x^{-2}$, evaluated at $\mu(y)$,
\begin{align}
\mathbb{E}[\log \widehat{\pdf}(y|\design)\mid y]
&=
\log \mu(y)
-
\frac{\sigma^2(y)}{2\mu(y)^2 M_1}
+
o(M_1^{-1}),
\label{eq:log_p_bias}
\\
\Var[\log \widehat{\pdf}(y|\design)\mid y]
&=
\frac{\sigma^2(y)}{\mu(y)^2 M_1}
+
o(M_1^{-1}).
\label{eq:log_p_var}
\end{align}

\paragraph{Step 2: Apply the lemma to $(\log\widehat{\pdf}(y|\design))^2$.}
With $\psi(x)=x^2$, so that $\nabla\psi(x)=2x$ and $\nabla^2\psi(x)=2$.
Since $\log\widehat{\pdf}(y|\design)$ has a nonzero bias of order $M_1^{-1}$
from \cref{eq:log_p_bias}, its mean is
$\log\mu(y) - \sigma^2(y)/(2\mu(y)^2 M_1)$, and we evaluate the gradient
in \cref{eq:delta_var} at this mean for greater accuracy:
\begin{align}
\mathbb{E}\!\left[(\log \widehat{\pdf}(y|\design))^2 \mid y\right]
&=
(\log \mu(y))^2
+
\frac{(1-\log \mu(y))\,\sigma^2(y)}{\mu(y)^2 M_1}
+
o(M_1^{-1}),
\label{eq:log_p_sq_bias}
\\
\Var\!\left[(\log \widehat{\pdf}(y|\design))^2 \mid y\right]
&=
4\left[\log\mu(y) - \frac{\sigma^2(y)}{2\mu(y)^2 M_1}\right]^2
\frac{\sigma^2(y)}{\mu(y)^2 M_1}
+
o(M_1^{-1})
\nonumber\\
&=
\frac{4(\log \mu(y))^2\,\sigma^2(y)}{\mu(y)^2 M_1}
+
o(M_1^{-1}),
\label{eq:log_p_sq_var}
\end{align}
where in the last line the correction from the bias term is of order
$M_1^{-2}$ and absorbed into the remainder.

\paragraph{Step 3: Aggregate across outer samples.}
Since the outer samples are independent, applying the law of total variance
gives
\begin{align}
\Var[\widehat{M}_{2,a}(\design)]
&=
\frac{1}{N}
\Var_Y\!\left[
\mathbb{E}\!\left[(\log\widehat{\pdf}(Y|\design))^2\mid Y\right]
\right]
+
\frac{1}{N}
\mathbb{E}_Y\!\left[
\Var\!\left[(\log\widehat{\pdf}(Y|\design))^2\mid Y\right]
\right]
+
o(N^{-1})
\nonumber\\
&=
\frac{A_{5}(\design)}{N}
+
\frac{B_{5}(\design)}{NM_1}
+
o(N^{-1})+o((NM_1)^{-1}),
\end{align}
where the $N^{-1}$ term arises from the outer variability in
$(\log\mu(Y))^2$ via \cref{eq:log_p_sq_bias}, and the $(NM_1)^{-1}$ term
from the inner variability via \cref{eq:log_p_sq_var}. The bias follows
directly from \cref{eq:log_p_sq_bias} by taking the expectation over $Y$:
\begin{align}
\mathbb{E}[\widehat{M}_{2,a}(\design)-M_{2,a}(\design)]
&=
\frac{E_{5}(\design)}{M_1}
+
o(M_1^{-1}).
\end{align}

\subsection{Bias and variance of $\widehat{M}_{2,b}(\design)$}
\label{app:esti_util_mu2_b}

Recall from \cref{e:M2b_hat} that
\[
\widehat{M}_{2,b}(\design)
=
-\frac{2}{N}
\sum_{i=1}^N
\log \pdf(y^{(i)}|\param^{(i)},\design)
\log \widehat{\pdf}(y^{(i)}|\design).
\]
Each summand is a product of the exact term
$\log\pdf(y^{(i)}|\param^{(i)},\design)$ and the estimated term
$\log\widehat{\pdf}(y^{(i)}|\design)$. Since $\log\pdf(y|\param,\design)$ is
exact given $\theta$ and $y$, the only stochasticity in the inner loop enters
through $\log\widehat{\pdf}(y|\design)$, whose conditional mean and variance
are given by \cref{eq:log_p_bias} and \cref{eq:log_p_var}, respectively.

\paragraph{Step 1: Compute the conditional mean and variance given $\theta, y$.}
Since $\log\pdf(y|\param,\design)$ is deterministic given $\theta$ and $y$,
\begin{align}
&\mathbb{E}\!\left[
\log\pdf(y|\param,\design)\log\widehat{\pdf}(y|\design)
\mid\theta,y
\right]
\nonumber\\
&\quad=
\log\pdf(y|\param,\design)
\left[
\log\mu(y)
-
\frac{\sigma^2(y)}{2\mu(y)^2 M_1}
\right]
+
o(M_1^{-1}),
\label{eq:m2b_cond_mean}
\\
&\Var\!\left[
\log\pdf(y|\param,\design)\log\widehat{\pdf}(y|\design)
\mid\theta,y
\right]
\nonumber\\
&\quad=
[\log\pdf(y|\param,\design)]^2
\frac{\sigma^2(y)}{\mu(y)^2 M_1}
+
o(M_1^{-1}).
\label{eq:m2b_cond_var}
\end{align}

\paragraph{Step 2: Aggregate across outer samples.}
Applying the law of total variance to the $N$ independent outer samples,
\begin{align}
\Var[\widehat{M}_{2,b}(\design)]
&=
\frac{4}{N}
\Var_{\Param,Y}\!\left[
\mathbb{E}\!\left[
\log\pdf(Y|\Param,\design)\log\widehat{\pdf}(Y|\design)
\mid\Param,Y
\right]
\right]
\nonumber\\
&\quad+
\frac{4}{N}
\mathbb{E}_{\Param,Y}\!\left[
\Var\!\left[
\log\pdf(Y|\Param,\design)\log\widehat{\pdf}(Y|\design)
\mid\Param,Y
\right]
\right]
+o(N^{-1})
\nonumber\\
&=
\frac{A_{6}(\design)}{N}
+
\frac{B_{6}(\design)}{NM_1}
+
o(N^{-1})+o((NM_1)^{-1}),
\end{align}
where the $N^{-1}$ term arises from the outer variability in
$\log\pdf(Y|\Param,\design)\log\mu(Y)$ via \cref{eq:m2b_cond_mean}, and the
$(NM_1)^{-1}$ term from the inner variability via \cref{eq:m2b_cond_var}.
The bias follows from \cref{eq:m2b_cond_mean} by taking the expectation over
$\Param$ and $Y$:
\begin{align}
\mathbb{E}[\widehat{M}_{2,b}(\design)-M_{2,b}(\design)]
&=
\frac{E_{6}(\design)}{M_1}
+
o(M_1^{-1}).
\end{align}

\subsection{Bias and variance of $\widehat{M}_{2,c}(\design)$}
\label{app:esti_util_mu2_c}

Recall from \cref{e:M2c_hat} that
$\widehat{M}_{2,c}(\design) = N^{-1}\sum_{i=1}^N \widehat{m}(y^{(i)})^2$,
where
\[
\widehat{m}(y)
=
\frac{\widehat{a}(y)}{\widehat{b}(y)},
\quad
\widehat{a}(y)=\frac{1}{M_2}\sum_{k=1}^{M_2}Q(y,\Param^{(k)}),
\quad
\widehat{b}(y)=\frac{1}{M_1}\sum_{j=1}^{M_1}R(y,\Param^{(j)}),
\]
with
\begin{align}
Q(y,\Param)&=\pdf(y|\Param,\design)\log \pdf(y|\Param,\design),
\\
R(y,\Param)&=\pdf(y|\Param,\design),
\end{align}
and independent prior samples $\Param^{(k)},\Param^{(j)}\sim\pdf(\param)$.
Define
\[
a(y)=\mathbb{E}[Q(y,\Param')\mid y],
\quad
b(y)=\pdf(y|\design),
\quad
\sigma_Q^2(y)=\Var[Q(y,\Param')\mid y],
\quad
\sigma_R^2(y)=\Var[R(y,\Param^\ast)\mid y].
\]
Both $\widehat{a}(y)$ and $\widehat{b}(y)$ are unbiased for $a(y)$ and
$b(y)$ respectively, with conditional variances $\sigma_Q^2(y)/M_2$ and
$\sigma_R^2(y)/M_1$.

\paragraph{Step 1: Apply the bivariate lemma to $\widehat{m}(y)=\widehat{a}(y)/\widehat{b}(y)$.}
With $\phi(a,b)=a/b$, we have $\partial\phi/\partial a=b^{-1}$,
$\partial\phi/\partial b=-ab^{-2}$, $\partial^2\phi/\partial a^2=0$,
$\partial^2\phi/\partial b^2=2ab^{-3}$, and $\partial^2\phi/\partial
a\partial b=-b^{-2}$. Since $\widehat{a}$ and $\widehat{b}$ use independent
samples the cross term vanishes, and since $\partial^2\phi/\partial a^2=0$
only the curvature in $\widehat{b}$ contributes to the bias:
\begin{align}
\mathbb{E}[\widehat{m}(y)\mid y]
&=
\frac{a(y)}{b(y)}
+
\frac{a(y)\sigma_R^2(y)}{b(y)^3 M_1}
+
o(M_1^{-1})+o(M_2^{-1}),
\label{eq:mhat_bias}
\\
\Var[\widehat{m}(y)\mid y]
&=
\frac{\sigma_Q^2(y)}{b(y)^2 M_2}
+
\frac{a(y)^2\sigma_R^2(y)}{b(y)^4 M_1}
+
o(M_1^{-1})+o(M_2^{-1}).
\label{eq:mhat_var}
\end{align}

\paragraph{Step 2: Apply the lemma to $\widehat{m}(y)^2$.}
With $\psi(x)=x^2$, so that $\nabla\psi(x)=2x$, and evaluating the gradient
in \cref{eq:delta_var} at $\mathbb{E}[\widehat{m}(y)\mid y] =
a(y)/b(y) + a(y)\sigma_R^2(y)/(b(y)^3 M_1)$ for greater accuracy,
\begin{align}
\mathbb{E}[\widehat{m}(y)^2\mid y]
&=
\left(\frac{a(y)}{b(y)}\right)^2
+
\frac{3a(y)^2\sigma_R^2(y)}{b(y)^4 M_1}
+
\frac{\sigma_Q^2(y)}{b(y)^2 M_2}
+
o(M_1^{-1})+o(M_2^{-1}),
\label{eq:mhat_sq_bias}
\\
\Var[\widehat{m}(y)^2\mid y]
&=
4\left[\frac{a(y)}{b(y)}
+
\frac{a(y)\sigma_R^2(y)}{b(y)^3 M_1}\right]^2
\left[
\frac{\sigma_Q^2(y)}{b(y)^2 M_2}
+
\frac{a(y)^2\sigma_R^2(y)}{b(y)^4 M_1}
\right]
+
o(M_1^{-1})+o(M_2^{-1})
\nonumber\\
&=
\frac{4a(y)^2\sigma_Q^2(y)}{b(y)^4 M_2}
+
\frac{4a(y)^4\sigma_R^2(y)}{b(y)^6 M_1}
+
o(M_1^{-1})+o(M_2^{-1}),
\label{eq:mhat_sq_var}
\end{align}
where in \cref{eq:mhat_sq_bias} the factor of 3 arises from combining the
gradient term $2(a/b)\cdot a\sigma_R^2/(b^3 M_1)$ with the trace term
$a\sigma_R^2/(b^3 M_1)$ from \cref{eq:delta_bias}, and in
\cref{eq:mhat_sq_var} the higher-order correction from the bias in the
gradient evaluation is of order $M_1^{-2}$ and absorbed into the remainder.

\paragraph{Step 3: Aggregate across outer samples.}
Since the outer samples are independent, the law of total variance gives
\begin{align}
\Var[\widehat{M}_{2,c}(\design)]
&=
\frac{1}{N}
\Var_Y\!\left[
\mathbb{E}[\widehat{m}(Y)^2\mid Y]
\right]
+
\frac{1}{N}
\mathbb{E}_Y\!\left[
\Var[\widehat{m}(Y)^2\mid Y]
\right]
+o(N^{-1})
\nonumber\\
&=
\frac{A_{7}(\design)}{N}
+
\frac{B_{7}(\design)}{NM_1}
+
\frac{C_{7}(\design)}{NM_2}
+
o(N^{-1})+o((NM_1)^{-1})+o((NM_2)^{-1}),
\end{align}
where the $N^{-1}$ term arises from the outer variability in
$(a(Y)/b(Y))^2$, while the $(NM_1)^{-1}$ and $(NM_2)^{-1}$ terms
arise from the inner variability via \cref{eq:mhat_sq_var}. The bias
follows from \cref{eq:mhat_sq_bias} by taking the expectation over $Y$:
\begin{align}
\mathbb{E}[\widehat{M}_{2,c}(\design)-M_{2,c}(\design)]
&=
\frac{E_{7}(\design)}{M_1}
+
\frac{F_{7}(\design)}{M_2}
+
o(M_1^{-1})+o(M_2^{-1}).
\end{align}